\ifdefined\PRformat
\documentclass[usenatbib, twocolumn, nofootinbib, superscriptaddress, longbibliography, reprint, prd]{revtex4-1}
\else
\documentclass[usenatbib, twocolumn, twocolappendix, nofootinbib, 
]{aastex701}
\fi

\usepackage{lipsum}
\usepackage{orcidlink}
\usepackage{fdsymbol}
\usepackage{multirow}
\usepackage{bigints}
\usepackage{natbib}
\usepackage{color}
\usepackage{xcolor}
\usepackage{wrapfig}
\usepackage{amsmath,amsfonts,bm}
\usepackage{hyperref}
\usepackage[T1]{fontenc}
\usepackage{graphicx}   
\usepackage[mathlines]{lineno}
\usepackage{wrapfig}
\usepackage{fontawesome}
\usepackage{enumitem}

\newcommand{\cmbsfour}{$\overline{\rm CMB-S4}$}

\newcommand{\sptthreeg}{SPT-3G}

\newcommand{\planck}{{\it Planck}}

\newcommand{\snr}{S/N}

\newcommand{\asob}{ASO-Baseline}
\newcommand{\asog}{ASO-Goal}
\newcommand{\sne}{SNIa}

\newcommand{\cltt}{C_{\ell}^{TT}}
\newcommand{\clee}{C_{\ell}^{EE}}
\newcommand{\clte}{C_{\ell}^{TE}}

\newcommand{\sz}{Sunyaev–Zeldovich}

\newcommand{\desfiveyearsne}{DES-Y5-\sne}

\newcommand{\lsttthreeyearsne}{LSST-Y3-\sne}
\newcommand{\whichlsstsnrsample}{\lsttthreeyearsne}
\newcommand{\howmanysneinlsst}{5800}
\newcommand{\desidata}{DESI-DR2-BAO}
\newcommand{\desidrthreedata}{DESI-DR3-BAO}
\newcommand{\lsstdata}{\whichlsstsnrsample}

\newcommand{\wde}{w_{0}}
\newcommand{\wa}{w_{a}}

\newcommand{\omk}{\Omega_{k}}
\newcommand{\omegam}{\Omega_{m}}
\newcommand{\omchsq}{\Omega_{c}h^{2}}
\newcommand{\ombhsq}{\Omega_{b}h^{2}}
\newcommand{\taure}{\tau_{\rm re}}

\newcommand{\As}{A_{s}}
\newcommand{\loga}{{\rm ln}(10^{10}\As)}
\newcommand{\rdh}{r_{d}H_{0}}

\newcommand{\lcdm}{\Lambda {\rm CDM}}

\newcommand{\wwacdm}{\wde \wa {\rm CDM}}

\newcommand{\mnulcdm}{\summnu \Lambda {\rm CDM}}

\newcommand{\wwamnulcdm}{\wde \wa \summnu {\rm CDM}}
\newcommand{\wwaomklcdm}{\wde \wa \omk {\rm CDM}}
\newcommand{\wwamnuomklcdm}{\wde \wa \summnu \omk {\rm CDM}}
\newcommand{\fsky}{f_{\rm sky}}
\newcommand{\sqdeg}{\rm deg^{2}}

\newcommand{\clpp}{C_{L}^{\phi \phi}}
\newcommand{\clppsignal}{C_{L, {\rm signal}}^{\phi \phi}}
\newcommand{\nlpp}{N_{L}^{\phi \phi}}
\newcommand{\elmin}{\ell_{\rm min}}
\newcommand{\elmax}{\ell_{\rm max}}
\newcommand{\Lmin}{L_{\rm min}}
\newcommand{\Lmax}{L_{\rm max}}
\newcommand{\ukam}{\ensuremath{\mathrm{\mu K\,arcmin}}}
\newcommand{\nzero}{N_{L}^{(0, \phi \phi)}}
\newcommand{\none}{N_{L}^{(1, \phi \phi)}}
\newcommand{\nonehalf}{N_{L}^{(3/2, \phi \phi)}}
\newcommand{\deltael}{\Delta_{\ell}}

\newcommand{\whichplanckgalmask}{\texttt{GAL090}}
\newcommand{\fskyso}{0.4}
\newcommand{\fskysfourfull}{0.67}
\newcommand{\fskysfour}{0.57}
\newcommand{\binningdeltael}{100}
\newcommand{\elminval}{300}
\newcommand{\elmaxval}{3500}
\newcommand{\Lminval}{30}
\newcommand{\Lmaxval}{3500}

\newcommand{\summnu}{\sum m_{\nu}}
\newcommand{\tteete}{TT/EE/TE}
\newcommand{\pp}{\phi \phi}
\newcommand{\tteetepp}{TT/EE/TE/\phi \phi}

\newcommand{\consdesmockomegam}{0.316\pm 0.013}
\newcommand{\conslcdmlsstunbinnedomegam}{0.3152\pm 0.0048}
\newcommand{\conslcdmlsstbinnedomegam}{0.3154 \pm 0.0069}

\newcommand{\abstracttext}{
Observations of Type Ia supernovae (\sne), which probe the late Universe, together with baryon acoustic oscillations (BAO) and the cosmic microwave background (CMB), which probe the intermediate and early epochs, provide complementary constraints on the expansion history of the Universe.
In this work, we forecast constraints on dark energy and other extensions to the standard cosmological model by combining the \sne{} sample expected from the Vera C. Rubin Observatory’s Legacy Survey of Space and Time (LSST), data from current and forthcoming CMB surveys, and BAO measurements from the Dark Energy Spectroscopic Instrument (DESI).
For the CMB, we use temperature, polarization, and lensing power spectra ($\tteetepp$) from South Pole Telescope, the planned Advanced Simons Observatory, and a CMB-S4–like experiment.
We derive constraints on $\lcdm$ and its extensions involving the dark energy equation of state parameters $(\wde, \wa)$ and the sum of neutrino masses $\summnu$, using a Markov Chain Monte Carlo (MCMC) sampling framework. 
We find that the LSST Year‑3 \sne{} sample can improve upon the DES Year‑5 dark energy constraints by a factor of $\times2-\times2.5$, with the gains driven primarily by the significantly higher \sne{} density in the LSST sample. 
Similarly, DESI-DR3 shows up to a $\times1.8$ improvement on dark energy parameters over DR2, driven largely by the substantial increase in low‑redshift sample. 
Combining CMB with \whichlsstsnrsample{} and \desidrthreedata{} yields $\sigma(\wde) = 0.028$ and $\sigma(\wa) = 0.11$ for $\wwacdm$ cosmology with the results being largely independent of the CMB dataset. 
The constraints weaken by 10\%-30\% when freeing $\summnu$ and spatial curvature. 
Moreover, the joint analysis of the three datasets can enable a $2-3\sigma$ detection of $\summnu$.
}

\begin{document}

\title{Probing Physics Beyond the Standard Model through Combined Analyses of Next-Generation Type Ia Supernova, CMB, and BAO Surveys}

\author[0000-0003-1405-378X]{Srinivasan Raghunathan}
\affiliation{Department of Physics, University of California, One Shields Avenue, Davis, CA, 95616, USA}
\affiliation{Center for AstroPhysical Surveys, National Center for Supercomputing Applications, Urbana, IL, 61801, USA}
\email{sriniraghuna@gmail.com}

\author[0000-0002-9436-8871]{Ayan Mitra}
\affiliation{Center for AstroPhysical Surveys, National Center for Supercomputing Applications, Urbana, IL, 61801, USA}
\email{ayan@illinois.edu}

\author[0000-0001-7301-6415]{Nikolina Šarčević}
\affiliation{Department of Physics, Duke University, Durham NC 27708, USA}
\email{nikolina.sarcevic@gmail.com}

\author[0000-0002-3833-8133]{Fei Ge}
\affiliation{California Institute of Technology, 1200 East California Boulevard., Pasadena, CA, 91125, USA}
\affiliation{Kavli Institute for Particle Astrophysics and Cosmology, 382 Via Pueblo Mall, Stanford, California 94305-4060, USA}
\affiliation{SLAC National Accelerator Laboratory, 2575 Sand Hill Road, Menlo Park, California 94025, USA}
\email{feige@caltech.edu}

\author[0000-0002-3500-6635]{Corentin Ravoux}
\affiliation{Université Clermont Auvergne, CNRS, LPCA, 63000 Clermont-Ferrand, France}
\email{corentin.ravoux@clermont.in2p3.fr}

\author[0000-0002-7950-6076]{Christos Georgiou}
\affiliation{Institut de Física d’Altes Energies (IFAE), The Barcelona Institute of Science and Technology, Campus UAB, 08193 Bellaterra (Barcelona), Spain}
\email{cgeorgiou@ifae.es}

\author[0000-0002-0965-7864]{Renée Hložek}
\affiliation{Dunlap Institute for Astronomy and Astrophysics, University of Toronto, Toronto, ON, CA}
\affiliation{David A. Dunlap Department of Astronomy and Astrophysics, University of Toronto, Toronto, ON, CA}
\email{hlozek@dunlap.utoronto.ca}

\author[0000-0003-3221-0419]{Richard Kessler}
\affiliation{Kavli Institute for Cosmological Physics,
University of Chicago, Chicago, IL 60637, USA}
\affiliation{Department of Astronomy and Astrophysics,
University of Chicago, 5640 South Ellis Avenue, Chicago, IL 60637, USA}
\email{kessler@kicp.uchicago.edu}

\author[0000-0001-6022-0484]{Gautham Narayan}
\affiliation{Department of Astronomy, University of Illinois Urbana-Champaign, 1002 West Green Street, Urbana, IL, 61801, USA}
\affiliation{Center for AstroPhysical Surveys, National Center for Supercomputing Applications, Urbana, IL, 61801, USA}
\email{gsnarayan@gmail.com}

\author[0000-0002-1408-6904]{Paul Rogozenski}
\affiliation{McWilliams Center for Cosmology and Astrophysics, Department of Physics,
Carnegie Mellon University, Pittsburgh, PA 15213, USA}
\email{progozen@andrew.cmu.edu}

\author[0000-0002-8000-6642]{Paul Shah}
\affiliation{Department of Physics \& Astronomy, University College London,
Gower Street, London, WC1E 6BT, UK}
\email{paul.shah.19@ucl.ac.uk}

\author[0000-0003-0805-1470]{Georgios Valogiannis}
\affiliation{Department of Astronomy and Astrophysics,
University of Chicago, 5640 South Ellis Avenue, Chicago, IL 60637, USA}
\affiliation{Kavli Institute for Cosmological Physics,
University of Chicago, Chicago, IL 60637, USA}
\email{gvalogiannis@uchicago.edu}

\author[0000-0001-7192-3871]{Joaquin Vieira}
\affiliation{Department of Astronomy, University of Illinois Urbana-Champaign, 1002 West Green Street, Urbana, IL, 61801, USA} \affiliation{Department of Physics, University of Illinois Urbana-Champaign, 1110 West Green Street, Urbana, IL, 61801, USA}
\affiliation{Center for AstroPhysical Surveys, National Center for Supercomputing Applications, Urbana, IL, 61801, USA}
\email{jvieira@illinois.edu}

\collaboration{all}{(the LSST Dark Energy Science Collaboration)}

\correspondingauthor{Srinivasan Raghunathan}\email{sriniraghuna@gmail.com}
\resetfootnotetrue 

\begin{abstract}
\abstracttext{}
\end{abstract}

\ifdefined\PRformat
\maketitle
\fi

\section{Introduction}
\label{sec_intro}
The standard model of cosmology $\lcdm$, has been extremely successful in describing the observations made by a variety of cosmological surveys. 
These observations include the temperature and polarization anisotropies of the cosmic microwave background (CMB, \citealt{planck20cosmo, balkenhol21, louis25, camphuis25}), CMB lensing \citep{omori17, pan23, madhavacheril24, qu25}, baryonic acoustic oscillations (BAO, \citealt{alam21, abdulkarim25}), 
\sne{} \citep{feng05, riess22, des24_sne_cosmology, popovic25}, 
galaxy clustering \citep{alam17, heymans20, abbott22a, miyatake23, adame25}, galaxy weak lensing \citep{joudaki20, abbott23c}, and also cross-correlations between the above datasets \citep{bleem12, giannantonio16, krolewski21, omori23, shaikh24, sabogal25}.
However, the properties of the essential components that make up the model, namely dark matter and dark energy, remain poorly understood. 
Additionally, evidence for tensions between different cosmological probes \citep{joudaki20, riess22, khalife23, abdulkarim25, ong26} has emerged over the past decade, driven by significant improvements in the measurement noise.
Enormous efforts are underway across the community — from the observational side, aiming to identify systematics \citep[][for example]{aylor19, riess22, roychoudhury24}, to the theoretical side, exploring physics beyond the standard model \citep[][for example]{feng05, karwal16, poulin19, lin19, schoneberg19, knox20, roychoudhury24, roychoudhury25, cai25, elbers25a, giare25} — to understand the origin of these tensions, albeit with only partial success.
Consequently, it is critical to use the observations from the current and upcoming state-of-the-art cosmological surveys to test the standard cosmological model to look for extensions as well as deviations. 

In this work, we forecast the cosmological constraints from three probes of the cosmic expansion history expected from next‑generation surveys. These include the Type Ia supernova (\sne) sample anticipated from the Vera C. Rubin Observatory’s Legacy Survey of Space and Time (LSST; \citealt{lsst09, lsstdesc12}); baryon acoustic oscillation (BAO) measurements from the Dark Energy Spectroscopic Instrument (DESI; \citealt{desi16}); and measurements of the CMB from current and forthcoming surveys. 
Because they probe late‑, intermediate‑, and early‑universe epochs, the three datasets provide distinct but highly complementary views of the expansion history.

For the \sne{} dataset, we use the simulated LSST-Y3 sample containing the expected statistical and systematic errors for the distance modulus measurements designed for the Rubin LSST Dark Energy Science Collaboration (DESC\footnote{\url{https://lsstdesc.org/}}, \citealt{mitra22}). 
For the CMB dataset, we use the temperature, polarization, and lensing power spectra ($\tteetepp$) from the South Pole Telescope (SPT-3G, \citealt{benson14, bender18}); the enhanced configuration of the Simons Observatory (SO; \citealt{SO18}), referred to as Advanced-SO (ASO; \citealt{ASO25}); and also a futuristic Stage-IV survey with specifications similar to the recently canceled CMB-S4 experiment \citep{cmbs4-sb1, cmbs4collab19}. 
We consider multiple survey options for each of the above CMB surveys, which we describe in the datasets section. 
Finally, for BAO, we consider multiple distance measurements from both the recently released DESI DR2 dataset \citep{abdulkarim25} and the upcoming mock DR3 data based on \citet{desidr3}. 

The goal of this work can be decomposed into the following points:
\begin{itemize}
    \item{Compare forecasted constraints from the standard Fisher formalism with the results from Markov Chain Monte Carlo (MCMC) sampling for \sne{} and CMB datasets.}
    \item{Study the impact of binning on cosmological constraints. Specifically, (a) to compare the constraints obtained when the CMB power spectra are binned with different values of $\Delta_{\ell}$ with the unbinned case, and similarly (b) to compare the constraints obtained from the unbinned sample with the results when the \sne{} sample in grouped in multiple redshift bins.}
    \item{Forecast and compare cosmological constraints using different combinations of CMB, \sne, and BAO datasets both for the standard $\lcdm$ model and multiple extensions to it namely:}
        \begin{itemize}
            \item {$\wwacdm$ model \citep{chevallier01, linder03} that includes two more parameters for dark energy namely the equation of state ($\wde$) and its evolution ($\wa$);} 
            \item {a model that includes the sum of neutrino masses $\mnulcdm$.}
            \item{We also explore combinations of the above cosmologies along with the case when the spatial curvature ($\omk$) is also varied in order to study the impact of correlations between the parameters in the extended scenarios.}
            \item{Finally, we also show the improvements in constraining power when swapping the \sne{} sample from \desfiveyearsne{} to \whichlsstsnrsample; and the BAO measurements from \desidata{} to \desidrthreedata.}
        \end{itemize}
\end{itemize}~\\
Note that in this work we only model \sne{} and CMB datasets, and use the covariance matrix for BAO measurements as released by the DESI team. 
As a result, we do not model any source of systematic uncertainties in the BAO measurements and only use the BAO dataset to asses the expected improvements in cosmological constraints when they are combined with \sne{} and CMB data. 
We also do not attempt to compare the differences between Fisher vs MCMC nor study the choices of binning for BAO measurements.

The paper is organized as follows. 
In \S\ref{sec_data_exp}, we explain the process to generate mock-data vector and covariance matrices for SNe-Ia, CMB, and BAO datasets along with the noise and foreground modeling for CMB surveys. 
In \S\ref{sec_methods}, we describe the MCMC sampling methods and the Fisher formalism. 
The results and discussion are given in \S\ref{sec_results} followed by the conclusion in \S\ref{sec_conclusion}.

\section{Datasets and experimental setup}
\label{sec_data_exp}

In this section, we describe the mock data vectors for \sne, the CMB power spectra, and BAO, along with a discussion of the corresponding covariance matrices for the three datasets across the experimental surveys considered. 
We additionally incorporate the original data vectors released by \desfiveyearsne{} and \desidata{} for comparison, and we describe these datasets in the corresponding sections below.

\subsection{Mock data vectors and baseline cosmological model}
\label{sec_mock_data_vectors}

\begin{deluxetable*}{| c | c | c | c |}
\tabletypesize{\small}
\def\arraystretch{1.1}
\tablecaption{Fiducial values of the parameters and the associated priors used in this work \citep{planck20_2018cosmo}. 
The priors are either flat in the range $\mathcal{U}[{\rm min, max}]$ or Gaussian, $\mathcal{N}(\mu, \sigma^{2})$, with widths $\sigma$ centered at the mean $\mu$. 
The last column marks the dataset that are sensitive to the parameter under consideration.
}
\label{tab_parameters}
\tablehead{
\multirow{2}{*}{\hspace{3cm}Parameter\hspace{3cm}} & \multirow{2}{*}{Fiducial} & \multirow{2}{*}{Prior} & Constraining \\
& & & dataset
}
\startdata
\hline\hline
\multicolumn{4}{l}{\it Cosmological parameters ($\lcdm$):}\\\hline
Amplitude of scalar fluctuations $\loga$ & 3.044 & $\mathcal{U}[1.6, 3.9]$ & CMB, BAO\\\hline
Dark matter density $\omchsq$ & 0.1200 &  $\mathcal{U}[0.001, 0.99]$ & CMB, \sne{}, BAO \\\hline
Baryon density $\ombhsq$ & 0.02237 & $\mathcal{U}[0.005, 0.1]$ & CMB, \sne, BAO  \\\hline
Scalar spectral index $n_{s}$ & 0.9649 & $\mathcal{U}[0.8, 1.2]$ & \multirow{2}{*}{CMB, BAO}\\\cline{1-3}
Hubble parameter $h$ & 0.6732 & $\mathcal{U}[0.4, 1]$ & \\\hline
Reionization optical depth $\taure$ & 0.0544 & $\mathcal{N}(0.0544, 0.007^{2})$ & CMB \\\hline\hline 
\multicolumn{4}{l}{\it Extensions:}\\\hline
Sum of neutrino masses $\summnu$ & \multirow{2}{*}{0.06 [eV]} & \multirow{2}{*}{$\mathcal{U}[0, 0.5]$} & \multirow{2}{*}{CMB, \sne, BAO} \\
(Normal hierarchy) & & & \\\hline
Equation of state (EoS) of dark energy $\wde$ & -1 & $\mathcal{U}[-10, 10]$ & \multirow{2}{*}{CMB, \sne{}, BAO} \\\cline{1-3}
Time varying dark energy EoS $\wa$ & 0 &  $\mathcal{U}[-30, 20]$ & \\\hline
Spatial curvature $\omk$ & 0 &  $\mathcal{U}[-0.3, 0.3]$ & CMB, \sne, BAO \\\hline\hline
\multicolumn{4}{l}{\it Nuisance parameters:}\\\hline
 SNIa absolute magnitude $\mathcal{M}$ & -19.5 & $\mathcal{U}[-15, -25]$ & \sne \\\hline 
 CMB temperature calibration$^{\dagger}$ $T_{\rm cal}^{\nu}$ & 1 & $\mathcal{N}(1, 0.01^{2})$ & CMB \\\hline
 CMB polarization calibration$^{\dagger}$ $P_{\rm cal}^{\nu}$ & 1 & $\mathcal{N}(1, 0.01^{2})$ & CMB \\\hline
 \enddata
\tablecomments{\dagger\ - The total number of parameters depends on the number of frequency bands in the CMB experiment. The calibration factors are applied to the theory spectra during the sampling step as described by Eq.(\ref{eq_tcal_pcal}). The priors are chosen based on some of the recent works in the literature \citep{balkenhol21, louis25, camphuis25}.}
\end{deluxetable*}

We generate the mock data vectors assuming a standard 6-parameter $\lcdm$ model as the baseline model based on \planck{} 2018 measurements (TT, TE, EE + lowE + lensing in Table 2 of \citealt{planck20_2018cosmo}). 
The data vectors include the appropriate distance measurements for \sne{} and BAO, and also CMB power spectra measurements. 
These data vectors are discussed in more detail in the subsequent sections. 
To generate CMB power spectra, we use the Code for Anisotropies in the Microwave Background (CAMB, \citealt{lewis00}) software. 
In Table~\ref{tab_parameters}, we list the fiducial values of all the parameters used in this work. 
We also provide the values used for extended cosmology scenarios that include dark energy and sum of neutrino masses assuming a normal ordering with a fiducial value set to $\summnu = 0.06 {\rm eV}$ based on the lower limits from the neutrino oscillation experiments. 
We also assess the impact of allowing the spatial curvature to vary when considering the above extensions, although we do not focus exclusively on the resulting constraints on spatial curvature in this work. 
When using \sne{} data alone, we restrict the analysis to $\lcdm$ and $\wwacdm$ cases since \sne{} data 
do not constrain $\summnu$. 
The table also lists the priors used for MCMC formalism discussed in \S\ref{sec_mcmc}. 
Besides the fiducial cosmology discussed here, in Appendix~\ref{app_constraints_and_cosmo_dep} we briefly examine how the constraining power depends on the assumed cosmology by regenerating the constraints for the \desfiveyearsne{} \citep{des24_sne_cosmology} and DES-Dovekie \citep{popovic25} cosmologies.

\subsection{\whichlsstsnrsample{} data}
\label{sec_sne_data}
\begin{figure*}
\centering
\includegraphics[width=0.85\textwidth, keepaspectratio]{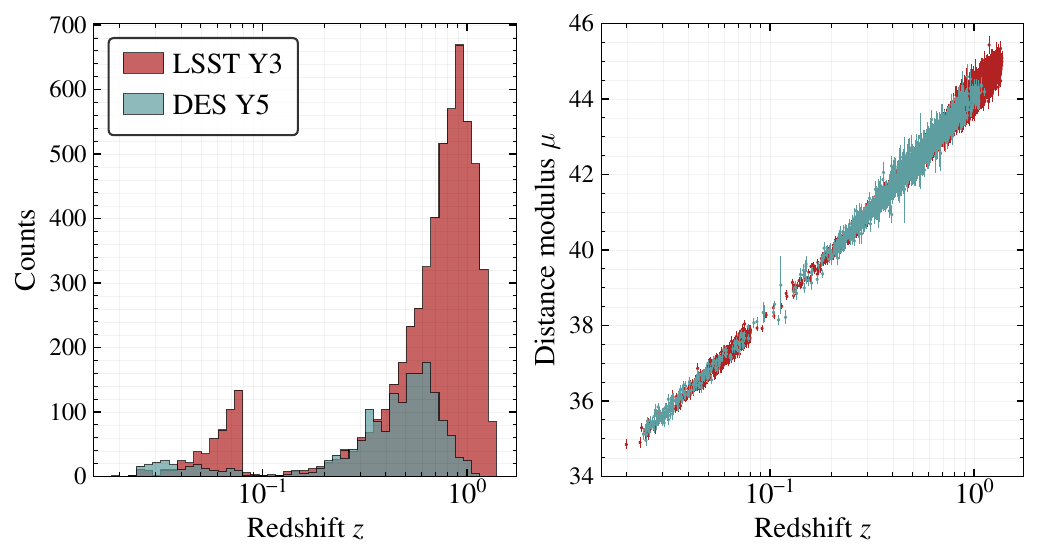}
\caption{Redshift distribution (left) and the distance modulus as a function of redshift (right) are presented for \sne{} samples used in this work. 
Red distribution represents the \whichlsstsnrsample{} mock dataset and we have also shown \desfiveyearsne{} in teal for comparison. 
The sample size of \whichlsstsnrsample{} is roughly $\times3.3$ larger than \desfiveyearsne. 
}
\label{fig_lsst_sne_details}
\end{figure*}

In this work, we use the mock \sne{} catalog from the first three years of LSST data (\whichlsstsnrsample) generated in \citet{mitra22} and we refer the reader to that work for more details. 
Here, we briefly summarize the steps to obtain the mock \sne{} catalog. 
The simulations used for generating the mock catalog are based on the time domain pipeline using {\tt PLAsTiCC} data \citep{kessler19, hlozek20} that were originally designed for the 
Rubin LSST-DESC.
The mock-data generation can be decomposed into three main stages:
(a) standardizing the SNe brightness with the light-curve fitting using \textsc{SALT2} \citep{guy07}; 
(b) generating simulations necessary for Bayesian Estimation Applied to Multiple Species (BEAMS) with Bias Corrections (BBC, \citealt{kessler17}); and finally
(c) constructing the bias-corrected Hubble diagram data along with the statistical and systematic covariance matrices.
The simulations adopt standard flat $\lcdm$ cosmology using the fiducial values listed in Table~\ref{tab_parameters}.  
The noise model for the simulated magnitudes includes contributions from photon noise, calibration errors, and uncertainties in the properties of host-galaxies of \sne, as well as an empirically determined signal-to-noise ($\snr$) ratio. 

The mock \sne{} catalog contains objects whose host-galaxy redshifts $z$ are available either from spectroscopy or from photometry; we refer to these subsamples as $z_{\rm spec}$ and $z_{\rm phot}$, respectively \citep{mitra22}.
Note that these labels refer solely to how the redshift is obtained, not to how the supernova type is determined. 
The $z_{\rm phot}$ sample, as expected, is significantly ($\sim60\%$) larger than the $z_{\rm spec}$ subset, and uses the photometric redshift estimate of the host-galaxy as a prior, adapted from \citet{kessler10}, during the light curve fitting process. 
The $z_{\rm phot}$ sample makes up most of high redshift $z > 0.08$ objects in the catalog although a minor fraction of spectroscopic events found in the {\tt PLAsTiCC} subset are also present in the $z > 0.08$ sample. 
The objects at low-$z$ ($z \le 0.08$) come from the spectroscopic sample, predicted to be observed by the 4MOST spectrograph \citep{frohmaier25} with an expected redshift precision of $\sigma_{z} \sim 10^{-5}$. 
The low-$z$ $z_{\rm spec}$ sample is simulated using the Data Challenge-2 (DC2, \citealt{sanchez22}) with a cadence corresponding to the Wide Fast Deep (WFD\footnote{\url{https://survey-strategy.lsst.io/baseline/wfd.html}}) observations; and the high-$z$ simulation is based on the cadence of observations for the Deep Drilling Fields (DDF\footnote{\url{https://survey-strategy.lsst.io/baseline/ddf.html}}). 
The details of the simulation can be found in \S III of \citet{mitra22}. 

In total, the \whichlsstsnrsample{} dataset used in this analysis contains 5784 \sne{} (3551 are photometric and 2233 are spectroscopic objects). 
We present the redshift distribution and the errors on the distance modulus as a function of redshift in red on the left and right panels of Fig.~\ref{fig_lsst_sne_details}. 
The errors include both the statistical and systematic contributions. 
The \whichlsstsnrsample{} consists of $\sim 1200$ (20\%) of the objects at $z \ge 1$. 

In the past, analyses using the BBC approach have relied on redshift‑binned \sne{} samples for cosmological inference to reduce computation time, anticipating much larger datasets in the future \citep[][for example]{popovic21}. 
However, \citet{brout21} showed that although the statistical covariance of a binned sample can be tuned to match that of an unbinned sample, binning generally leads to larger systematic errors. 
A subsequent study by \citet{kessler23} introduced a rebinning scheme based on redshift, stretch, and color, and demonstrated through simulations that this approach reduces systematic errors relative to redshift‑only binning. 
Nevertheless, a fully optimal binning strategy capable of reducing uncertainties to the level of the unbinned case is yet to be established.
Subsequently, other recent works \citep{vincenzi24, des24_sne_cosmology, popovic25} have worked with the unbinned sample for cosmological inference. 
Besides the fiducial unbinned \whichlsstsnrsample{} sample, we also forecast cosmological constraints using an alternative \whichlsstsnrsample{} sample binned following \citet{kessler23} into 14 redshift bins.\\

\noindent
{\bf Comparison with \desfiveyearsne:}
Beyond \whichlsstsnrsample{}, which constitutes the baseline \sne{} sample in this analysis, we additionally employ the \desfiveyearsne{} sample \citep{vincenzi24, des24_sne_cosmology, popovic21} for comparison. 
For this dataset, we use both the original \desfiveyearsne{} data vector and a mock data vector generated using the same prescription applied to the \whichlsstsnrsample{} sample described above. 
The covariance matrix is adopted directly from the original \desfiveyearsne{} release.
The \whichlsstsnrsample{} \sne{} sample is $\times3.3$ larger than the \desfiveyearsne{} \sne, which is also shown in the Fig.~\ref{fig_lsst_sne_details} in teal for comparison. 
The \desfiveyearsne{} consists of 1754 objects after removing the false positives (1829 in total) in the range $z \in [0.025, 1.12]$. 
The mean redshift $\bar{z}$ of the \desfiveyearsne{} sample is $\bar{z} = 0.46$ while for the \whichlsstsnrsample, it is $\bar{z} = 0.72$. 

\subsection{CMB data}
\label{sec_cmb_data}
\begin{deluxetable*}{| c || c || c | c | c | c | c | c |}
\tabletypesize{\small}
\tablecaption{Experimental beam, sky fraction, and noise levels for different surveys considered in this work. For polarization noise, we use $\Delta_{P} = \sqrt{2}\Delta_{T}\ \ukam$. 
} 
\label{tab_cmb_experiments}
\tablehead{
\multirow{2}{*}{\hspace{1cm}Survey\hspace{1cm}} & Sky fraction & \multicolumn{6}{c|}{Beam $\theta_{\rm FWHM}$ in arcminutes (Noise $\Delta_{T}$ in $\ukam$)} \\
\cline{3-8}
& $\fsky$ & 30 GHz & 40 GHz & 90 GHz & 150 GHz & 220 GHz & 280 GHz
}
\startdata
\hline \hline
\sptthreeg\ Main & 0.036 & \multirow{3}{*}{ - } & \multirow{3}{*}{ - } & 1.7 (2.5) & 1.2 (2.1) & 1 (7.6) & \multirow{3}{*}{ - } \\
\sptthreeg\ Summer & 0.065 & & & 1.7 (9.4) & 1.2 (8.7) & 1 (29.5) & \\
\sptthreeg\ Wide & 0.145 & & & 1.7 (14) & 1.2 (12) & 1 (42) & \\\hline\hline
\asob & \multirow{2}{*}{ 0.4 } & 7.4 (61) & 5.1 (30) & 2.2 (5.3) & 1.4 (6.6) & 1 (15) & 0.9 (35) \\
\asog & & 7.4 (44) & 5.1 (23) & 2.2 (3.8) & 1.4 (4.1) & 1 (10) & 0.9 (25) \\\hline\hline
\cmbsfour & 0.57 & 7.8 (27.1) & 5.3 (11.6) & 2.2 (2) & 1.4 (2) & 1 (6.9) & 0.9 (16.7) \\\hline\hline
\enddata
\end{deluxetable*}
\begin{deluxetable*}{| c || c | c | c | c | c | c |}
\tabletypesize{\small}
\tablecaption{Atmospheric noise specifications for $TT/EE$ power spectra measurements from SPT and \cmbsfour{} surveys. 
Atmospheric noise is uncorrelated between CMB temperature and polarization, and hence $TE$ is assumed to be zero. 
The values for SPT are derived from actual observations \citep{prabhu24} and for \cmbsfour, the values are obtained from \citet{cmbs4collab19}.
For ASO, we use the publicly available noise calculator (\url{https://github.com/simonsobs/so_noise_models}) as mentioned in the text.}
\label{tab_atmnoise_specs}
\tablehead{
\multirow{2}{*}{\hspace{2cm}Survey\hspace{2cm}} & \multicolumn{6}{c|}{$\ell_{{\rm knee}}, \alpha_{{\rm knee}}$} \\
\cline{2-7}
& 30 GHz & 40 GHz & 90 GHz & 150 GHz & 220 GHz & 280 GHz
}
\startdata
\hline \hline
\multicolumn{7}{l|}{\it CMB temperature $TT$ spectra:}\\\hline
\sptthreeg\ Main & \multirow{3}{*}{ - } & \multirow{3}{*}{ - } & 1200, -3 & 2200, -4 & 2300, -3 & \multirow{3}{*}{ - } \\\cline{4-6}
\sptthreeg\ Summer & & & \multirow{2}{*}{ 1600, -4.5} & \multirow{2}{*}{2600, -4} & \multirow{2}{*}{ 2600, -3.9} & \\
\sptthreeg\ Wide & & & & &  & \\\hline\hline
\cmbsfour & 1200, 4.2 & 1200, 4.2 & 1200, 4.2 & 1900, 4.1 & 2100, 4.1 & 2100, 3.9 \\
\hline \hline
\multicolumn{7}{l|}{\it CMB polarization $EE$ spectra:}\\\hline
\sptthreeg\ Main & \multirow{3}{*}{ - } & \multirow{3}{*}{ - } & 300, -1 & 300, -1 & 300, -1 & \multirow{3}{*}{ - } \\\cline{4-6}
\sptthreeg\ Summer & & & \multirow{2}{*}{300, -2.2} & \multirow{2}{*}{490, -2} & \multirow{2}{*}{500, -2.5} & \\
\sptthreeg\ Wide & & & & &  & \\\hline\hline
\cmbsfour & 150, 2.7 & 150, 2.7 & 150, 2.6 & 220, 2.2 & 200, 2.2 & 200, 2.2 \\\hline\hline
\enddata
\end{deluxetable*}

We consider three different experiments for CMB data: the ongoing South Pole Telescope (SPT-3G) survey \citep{benson14, bender18, prabhu24, vitrier25}; 
ASO \citep{ASO25}; and a future CMB survey matching the specifications of the CMB-S4 survey \citep{cmbs4collab19} which we represent as \cmbsfour. 
The observables for CMB are the temperature, polarization, and lensing power spectra $TT, EE, TE, \phi \phi$. \\

\noindent
{\bf SPT-3G:} For SPT-3G, we consider the Ext-10k survey described in \citet{prabhu24, vitrier25} that covers $\fsky = 0.25$ (10,000 $\sqdeg$) of the southern sky excluding regions contaminated by the galactic foregrounds. 
This Ext-10k survey is composed of three different surveys with different noise properties and sky fractions: (a) the deep SPT-3G Main field comprising of $\fsky = 0.036$ (1500 $\sqdeg$); (b) the summer fields covering roughly $\fsky = 0.065$ (2650 $\sqdeg$); and (c) the wide fields with a sky fraction of roughly $\fsky = 0.145$ (6000 $\sqdeg$). 
\\

\noindent
{\bf ASO:} For ASO, we perform forecasts for two configurations: \asob{} and \asog{} \citep{ASO25}.  
For both these configurations, we assume a sky fraction of $\fsky = \fskyso$, which is roughly 16,500 $\sqdeg$ after excluding the regions contaminated by the galactic foregrounds \citep{SO18, ASO25}.\\

\noindent
{\boldsymbol \cmbsfour:} For \cmbsfour, we only consider a wide-area survey expected to be carried out in Chile similar to the one originally planned by the CMB-S4 experiment with $\fsky = \fskysfourfull$ \citep{cmbs4collab19}. 
We remove the regions with high levels of galactic foreground emission, using \planck's \whichplanckgalmask{} mask and assume a remaining sky fraction of $\fsky = \fskysfour$ \citep{cmbs4collab19} for the subsequent calculations here.

\subsubsection{Noise and foreground modeling}
For all the above experiments, we use data from the multiple frequency bands. 
The noise in the individual bands receives contribution from the detector white noise and atmospheric noise, and the noise power spectrum is modeled as
\begin{equation}
 N_\ell^{XX} =  \Delta_{X}^2 \left[1 + \left(\frac{\ell}{\ell_{\rm knee}}\right)^{\alpha_{\rm knee}}\right]\ ,
\label{eq_cmb_noise_model}
\end{equation}
where $\Delta_{X}$ corresponds to the white noise levels in the temperature ($T$) and polarization ($P$) maps; and ($\ell_{\rm knee}, \alpha_{\rm knee}$) are used to model the contribution from the atmospheric noise. 
In Table~\ref{tab_cmb_experiments}, we list the values used for noise along with the experimental beam and sky fractions. 
In Table~\ref{tab_atmnoise_specs}, we give the values used for atmospheric noise modeling for \sptthreeg{} and \cmbsfour. 
These are based on \citet{cmbs4collab19, raghunathan23, prabhu24}.
For ASO, we use the publicly available noise model downloaded from this link\footnote{\url{https://github.com/simonsobs/so_noise_models}}.

Since the CMB maps receive contributions from extragalactic foreground signals we also include them in our analysis.
They can be decomposed into contributions from radio and dusty star-forming galaxies, as well as the kinematic and thermal \sz{} (kSZ and tSZ) effects. 
These signals are modeled using measurements from the SPT \citep{reichardt21}. 
In the modeling scheme, we have ignored the contribution from bright point sources with flux $S_{150} \ge 3 {\rm mJy}$ at 150 GHz and also clusters detected above a signal-to-noise $\snr \ge 5$ for all the experiments \citep{raghunathan22}.
We refer the reader to \citet{raghunathan23} for details about the details on modeling the temperature portion of the extragalactic foreground signals. 
For polarization, we assume a 3\% polarization fraction for radio and dusty galaxies based on \citet{datta18, gupta19} but ignore the polarized SZ contributions. 

We do not include the galactic foreground signals, since we are already applying a conservative cut in our sky fraction to remove the contaminated regions (See Table \ref{tab_cmb_experiments}). 
In the unmasked region, the galactic foregrounds are expected to be subdominant, and including them has a negligible effect on the cosmological constraints \citep{prabhu24}.
Nevertheless, we introduce a conservative scale cut $\ell_{\rm min} = 300$ to remove large-scale modes where the contribution from the galactic foreground signals could be generally significant.

\subsection{Binning, multipole ranges, lensing noise and covariance matrix} 
The information from all the bands are optimally combined, using internal linear combination (ILC) technique, to reduce the overall variance from noise and foreground signals \citep{cardoso08, raghunathan23}. 

During the ILC step, we also take the frequency-dependent temperature $T_{\rm cal}^{\nu}$ and polarization $P_{\rm cal}^{\nu}$ calibration factors into account. 
The calibration factors are obtained by cross-correlating the CMB maps from a given survey with \planck{} maps and defined as \mbox{$X_{\rm cal}^{\rm survey} = \sqrt{ C_{\ell}^{{\rm survey} \times \text{\planck} } / C_{\ell}^{\text{\planck}_{1} \times \text{\planck}_{2} }}$} with \mbox{$X \in [T,P]$} \citep[][for example]{camphuis25}.
The subscripts \planck$_{1}$ and \planck$_{2}$ indicate the two \planck{} datasets with mutually uncorrelated noise to avoid the noise bias in the denominator.
The calibration is performed separately for each frequency band, either by cross‑correlating with the corresponding \planck{} band or by determining a calibration factor for a single band (typically 150 GHz) using the \planck{} cross‑correlation, after which the remaining bands are calibrated internally.
As we will see below in \S\ref{sec_methods}, we allow the calibration parameters to vary during the fitting process and marginalize over them to obtain cosmological constraints. 
The calibration factors are applied to the theory spectra $\cltt, \clee, \clte$ as:
\begin{eqnarray}
    C_{\ell}^{TT, \nu_{1}\nu_{2}} = & \frac{C_{\ell}^{TT, \nu_{1}\nu_{2}}}{T_{\rm cal}^{\nu_{1}} T_{\rm cal}^{\nu_{2}}},\notag\\
    C_{\ell}^{EE, \nu_{1}\nu_{2}} = & \frac{C_{\ell}^{EE, \nu_{1}\nu_{2}}}{P_{\rm cal}^{\nu_{1}} P_{\rm cal}^{\nu_{2}}},\notag\\
    C_{\ell}^{TT, \nu_{1}\nu_{2}} = & \frac{2 C_{\ell}^{TE, \nu_{1}\nu_{2}}}{ T_{\rm cal}^{\nu_{1}} P_{\rm cal}^{\nu_{2}} + T_{\rm cal}^{\nu_{2}} P_{\rm cal}^{\nu_{1}}}
\label{eq_tcal_pcal}
\end{eqnarray}

The ILC technique is equivalent to the inverse variance combination of data from all the bands but after taking into account the correlation between different frequency bands, caused by foreground signals, into account. 
In the absence of correlation between frequency bands, this simplifies to $N_{\ell}^{\rm ILC} = \sum_{i} \dfrac{1}{N_{\ell}^{\nu_{i}}}$. 
The ILC residuals $N_{\ell}^{\rm ILC}$ contain residual signals from both noise and foregrounds components.

We combine the ILC residuals with the signal power spectra $C_{\ell}^{TT}\ /\ C_{\ell}^{EE}\ /\ C_{\ell}^{TE}$ to compute the covariance matrix, as described below. 
To obtain the final observed power spectra (the data vector) for different experiments, we use only the signal power spectra. 
Consequently, we do not marginalize over any foreground parameters. 
We note that this choice may lead to slightly optimistic constraints for CMB $TT$-only spectra, but it should have a negligible impact once $EE/TE/\phi\phi$ measurements and other external datasets are included.
The spectra are binned with $\Delta_{\ell} = \binningdeltael$ with uniform weights for all multipoles in a conservative $\ell$ range $[\elmin, \elmax] = [\elminval, \elmaxval]$ for all the experiments. 
We compare and quantify the constraining power with different binning schemes in \S\ref{sec_fisher_vs_mcmc}.

We adopt the same ranges for lensing reconstruction using the standard lensing quadratic estimator \citep{hu02, okamoto03} and compute the lensing noise. 
The lensing noise spectrum can be decomposed into \mbox{$\nlpp = \nzero + \none + \nonehalf + ... \equiv \nzero$} \citep{madhavacheril20b}. 
While the higher order biases are important to obtain an unbiased lensing power spectrum, particularly for the current and future low-noise CMB datasets, they are much smaller ($\lesssim 2\%$) compared to $\nzero$ which dominates the error budget. 
Subsequently, we only consider $\nzero$ as the source of lensing noise.
The sQE is known to be sensitive to the contamination from the non-Gaussian astrophysical foregrounds on small-scales \citep{madhavacheril18, ferraro18}, and several modified lensing estimators have been introduced to robustly reconstruct lensing at the expense of increased statistical errors \citep{namikawa13, osborne14, madhavacheril18, schaan19, sailer23, raghunathan23}. 
In this work, we do not use these modified lensing estimators but employ a stringent cut on the maximum multipole $\ell = \elmaxval$ used for lensing reconstruction to mitigate the contamination from foreground signals. 
We note that this choice of $\elmax = \elmaxval$ is conservative for polarization channels since the extragalactic foregrounds are largely unpolarized \citep{gupta19, datta18}.
Similar to the CMB power spectra, we bin the lensing power spectrum $\clpp = \clppsignal + \nlpp$ with $\Delta_{L} = \binningdeltael$ in the range $[\Lmin, \Lmax] = [\Lminval, \Lmaxval]$. 
Note that we use the subscript $L$ to represent the modes of the lensing power spectrum rather than the $\ell$ that is used for $\tteete$ spectra. 
This is to differentiate between the reconstructed lensing information using the lensing-induced correlation between two Fourier modes $\ell$ and $\ell^{\prime}$ with $L = \ell - \ell^{\prime}$ \citep{bernardeau96, okamoto03}. 
This also let us reconstruct the largest scale modes and hence we set $L_{\rm min} = 30$ for the lensing power spectrum.

The covariance matrix $\Sigma_{\ell, \ell^{\prime}}$ has the length $N_{\rm bins} \times N_{\rm obs}$ on each side with the total number of multipole bins $N_{\rm bins} = \sum_{_{XX}} N_{\rm bins}^{XX}$ and $N_{\rm obs}$ corresponds to the total number of power spectra \mbox{$XX \in [TT, EE, TE, \pp]$}.
The diagonal blocks of the covariance matrix at each multipole $\ell$ is given as
\begin{widetext}
    \begin{align}
        \Sigma_{\ell, \ell}^{XY} & = \dfrac{2}{\left(2\ell+1\right)f_{\rm sky}} \begin{pmatrix}
                (\tilde C_\ell^{TT})^2 & (C_\ell^{TE})^2 & \tilde C_\ell^{TT} C_\ell^{TE} & 0 \\
                (C_\ell^{TE})^2 & (\tilde C_\ell^{EE})^2 & \tilde C_\ell^{EE} C_\ell^{TE} & 0 \\
                \tilde C_\ell^{TT} C_\ell^{TE} & \tilde C_\ell^{EE} C_\ell^{TE} & \frac{\left[(C_\ell^{TE})^2 + \tilde C_\ell^{TT}\tilde C_\ell^{EE}\right]}{2} & 0 \\   
                0 & 0 & 0 & (\tilde C_{L}^{\pp})^2        
        \end{pmatrix}.
            \label{eq_cmb_covariance}
    \end{align}
\end{widetext}
where $C_{\ell}^{XX}$ and $C_{L}^{\pp}$ corresponds to the lensed CMB and CMB-lensing power spectra. 
We set the off-diagonal blocks, representing correlation between adjacent multipole bins, $\Sigma_{\ell, \ell^{\prime}}^{XY} = 0$. 
While there exists a non-zero correlation between lensed CMB and CMB-lensing spectra, they are expected to be small and hence we ignore them \citep{trendafilova23}.
We note that the exclusion of this term can result in slightly ($\sim 10\%$) optimistic parameter constraints but it has been found to have negligible impact when external datasets, as is the case in this work, are included \citep{peloton17}.
In Eq.(\ref{eq_cmb_covariance}), $f_{\textrm{sky}}$ is the sky fraction of the CMB survey given in Table~\ref{tab_cmb_experiments}. 
The fields with the tilde symbol represent the combination of signal and noise spectra as
\begin{eqnarray}
    \tilde C_\ell^{XX} = C_\ell^{XX} + N_\ell^{{\rm ILC}, XX}, \\ 
    \tilde C_L^{\pp} = C_L^{\pp} + \nzero.
    \label{eq_cl_plus_nl}
\end{eqnarray} where signal spectra correspond to $XX \in [TT, EE]$ and the noise spectra $N_{\ell}^{{\rm ILC},TT}$ and $N_{\ell}^{{\rm ILC}, EE}$ are the residual noise and extragalactic foreground signals after the ILC step. 
Note that we do not represent $C_\ell^{TE}$ with the tilde since (a) the noise is uncorrelated between the CMB temperature and polarization maps, (b) we are ignoring the impact of galactic foregrounds, and (c) the polarization fraction of the foreground signals has been found to be negligible \citep{datta18, gupta19}. 
Including the small $TE$ correlation due to the non-zero (3\%) polarization fraction assumed for extragalactic sources has no impact on our final results. 

\subsection{BAO data}
\label{sec_bao_data}
\begin{figure}
\centering
\includegraphics[width=0.48\textwidth, keepaspectratio]{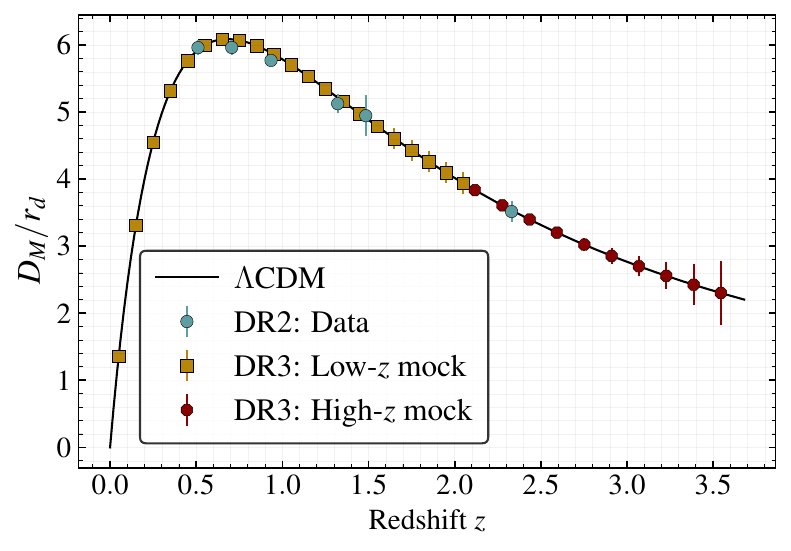}
\caption{Distance measurements $D_{M}/r_{d}$ from DESI DR2 release (teal circles, \citealt{abdulkarim25}) along with the expectations from the next DR3 release. 
For DR3, we show the low- (high-redshift) measurements from the clustering of galaxies and quasars (Ly-$\alpha$ forest) at $z \lesssim 2$ ($z>2$) in yellow squares (red hexagons). 
The DR2 values are actual measurements while the DR3 values are expectations from $\lcdm$ (black curve) along with the forecasted errors.
}
\label{fig_desi_dr2_dr3_data}
\end{figure}

For the BAO measurements, we use the recently released DESI DR2 measurements \citep[Table IV of][]{abdulkarim25} and also the forecast for the upcoming DR3 release \citep{desidr3} which aims to extend the measurements to significantly higher redshifts ($z_{\rm max}^{\rm DR3} = 3.55$) compared to DR2 ($z_{\rm max}^{\rm DR2} = 2.05$). 
The forecasted data vectors are based on clustering of galaxies and quasars at low redshifts $z \lesssim 2$ and the clustering of Ly-$\alpha$ forest at higher redshifts $z \ge 2$ identified from the spectra of quasars.

The main observables measured by BAO are the distance ratio $D_{\rm H}(z) / r_{\rm d} = \dfrac{c}{H(z) r_{\rm d}}$ and $D_{\rm M}(z) / r_{\rm d}$ \citep{abdulkarim25} where $H(z)$ is the Hubble function; and $D_{\rm H}$ and $D_{\rm M}$ are the Hubble and comoving transverse distances \citep{hogg99}, both expressed in terms of the scale of the sound horizon $r_{\rm d}$ in Mpc. 
If the galaxy clustering measurement shape is fully interpreted, it gives a redshift space distortion measurement of $f \sigma_{8}(z)$\footnote{Only forecasted for \desidrthreedata{} at the moment.} which is the product of the growth rate $f$ and the amplitude of matter density fluctuations at $8 h^{-1}{\rm Mpc}$ scale $\sigma_8$.
At any given redshift, there are significant correlations between all the above observables, and these are taken into account in our analysis using the covariance matrix \citep[see Table IV of][for example]{abdulkarim25}. 

In Fig.~\ref{fig_desi_dr2_dr3_data}, we show the actual measurements $D_{M}/r_{d}$
from DESI DR2 release as teal circles along with the forecasts matching the specifications for the upcoming DESI DR3 release. 
For DR3, the yellow squares correspond to the low redshift clustering from galaxies and quasars, and the red hexagons correspond to the measurements of the clustering of Ly-$\alpha$ forest at high redshifts. 
As it is evident from Fig.~\ref{fig_desi_dr2_dr3_data}, the DR3 release has a lot more high redshift Ly-$\alpha$ measurements compared to DR2 and these have the potential to significantly improve the cosmological constraints as we will see below. 
For the forecasting purposes, we use the covariance matrix released by the DESI collaboration, without any scaling, but create our own data vectors both for the DR2 and DR3 datasets assuming the fiducial cosmology \citep{planck20cosmo} to match the analysis choices of CMB and SNe datasets. 
The black curve in Fig.~\ref{fig_desi_dr2_dr3_data} shows the $\lcdm$ expectation. 

For reference, we list the forecasted data vectors used in this work in Appendix~\ref{app_desi_data_vectors}. 
The forecasted data vectors for \desidata{} are in Table~\ref{tab_desi_dr2_data_vectors} with the covariance being the same as listed in Table 1 of \citet{abdulkarim25}.
For \desidrthreedata, the data vectors are decomposed into low and high redshift measurements, and are listed in 
Table~\ref{tab_desi_dr3_data_vectors_highz} and Table~\ref{tab_desi_dr3_data_vectors_lowz}. 
Besides the data vectors, the tables also contain the correlations between observables $A$ and $B$ as 
$\rho(A,B)$ where $A, B \in [H_{z}r_{d}, D_{A}/r_{d}, f{\sigma_{8}}]$.

\section{Methodology}
\label{sec_methods}

\subsection{MCMC}
\label{sec_mcmc}

We use the Code for BAYesian Analysis (\texttt{cobaya}\footnote{\url{https://cobaya.readthedocs.io/en/latest/}}, \citealt{torrado19}) to sample the posteriors using the Metropolis-Hastings MCMC algorithm. 
For theory calculations, we make use of the CAMB  \citep{lewis00} software, the Einstein-Boltzmann solver, which is inbuilt within \texttt{cobaya}. 
To achieve chain-convergence, we use the 
Gelman-Rubin statistic \mbox{$R-1 = 0.01$}, as implemented in \citep{lewis13}.
The chains are analyzed using the \texttt{GetDist}\footnote{\url{https://getdist.readthedocs.io/en/latest/index.html}} package. 
We set the initial 30\% of the chains as the burn-in phase and disregard those samples in all cases.

In Table~\ref{tab_parameters}, we list the set of cosmological and nuisance parameters that are sampled during the MCMC. 
The fiducial values and the prior ranges are provided; and we also mark the datasets that are sensitive to the parameter under consideration in the last column. 
We choose wide priors on $\wde$ and $\wa$ based on the recent results \citep[see Table 1 of][]{steinhardt25} where the authors pointed out significant shifts in the inferred values of parameters like $\omchsq$ based on the chosen prior ranges. 
We note that this choice of prior allows the phantom-crossing \citep{vikman05, nojiri05} with $\wde < -1$ violating the null energy condition. 
The $\wde < -1$ scenario is handled using the parametrized post-Friedmann (PPF) approach in \texttt{CAMB} \citep{fang08, caldwell25}.
The likelihood code is publicly available for downloading through this link\footnote{\url{https://github.com/sriniraghunathan/CMB_BAO_SNe_likelihoods}}.

\subsection{Fisher formalism}
\label{sec_fisher}

Besides the MCMC approach, we also compute the parameter constraints using the standard Fisher formalism, in order to compare the constraints from both the methods. 
As mentioned earlier, this test is limited to CMB and \sne{} datasets only.
We build the Fisher matrices as described below.

\subsubsection{Fisher matrix for CMB power spectra}
\label{sec_cmb_fisher}
For CMB, we use the lensed CMB power spectra of the temperature and polarization fields, and form the Fisher matrix as \citep{jungman96, tegmark97, galli14}
\begin{equation}
    \mathcal{F}_{\alpha \beta}^{\rm CMB} = \sum_{XX,YY}\ \sum_{\ell = \ell_{\rm min}}^{\ell_{\rm max}}\ \dfrac{\partial C_\ell^{XX}}{\partial \theta_\alpha} \cdot \left(\Sigma_{\ell,\ell}^{XY}\right)^{-1} \cdot \dfrac{\partial C_\ell^{YY}}{\partial \theta_\beta},
    \label{eq_cmb_fisher_matrix}
\end{equation}
where $XX, YY \in [\tteete]$ power spectra with $(\elmin, \elmax) = (\elminval, \elmaxval)$; 
$\theta_{\alpha}$ represents the parameters being constrained; $\partial C_\ell / \partial \theta_{\theta}$ is the partial derivatives of the lensed CMB spectra with respect to the parameters $\theta$ and the derivatives are obtained using the finite difference method; and  $\Sigma_{\ell, \ell}$ is the covariance matrix of the lensed CMB power spectra (without including the CMB-lensing spectrum in this case) as given in Eq.(\ref{eq_cmb_covariance}). 

\subsubsection{Fisher matrix for \sne}
\label{sec_sne_fisher}
For \sne, we construct the Fisher matrix as,
\begin{equation}
    \mathcal{F}_{\alpha \beta}^{\rm \sne} = \sum_{ij}\ \dfrac{\partial \mu_{i}}{\partial \theta_\alpha} \cdot \Sigma_{{ij}}^{-1} \cdot \dfrac{\partial \mu_{j}}{\partial \theta_\beta},
    \label{eq_sne_fisher_matrix}
\end{equation}
where $\mu$ represents the distance modulus of the $i^{th}$ \sne{} at redshift $z_{i}$, $\frac{\partial \mu}{\partial \theta}$ is the partial derivative of the distance modulus with respect to parameter $\theta$ calculated using finite difference method, and the summation runs over all the \sne{} in catalog. 
In Eq.(\ref{eq_sne_fisher_matrix}), the covariance matrix $\Sigma_{{ij}}$ is the combination of statistical and systematic covariance matrix. 
The presence of systematic errors introduces correlation between different \sne{} and hence $\Sigma_{{ij}}$ is a non-diagonal matrix. 
In the absence of systematic errors, though, Eq.(\ref{eq_sne_fisher_matrix}) reduces to 
\begin{equation}
    \mathcal{F}_{\alpha \beta_{\rm \sne}} = \sum_{i}\ \dfrac{\partial \mu_{i}}{\partial \theta_\alpha} \cdot \dfrac{1}{\sigma_{\mu_{i}}^2} \cdot \dfrac{\partial \mu_{i}}{\partial \theta_\beta}.
    \label{eq_sne_fisher_matrix_simple}
\end{equation}
Details about the statistical and systematic errors for \sne{} data are described in \S\ref{sec_sne_data}. 

\section{Results and Discussion}
\label{sec_results}
We start with the comparison of constraints (a) between the recently released \desfiveyearsne{} and LSST \sne{} sample, and similarly (b) when swapping the BAO measurements from \desidata{} to \desidrthreedata.
These are followed by a comparison of the results from the MCMC and Fisher analyses, as well as an examination of the degradation of constraints due to binning, both of which are restricted to the CMB and \sne{} datasets.
Next, we present constraints on $\lcdm$ and the extensions, from CMB, \sne, and BAO, both individually and jointly. 
As a reminder, all results apart from the comparisons presented in \S\ref{sec_fisher_vs_mcmc} are derived from the MCMC analysis.

\subsection{Expected improvement from \lsstdata{} compared to \desfiveyearsne}
\label{sec_des_lsst_comparison}
\begin{figure}
\centering
\includegraphics[width=0.49\textwidth, keepaspectratio]{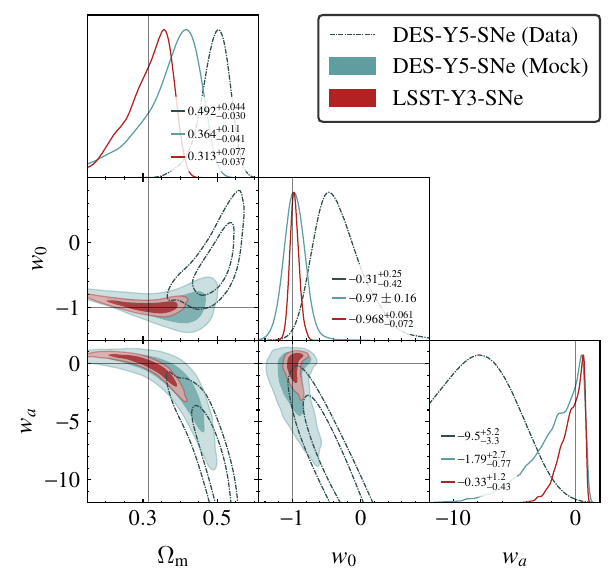}
\caption{Constraints on dark energy EoS parameters and the matter density: Teal color shows the constraints from \desfiveyearsne{} sample: dash-dotted corresponds to the actual constraints similar to the original works \citep{des24_sne_cosmology, vincenzi24, steinhardt25} while the solid contours correspond to a mock dataset with \desfiveyearsne-like errors, but simulated assuming the fiducial cosmology as specified in Table~\ref{tab_parameters}, that is different from the one derived using \desfiveyearsne.
The expected constraints from \whichlsstsnrsample{} are shown in red. 
We find that the uncertainties on dark energy parameters can be reduced by $\times 2-\times2.5$ with  \whichlsstsnrsample{} relative to DES. 
The majority of this constraining power comes from the increased number density in the LSST sample compared to DES, with roughly $\times 1.3 - \times 1.5$ contributed by objects at $z \ge 1$.
}
\label{fig_w0wa_lsst_des}
\end{figure}

To compare the constraints from \desfiveyearsne{} and the ones expected from \whichlsstsnrsample, we consider two cosmologies: (A) standard $\lcdm$, and (B) $\wwacdm$. 
We present the results for the latter in Fig.~\ref{fig_w0wa_lsst_des}. 
For this test, we do not use the official DES chains; instead, we run our own chains.
As mentioned earlier, we run chains both using the original \desfiveyearsne{} and the mock data vector generated in this work. 
In the figure, red contours correspond to \whichlsstsnrsample{} and the solid (dash-dotted) teal contours are for \desfiveyearsne{} mock (original data). 
The dash-dotted teal contours agree \citep[See Fig. 8 of][]{des24_sne_cosmology} well with the original results \citep{abbott24, vincenzi24, steinhardt25}.

For case (A), we only allow the physical matter density $\omchsq$ (but the value of $H_{0}$ is fixed) and the absolute magnitude nuisance-parameter to vary with the rest of the parameter values fixed using \planck{} $\lcdm$ values as described in Table~\ref{tab_parameters}. 
In this case, we obtain \mbox{$\omegam = \consdesmockomegam$} for \desfiveyearsne{} mock and \mbox{$\omegam = \conslcdmlsstunbinnedomegam$} for \whichlsstsnrsample, indicating a $\times 2.7$ improvement in the constraining power.

On the other hand, for case (B), $\wwacdm$ cosmology, we find that \whichlsstsnrsample{} can improve the uncertainties on $\theta \in [\wde, \wa]$ by roughly $\times 2-\times 2.5$ compared to the \desfiveyearsne{} mock sample.
We perform two tests to understand where the improvement comes from when switching to \whichlsstsnrsample.

The first test isolates the contribution from SNe with $z \ge z_{\rm max}^{\rm DES}$ by imposing a $z_{\rm max} > 1$ cut on the \whichlsstsnrsample{}, roughly matching the DES redshift range. 
For this case, we find that the errors on $\sigma(\wde)$ ($\sigma(\wa)$) can be improved by a $\times1.9$ ($\times1.6$) by LSST compared to DES. 
This suggests that ignoring high redshift ($z \ge 1$) \sne{} in LSST reduces the constraining power on dark energy parameters by $\sim \times 1.3-1.5$. 
We demonstrate this using yellow contours in Fig.~\ref{fig_w0wa_lsstzmaxcutanddeslike_des_appendix}.

In the second test, we assess the improvement due to calibration errors in LSST. For this purpose, we select a subset of the \whichlsstsnrsample{} with a redshift distribution $dN/dz$ matched to that of the DES sample. 
We repeat this test five times and find that dark energy constraints to be not significantly different compared to the constraints from \desfiveyearsne{} mock sample. 
This is shown as red contours in Fig.~\ref{fig_w0wa_lsstzmaxcutanddeslike_des_appendix} of the appendix.

These two tests indicate that the majority of the improvement from \whichlsstsnrsample{} arises from the increased number density and the inclusion of high‑redshift \sne{} in the \whichlsstsnrsample, and not due to changes/improvements in calibration errors. 
This is also in agreement with the errors shown in Fig.~\ref{fig_lsst_sne_details}, which do not differ significantly between the LSST and DES samples.

As shown in Table~\ref{tab_parameters}, we use wide priors for both $\wde$ and $\wa$. 
Restricting them to a narrower range $\mathcal{U}[-3, 1]$ for $\wde$ and $\mathcal{U}[-3, 3]$ for $\wa$ artificially shifts the inferred values significantly. 
In this case, we obtain the following values with the original \desfiveyearsne{} \sne{} dataset: \mbox{$\omegam = 0.379^{+0.064}_{-0.027}$;} \mbox{$\wde = -0.82^{+0.14}_{-0.11}$;} and \mbox{$\wa < 0.0520$ (95 \% C.L.)}.
These results are similar to the values reported by \citet{steinhardt25}, who emphasized the importance of the assumed priors when combining multiple cosmological datasets.

\subsection{Expected improvement from \desidrthreedata{} compared to \desidata}
\label{sec_desi_comparison}
\begin{figure}
\centering
\includegraphics[width=0.49\textwidth, keepaspectratio]{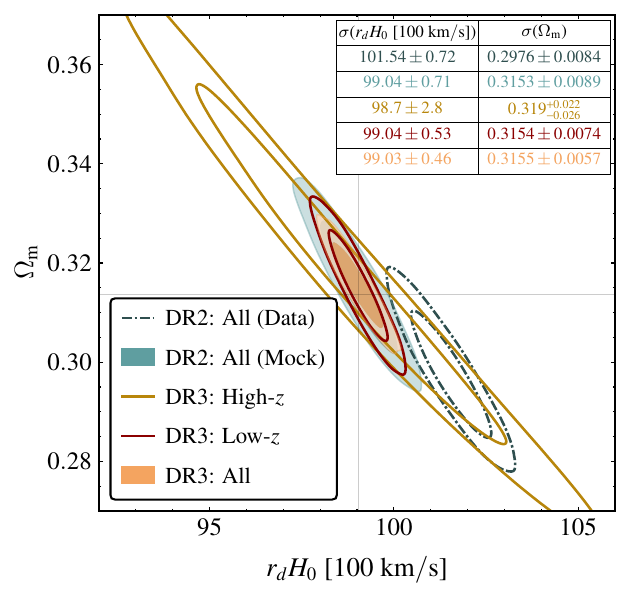}
\caption{Comparing the constraints in the $r_{d}H_{0} - \Omega_{m}$ plane for $\lcdm$ cosmology between DESI DR2 (teal) \citep{abdulkarim25} and the upcoming DR3 dataset (orange). 
For DR2, we present constraints derived from the actual DESI DR2 dataset \citep{abdulkarim25}, shown with dash‑dotted curves, and from the mock data vectors, drawn with a different cosmology given in Table~\ref{tab_parameters}, shown with solid curves.
For DR3, we show the constraints for the sample containing BAO measurements from galaxies/quasars at $z \lesssim 2$ in red and the ones from Ly-$\alpha$ forest at $z > 2$ in yellow. 
We see $\times1.5$ improvement in the uncertainties on both parameters when switching from DR2 to DR3. While not shown in this figure, the uncertainties in $\wde, \wa$ improve by $\times 1.8$ with DR3 relative to the DR2 measurements.
Such improvements will be crucial in addressing the growing tension between CMB and BAO. 
}
\label{fig_desi_dr2_dr3_cosntraints}
\end{figure}

In this subsection, we present the expected improvement from the upcoming DESI DR3 release \citep{desidr3} compared to the recent DR2 \citep{abdulkarim25} dataset. 
Specifically, we present the constraints for the $r_{d}H_{0} - \Omega_{m}$ plane, without including any other priors, for the standard $\lcdm$ cosmology where all the parameters are allowed to vary in the prior range given in Table~\ref{tab_parameters}.
Note that the BAO measurements need external priors on the scale of the sound horizon since they measure the product of $r_{d} H_{0}$. 

In Fig.~\ref{fig_desi_dr2_dr3_cosntraints}, we show the results for \desidata{} in teal: dash-dotted curves are for the actual dataset from \citet{abdulkarim25}, whereas the corresponding mock‑dataset results are shown using solid curves.
This can be compared to Figs: 7 and 8 of \citep{abdulkarim25} and our results agree with the original work. 
The constraints from the low redshift clustering of galaxies and quasars in the DR3 mock dataset is represented using the open red contour and the ones from the clustering of high redshift Ly-$\alpha$ forest measurements is show using the open yellow contour. 
The combined DR3 constraints is shown in orange.
The constraints are also listed as a table in the figure for reference. 
We obtain $\times1.5$ improvement in the uncertainties for both $\rdh$ and $\omegam$ with the DR3 sample compared to the DR2 release. 

The majority of the constraining power comes from the improvement in the low-$z$ sample. 
On the other hand, the \mbox{high-$z$} also contributes a little, particularly due to a small change in the direction of degeneracy compared to the low-$z$ sample, as can be seen in the figure. 
While not shown in the figure, for $\wwacdm$ cosmology, the uncertainties on $\wde, \wa$ reduce  by $\times1.8$ when switching from the DR2 to DR3 sample. 
Apart from separating the high-$z$ Ly-$\alpha$ forest BAO measurements from the galaxy/quasar measurements at $z \lesssim 2$, we do not investigate alternative BAO binning choices and leave such explorations to future work.

The improvements expected from the DR3 release will be crucial in addressing the tensions between BAO and other observables, like the CMB, that is starting to grow as pointed out recently by many works \citep[for example]{camphuis25, ferreira25, ong26}.
Furthermore, given the improvements in the high redshift measurements, comparing constraints from low-$z$ and high-$z$ (not necessarily as defined in this work) can provide useful systematic checks since the non-linearities in structure formation can be more important for the low-$z$ sample compared to the high-$z$.

\subsection{Comparison of Fisher‑forecast and MCMC‑derived constraints, and the impact of binning for both \sne{} and CMB}
\label{sec_fisher_vs_mcmc}
\begin{figure}
\centering
\includegraphics[width=0.49\textwidth, keepaspectratio]{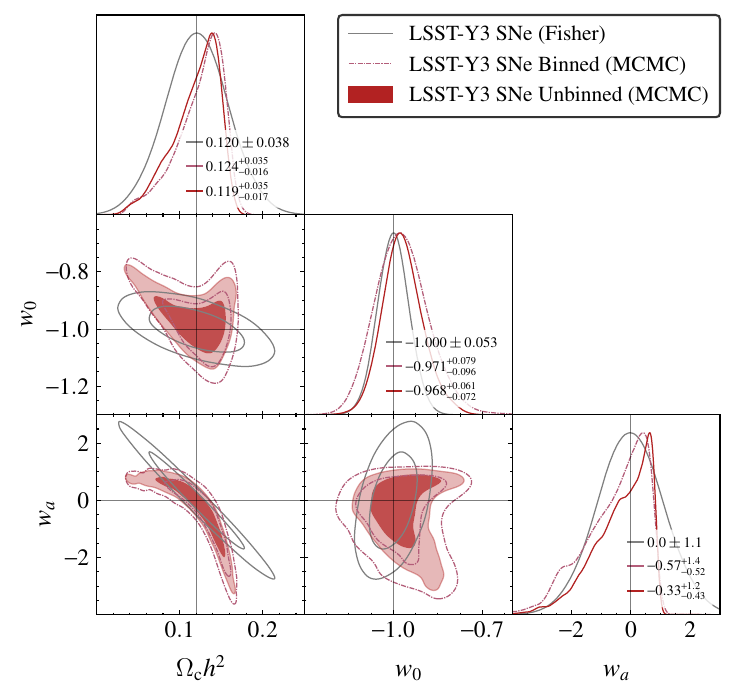}
\caption{Comparison of the constraints derived using the Fisher formalism in open gray contours and MCMC in red for the \whichlsstsnrsample. 
For MCMC, we present the results for both the unbinned (filled contours) and binned (open contours). The filled red contours are the same as in Fig.~\ref{fig_w0wa_lsst_des}. The posteriors are significantly non-Gaussian which are not captured by the Fisher formalism. We find that binning the \sne{} sample reduces the constraining power by $\sim 30\%$ for the dark energy parameters.}
\label{fig_fisher_vs_mcmc_sne}
\end{figure}

\noindent
{\bf \sne:} For SNe, we compare the results from \whichlsstsnrsample{} for $\wwacdm$ cosmology and they are presented in Fig.~\ref{fig_fisher_vs_mcmc_sne}. 
Since the \sne{} distance modulus measurements are most  sensitive to only a subset of cosmological parameters $\theta \in [\omchsq, \wde, \wa]$, we only compare the constraints for these ones. 
The fiducial values for the other parameters are 
fixed to the values indicated in Table~\ref{tab_parameters}.
The Fisher constraints are shown as open gray contours and the equivalent constraints from MCMC are in red: solid is for the unbinned sample (same as Fig.~\ref{fig_w0wa_lsst_des}) and dash-dotted is for the binned sample. 
As evident from the figure, the posteriors are strongly non‑Gaussian, a feature that the Fisher formalism cannot capture. 
Due to the non‑Gaussian nature of the posterior distribution and strong parameter degeneracies, projection effects appear in the MCMC results, as indicated by the red contours. 
This, however, does not introduce significant shifts in the inferred values.
As discussed later in \S\ref{sec_dark_energy}, once the \sne{} sample is combined with BAO and/or CMB, the posteriors become sufficiently Gaussian compared to the \sne‑only case.

In the figure, we also show the MCMC results obtained using the binned \sne{} sample as open red contours. 
As a reminder, for the binned case, the \howmanysneinlsst{} \sne{} in the \whichlsstsnrsample{} are binned based on the rebinning approach in \citet{kessler23} into 14 redshift bins.
For $\wwacdm$ cosmology, we find that the binned case degrades the uncertainties on $\wde, \wa$ by $\sim \times1.3$ compared to the unbinned case. 
In the case of $\lcdm$ cosmology (not shown in Fig.~\ref{fig_w0wa_lsst_des}), when only the matter density $\omegam$ parameter is allowed to vary, we get $\omegam = \conslcdmlsstunbinnedomegam$ ($\conslcdmlsstbinnedomegam$) for the unbinned (binned) analysis, corresponding to a slightly higher degradation than for the $\wwacdm$ case.\\

\begin{figure}
\centering
\includegraphics[width=0.49\textwidth, keepaspectratio]{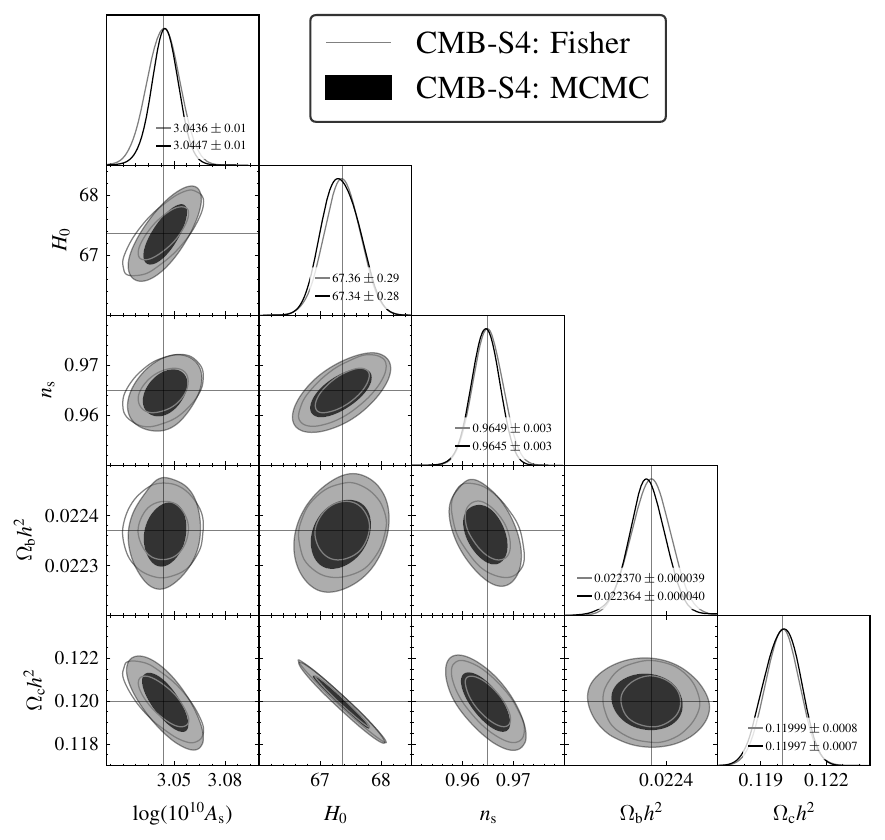}
\caption{Similar to Fig.~\ref{fig_fisher_vs_mcmc_sne} but showing the comparison between Fisher-based (open gray contours) constraints and MCMC (filled black contours) for \cmbsfour. 
Unlike \sne, the CMB posteriors are sufficiently Gaussian, and the differences in the parameter uncertainties are marginal.
In both cases, we have adopted our fiducial $\Delta_{\ell} = \binningdeltael$ power spectra binning.
}
\label{fig_fisher_vs_mcmc_cmb}
\end{figure}

\noindent
{\bf CMB:} In Fig.~\ref{fig_fisher_vs_mcmc_cmb}, we compare the $\lcdm$ constraints obtained using the $\tteete$ power spectra measurements expected from \cmbsfour, the most constraining CMB dataset in this work. 
For this exercise of comparison, we fix the nuisance (calibration) parameters, given in Eq.(\ref{eq_tcal_pcal}), to be maximally sensitive to any differences between Fisher and MCMC-based constraints for cosmological inference. 
In contrast to the \sne{} case above (see Fig.~\ref{fig_fisher_vs_mcmc_sne}), the MCMC results (filled black contours) agree well with the Fisher predictions open gray contours), and the resulting posteriors appear Gaussian. 
The difference in the uncertainties between the two approaches is marginal.
We use the same power‑spectrum binning scheme, $\Delta_{\ell} = \binningdeltael$, for the Fisher forecasts as in the MCMC analysis. 

\begin{figure}
\centering
\includegraphics[width=0.47\textwidth, keepaspectratio]{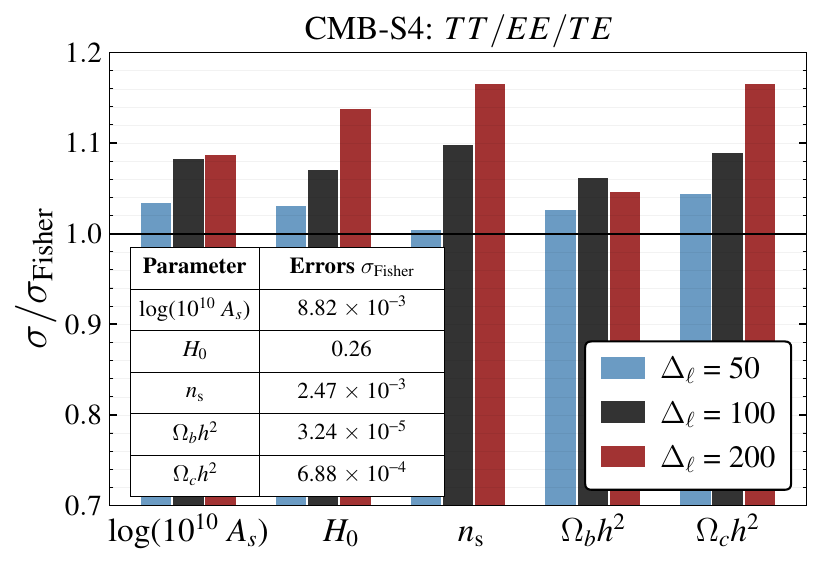}
\caption{
68 \% C.L. parameters constraints for the $\lcdm$ model from the CMB $\tteete$ power spectra derived using the MCMC sampling for the \cmbsfour{} survey. The colors represent the constraints for three different binning schemes: $\deltael = 50$ in blue; $\deltael = 100$ in black; and $\deltael = 200$ in red. The bars always correspond to the uncertainties relative to the ones obtained using the Fisher formalism with the unbinned spectra (given in the table). Binning increases the uncertainties of all the parameters by (a) $\le 5\%$ for $\deltael = 50$, (b) $5-10\%$ for $\deltael = 100$, which is baseline case; and (c) $10\%-15\%$ for $\deltael = 100$. 
}
\label{fig_fisher_vs_mcmc_binning_for_cmb}
\end{figure}

In Fig.~\ref{fig_fisher_vs_mcmc_binning_for_cmb} we present the results for different power-spectral binning and, like Fig.~\ref{fig_fisher_vs_mcmc_cmb}, these correspond to the results for the \cmbsfour{} survey. 
The figure shows the ratio of constraints from MCMC with binned spectra to the results obtained for the unbinned ($\deltael = 1$) spectra using the Fisher formalism: $\deltael = 50$ in blue, $\deltael = \binningdeltael$ in black, and $\deltael = 200$ in red. 
We note a mild reduction at $5-10\%$ in the constraining power for all parameters for $\deltael = 100$, our baseline choice. 
The degradation is less ($\le 5\%$) for $\deltael = 50$ but larger ($10\%-20\%$) for $\deltael = 200$. 
We emphasize that in this subsection we keep the bin width $\deltael$ fixed across all angular scales. Although optimal binning strategies tailored to the expected power spectrum can be employed, as is standard for CMB analyses \citep[e.g,][]{louis25, camphuis25}, we do not implement them here.
For $\ombhsq$, we obtain a slightly lower constraint for $\deltael = 200$ compared to $\deltael = 100$.
Although this should not occur in principle, the difference is marginal and consistent with the inherent dispersion in the MCMC process.

\subsection{Dark energy constraints}
\label{sec_dark_energy}
\begin{figure*}
\centering
\includegraphics[width=0.9\textwidth, keepaspectratio]{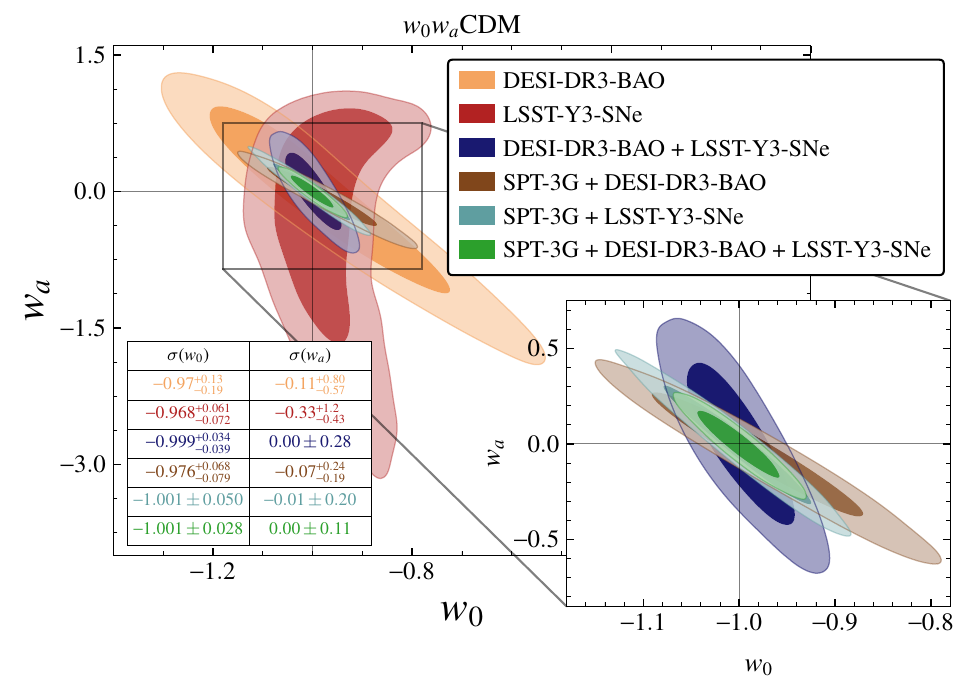}
\caption{Marginalized constraints on dark energy equation of state parameters for $\wwacdm$ cosmology. 
The constraints from \desidrthreedata{} (orange), \lsstdata{} (red), the combination without 
(navy)
and with (green) CMB $\tteetepp$ data. 
We also show the constraints when combining \desidrthreedata{} and \lsstdata{} individually with \sptthreeg{} in brown and 
teal contours. 
The combined constraints shown in green results in $\times1.5-\times2.5$ improvement when the \sptthreeg{} (green) is combined with \desidata{} and \whichlsstsnrsample.
We find minimal differences between the choice of CMB dataset as the constraints are primarily driven by the breaking of degeneracies in the $\wa-\wde$ probed by each observable (\desidrthreedata, \lsstdata, and CMB).
}
\label{fig_w0walcdm_for_w_wa}
\end{figure*}

\begin{figure}
\centering
\includegraphics[width=0.48\textwidth, keepaspectratio]{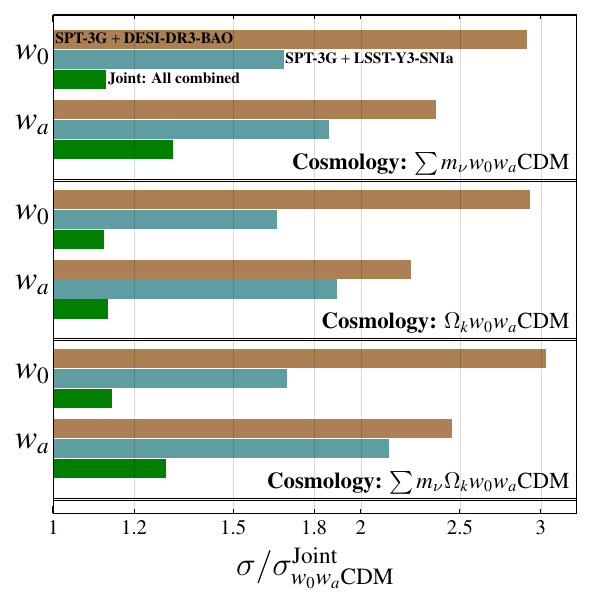}
\caption{Impact on dark energy equation of state parameters due to extended cosmology scenarios relative to a $\wwacdm$ cosmology. 
We present the impact using the ratio of constraints obtained from chains for the extended cosmology scenario under consideration over the ones obtained for $\wwacdm$ by combining all the three datasets: $\sigma(\wde) = 0.028$ and $\sigma(\wa) = 0.11$ shown using green color in Fig.~\ref{fig_w0walcdm_for_w_wa}. 
For the joint dataset, the constraints degrade by $<10\%$ ($10\%-30\%$) when jointly fitting dark energy parameters with spatial curvature (neutrino mass).}
\label{fig_w0walcdm_whisker_for_w_wa}
\end{figure}

\begin{figure*}
\centering
\includegraphics[width=\textwidth, keepaspectratio]{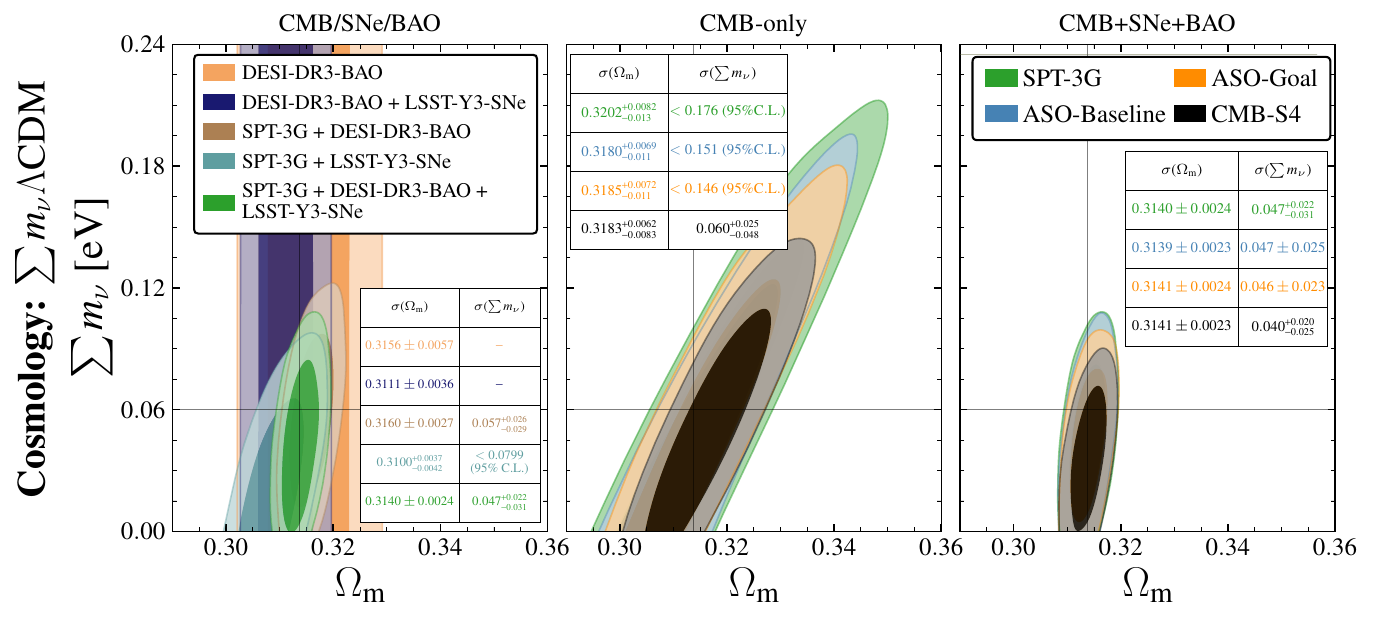}
\caption{Marginalized constraints on $\summnu - \omegam$ for $\mnulcdm$ cosmology. 
In the left panel, we show constraints from \desidrthreedata{} alone in orange; its combination with \whichlsstsnrsample{} in 
navy; and the further combination with CMB from \sptthreeg{} in green.
In the middle panel, we show the constraints from various CMB experiments, excluding data from \desidrthreedata{} and \whichlsstsnrsample{}, to highlight the impact of CMB-only measurements. 
In the right panel, we present the joint constraints obtained by combining CMB with \whichlsstsnrsample{} and \desidrthreedata{} for different CMB datasets. 
These results indicate that the future datasets can enable to $2-3\sigma$ detection of sum of neutrino masses.
}
\label{fig_w0wamnulcdm_for_h0_omegam_w_wa_mnu}
\end{figure*}

In Fig.~\ref{fig_w0walcdm_for_w_wa}, we present the constraints on the dark energy equation of state parameters $\theta \in [\wde, \wa]$. 
We report both the individual (except for CMB) and some combination of joint constraints from different datasets: \whichlsstsnrsample, \desidata, and CMB $\tteetepp$. 
For reference, we include a table within the figure listing the best-fit values along with their associated uncertainties (68\% C.L.).

The orange contours correspond to \desidrthreedata{} (similar to \citealt{abdulkarim25, lodha25} for \desidata); red corresponds to \whichlsstsnrsample, and 
navy corresponds to the joint constraints from the two. 
As listed in the table within in the figure, the uncertainties on $\wde, \wa$ improve by $\sim\times1.5-\times3.5$ from the joint dataset compared to the individual BAO or \sne{} results.

We find another $\times1.5-\times2.5$ improvement in both $\wde, \wa$ when the CMB data from \sptthreeg{} (green) is combined with \desidata{} and \whichlsstsnrsample. 
We also show the combination of \whichlsstsnrsample{} and \desidrthreedata{} with \sptthreeg{} as brown and 
teal contours. 
Since the CMB data alone does not constrain $\wde, \wa$ tightly, we do not show the individual constraints for CMB. 
Throughout this section, we use \sptthreeg{} as the CMB dataset since the constraining power does not differ significantly between CMB datasets. 
In fact, the uncertainties improve by $\sim <10\%$ when \sptthreeg{} is replaced by ASO or \cmbsfour. 
This is due to the different direction of degeneracy probed by each of the dataset in the $\wde-\wa$ plane. 
For clarity, we show a zoomed-in version of the figure to highlight the improvement in the constraints when \sptthreeg{} is included.

Although \desidrthreedata{} has a higher individual constraining power on $\wa$ than \whichlsstsnrsample{}, as shown by the orange and red contours in Fig.~\ref{fig_w0walcdm_for_w_wa}, the combination of CMB with \whichlsstsnrsample{} yields tighter constraints than CMB with \desidrthreedata{}. 
This improvement arises from the nearly orthogonal degeneracy directions in the former combination.

\subsubsection{Impact due to extended cosmologies}
\label{sec_dark_energy_ext_cosmologies}
Next, we investigate the impact on dark energy constraints for other extended cosmologies. 
In particular, we consider: $\wwamnulcdm$ and $\wwaomklcdm$, given the degeneracies between dark energy and spatial curvature or sum of neutrino masses.

In Fig.~\ref{fig_w0walcdm_whisker_for_w_wa}, we present the degradation in $\sigma(\wde)$ and $\sigma(\wa)$ for the extended cosmological models relative to $\wwacdm$ cosmology, which are explicitly labeled.
The plotted values represent the ratio of the current constraints to those obtained for the $\wwacdm$ model by combining all the three datasets. 
The colors indicate different dataset combinations (same as in Fig.~\ref{fig_w0walcdm_for_w_wa}): brown corresponds to CMB + \desidrthreedata, teal to CMB + \whichlsstsnrsample, and green to the combination of all three. 
Similar to Fig.~\ref{fig_w0walcdm_for_w_wa}, we choose to present only the constraints from \sptthreeg{} here, as the results from other CMB datasets do not show significant differences.

We find that freeing $\summnu$ has a slightly greater impact on dark energy constraints than freeing spatial curvature $\omk$. 
Jointly fitting for dark energy with neutrino mass degrades constraints on $\sigma(\wde)$ and $\sigma(\wa)$ by $\sim10\%-30\%$ (green bars in the top panel for the $\wwamnulcdm$ model), compared to $< 10\%$ when allowing the $\omk$ to vary (green bars in the middle panel for the $\wwaomklcdm$ model). 
When both neutrino mass and spatial curvature ($\wwamnuomklcdm$ model) are allowed to vary, the degradation 
is roughly similar to the case when only neutrino mass ($\wwamnulcdm$ model) is allowed to vary (green bars in the lower panel).
The degradation is more pronounced when CMB is combined only with \desidrthreedata{} ($\times2.3-\times3$ shown as brown bars in all the panels), than with \whichlsstsnrsample{} ($\times1.7-\times1.9$ shown as teal bars), because of the different degeneracy directions as described above.



\subsection{Neutrino mass constraints}
\label{sec_neutrino_mass}
The sum of neutrino masses, $\summnu$, has recently gathered significant attention due to a consistent $2$–$3\sigma$ preference for an unphysical negative value across multiple cosmological datasets that stands in tension with neutrino oscillation experiments \citep{craig24, loverde24, green25, lynch25, giare25b}. 
Several mechanisms have been proposed to explain this anomaly.
For example, excess lensing \citep{cozzumbo25}, modified value of the optical depth \citep{sailer25, elbers25c} compared to the large-scale polarization measurements from \planck{} \citep{planck20cosmo}, and modified forms of dark energy \citep{elbers25a}, among others. 
In this subsection, we forecast the constraints expected from the combination of \desidrthreedata{}, \whichlsstsnrsample{}, and CMB datasets.

\begin{figure}
\centering
\includegraphics[width=0.48\textwidth, keepaspectratio]{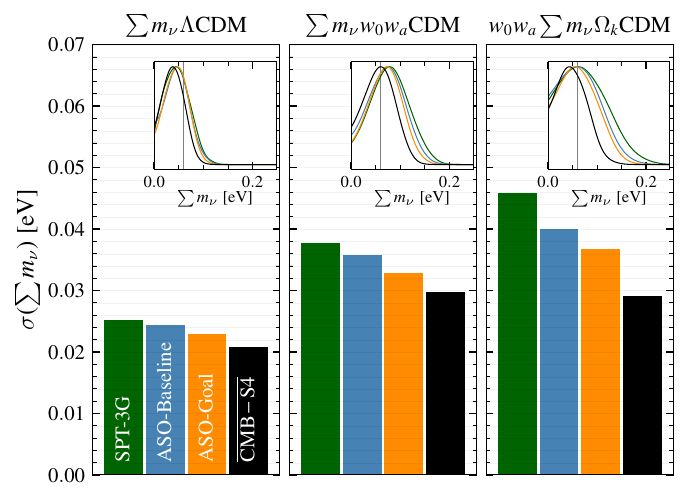}
\caption{Impact on the constraints on $\sigma(\summnu)$ due to extended cosmologies: Left panel is for $\mnulcdm$ and it is the same as Fig.~\ref{fig_w0wamnulcdm_for_h0_omegam_w_wa_mnu}; middle panel is for $\wwamnulcdm$; and right panel is for $\wwamnuomklcdm$. 
We note that constraints degrade by $\times1.5 - \times1.8$ for the extended cosmologies compared to the left panel. 
The reported constraints are for the combined CMB, \whichlsstsnrsample, and \desidrthreedata{} datasets.
}
\label{fig_mnu_constraints_extended_cosmologies}
\end{figure}

In Fig.~\ref{fig_w0wamnulcdm_for_h0_omegam_w_wa_mnu} we present the marginalized constraints for $\summnu-\omegam$ for the $\mnulcdm$ cosmology. 
Similar to previous figures, we present the constraints individually and for different dataset combinations. 
In the left panel, the \desidrthreedata{} only constraints are in orange; the combination of \desidrthreedata{} with \whichlsstsnrsample{} is shown in 
navy; and further combined with \sptthreeg{} data in green.
For reference, we also display the constraints obtained by combining \sptthreeg{} with \desidrthreedata{} (brown) and with the \whichlsstsnrsample{} (teal).
As shown in the table, including CMB data significantly improves the constraints on $\summnu$, shifting from no sensitivity to a $\sim2\sigma$ measurement. 
For the case when \sptthreeg{} and \whichlsstsnrsample{} are combined (teal), we obtain an upper limit of $\summnu < 0.08$ at 95\% C.L. 
In contrast, the \sptthreeg{}+\desidrthreedata{} combination yields a projected uncertainty of $\sigma(\summnu) \simeq 0.028$, which is further reduced by approximately 10\% when all three datasets are combined. 
This improvement arises from the inclusion of CMB lensing and the differing degeneracy directions between \desidrthreedata{} and \whichlsstsnrsample{} when combined with CMB data. 
The middle panel illustrates this by showing constraints from various CMB datasets: \sptthreeg{} (green), \asob{} (blue), \asog{} (orange), and \cmbsfour{} (black). 
In the right panel, we show the joint constraints by combining CMB with \whichlsstsnrsample{} and \desidrthreedata{} for different CMB datasets.
While CMB data alone cannot detect $\summnu$, they significantly tighten the upper limits compared to current surveys \citep{ge25, qu25, camphuis25, roychoudhury25}. 
In contrast, combining CMB data with \desidrthreedata{} and \whichlsstsnrsample{} enables a potential $2-3\sigma$ detection of $\summnu$ assuming a normal hierarchy.
Note that these forecast use a \planck-like prior on the optical depth to reionization $\sigma(\taure) = 0.007$.
Unlike for the case of $\wwacdm$, we see some improvements in the $\sigma(\summnu)$ constraints for ASO and \cmbsfour{} compared to \sptthreeg{} due to the reduced lensing bandpower errors resulting from the higher sky fraction for those experiments. 
This is particularly true for extended cosmologies where $\summnu$ is left free along with dark energy and/or spatial curvature. 
For these cases, the constraints degrade by $\times1.5 - \times1.8$ for all CMB experiments when combined with \whichlsstsnrsample{} and \desidrthreedata{} datasets as illustrated in Fig.~\ref{fig_mnu_constraints_extended_cosmologies}. 

\section{Conclusion}
\label{sec_conclusion} 

In this work, we forecasted the cosmological constraints expected from the combination of upcoming \sne, CMB, and BAO datasets. 
Our baseline results were obtained using an MCMC framework. 
We primarily focused on constraints on dark energy and neutrino masses, adopting a two‑parameter extension to $\lcdm$ for the former and a single‑parameter extension for the latter. 
We also studied the impact of combined extensions involving both dark energy and neutrino mass, as well as models that allow for nonzero spatial curvature. 
Besides the above forecasts, we also (a) compared the improvements moving from the recent \desfiveyearsne{} \sne{} sample to \whichlsstsnrsample{}, and \desidata{} to the upcoming \desidrthreedata{} release; (b) studied the differences in expected constraints between MCMC and the commonly used Fisher-formalism; and (c) studied the impact of binning the \sne{} sample and binning the CMB power spectra. 
We list our findings below.

\begin{itemize}
    \item{When moving from the recent \desfiveyearsne{} sample to \whichlsstsnrsample,  we found that the uncertainties on both $\wde$ and $\wa$ are reduced by approximately a factor of $\times 2 - \times 2.5$. This improvement is driven primarily by the substantially increased number density of \sne{} in the LSST sample, as well as by the inclusion of objects at $z \ge 1$. 
    }
    \item{When swapping the \desidata{} measurements with the forthcoming \desidrthreedata, we found a $\times 1.5$ reduction in the uncertainties on both $\rdh, \omegam$ for $\lcdm$ cosmology and  $\times 1.8$ reduction for $\wde, \wa$ for the $\wwacdm$ extension. The majority of this improvement arises from the expanded BAO sample at $z \le 2$, while clustering measurements of the Ly-$\alpha$ forest at $z > 2$ provide an additional, though smaller, contribution.}
    \item{We also found that the traditional Fisher formalism yields overly optimistic constraints and fails to capture the non‑Gaussianities in the posterior distribution for \sne{} cosmology when using the \whichlsstsnrsample{} sample. For the CMB, however, the differences between Fisher forecasts and MCMC-based approaches are only marginal.}
    \item{Turning to the impact of binning, we find that binning the \sne{} sample following the rebinning approach in \citep{kessler23} degrades the uncertainties on $(\wde, \wa)$ by $\sim 30\% - 50\%$. In contrast, for the CMB, binning the power spectra with $\Delta_{\ell} = \binningdeltael$ weakens the constraints on all parameters by only $\leq 10\%$ compared to the unbinned case.}
    \item{Upon combining CMB data $\tteetepp$ with \whichlsstsnrsample{} and \desidrthreedata, we found that we can achieve $\wde = -0.995 \pm 0.029$ and $\wa = -0.06 \pm 0.11$, with only minor differences across CMB experiments. This represents roughly $\times6-\times8$ ($\times2-\times5$) in $\sigma(\wa)$ ($\sigma(\wde)$) compared to \sne{} and BAO datasets individually. When extended cosmologies like neutrino masses $\summnu$ or spatial curvature $\omk$ are considered, the uncertainties on dark energy parameters increase by $\times1.1-1.3$ for the joint datasets. The results are insensitive to the choice of CMB data that is being included.}
    \item{The combined \desidrthreedata, \whichlsstsnrsample, and CMB datasets also can enable a $2-3\sigma$ detection of $\summnu$ assuming a normal Hierarchy depending on the choice of CMB (lensing) dataset that is being considered. 
    \planck-like prior on $\sigma(\taure) = 0.007$ has been assumed for these forecasts. 
    The sensitivity drops by $\times1.5-\times1.8$ for extended cosmologies that include dark energy and spatial curvature.}
\end{itemize}

As discussed in the text, the improvements in the cosmological constraints will be essential for resolving the mild tensions that have recently emerged, such as the CMB–BAO discrepancy under the assumption of the $\lcdm$ model.
The results presented here are expected to improve further with upcoming data releases, such as DESI-DR4-BAO, DESI-II, and LSST Year-5 and Year-10 supernova samples, as well as through the inclusion of additional cosmological probes, including the Sunyaev–Zeldovich effects, cluster abundances, $N\times2$ observables, fast radio bursts, standard sirens, among others \citep[][for example]{madhavacheril19, wang20, zuntz21, omori23, bocquet24, glowacki26}. 
These will be further 
complemented 
by advancements in survey design for upcoming and next-generation experiments, including SPT-3G$+$ \citep{anderson22}, CMB-HD \citep{sehgal22}, Nancy Grace Roman Space Telescope \citep{hounsell23}, the Wide-field Spectroscopic Telescope (WST, \citealt{mainieri24}), the all-sky Stage-5 Spectroscopic Survey (Spec-S5, \citealt{besuner25}). 

\section*{Data products and availability}
The data products and the likelihood codes are publicly available for download and usage in this github link: \url{https://github.com/sriniraghunathan/CMB_BAO_SNe_likelihoods}.

\section*{Acknowledgments}
SR dedicates this work to the loving memory of Eric Baxter.

We thank Colin Hill for useful discussions about the noise modeling for the Advanced-SO surveys; Brad Benson for feedback on the manuscript; Maciej Bilicki for help with the LSST-DESC publication process; members of the LSST-DESC and SPT collaborations for their valuable discussions throughout the course of this work. 
This paper has undergone internal review by the LSST-DESC.
We also thank the anonymous referee for their useful suggestions which has helped in shaping this manuscript better.

Author contributions are listed below: 
S. Raghunathan: Conceived and led the project and manuscript preparation; 
A. Mitra: Provided the \sne{} sample; 
N. Šarčević: Assisted with the LSST-DESC products; F. Ge and C. Ravoux: Contributed to the DESI BAO constraints. 

SR acknowledges support of Michael and Ester Vaida, and the National Science Foundation via award OPP-1852617; the support by the Illinois Survey Science Fellowship from the Center for AstroPhysical Surveys at the National Center for Supercomputing Applications; and also the support from Universities Research Association’s Visiting Scholars Program fellowship. 
CG is funded by the MICINN project PID2022-141079NB-C32. 
IFAE is partially funded by the CERCA program of the Generalitat de Catalunya.

This work made use of the following computing resources: Illinois Campus Cluster, a computing resource that is operated by the Illinois Campus Cluster Program (ICCP) in conjunction with the National Center for Supercomputing Applications (NCSA) and which is supported by funds from the University of Illinois at Urbana Champaign; 
and the computing resources provided on Crossover, a high-performance computing cluster operated by the Laboratory  Computing Resource Center at Argonne National Laboratory.

The DESC acknowledges ongoing support from the Institut National de 
Physique Nucl\'eaire et de Physique des Particules in France; the 
Science \& Technology Facilities Council in the United Kingdom; and the
Department of Energy, the National Science Foundation, and the LSST 
Corporation in the United States.  DESC uses resources of the IN2P3 
Computing Center (CC-IN2P3--Lyon/Villeurbanne - France) funded by the 
Centre National de la Recherche Scientifique; the National Energy 
Research Scientific Computing Center, a DOE Office of Science User 
Facility supported by the Office of Science of the U.S.\ Department of
Energy under Contract No.\ DE-AC02-05CH11231; STFC DiRAC HPC Facilities, 
funded by UK BIS National E-infrastructure capital grants; and the UK 
particle physics grid, supported by the GridPP Collaboration.  This 
work was performed in part under DOE Contract DE-AC02-76SF00515.

\appendix
\restartappendixnumbering

\section{Cosmological constraints from \whichlsstsnrsample{} and \desfiveyearsne{} samples}
\label{app_lsst_des_sne_constraints}

\begin{figure}
\centering
\includegraphics[width=0.48\textwidth, keepaspectratio]{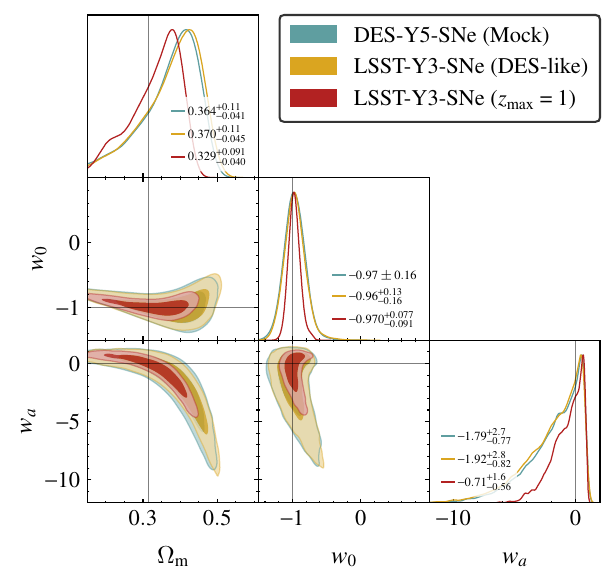}
\caption{Constraints on dark energy EoS parameters, analogous to Fig.~\ref{fig_w0wa_lsst_des}, for different subsamples of \whichlsstsnrsample: 
yellow contours corresponds to an LSST subsample with the same number counts and redshift distribution as the \desfiveyearsne{} sample; and red contours correspond to the LSST sample with high‑redshift SNe removed $z > z_{\rm max}^{\rm DES} (\sim 1)$. 
Teal contours correspond to \desfiveyearsne-mocks (same as in Fig.~\ref{fig_w0wa_lsst_des}); 
The yellow and teal contours are nearly identical, demonstrating that the reduced calibration uncertainties in the LSST sample alone do not significantly improve the dark energy constraints relative to DES. 
In contrast, removing the high‑redshift SNe (red) weakens the constraints by only about 30–50\%. 
These tests show that the factor of $\times2-\times2.5$ improvement in the dark‑energy constraints for the full \whichlsstsnrsample{} sample is driven primarily by the larger number of objects relative to the \desfiveyearsne{} sample.}
\label{fig_w0wa_lsstzmaxcutanddeslike_des_appendix}
\end{figure}

In this subsection, we derive the expected cosmological constraints for different subsamples of \whichlsstsnrsample{} catalog. 
As mentioned in the main text (\S\ref{sec_des_lsst_comparison}), this test helps in understanding the improvements in the constraining power when moving from \desfiveyearsne{} to \whichlsstsnrsample. 
We summarize the results in Fig.~\ref{fig_w0wa_lsstzmaxcutanddeslike_des_appendix}. 
In the figure, the teal contours correspond to \desfiveyearsne-mocks (same as in Fig.~\ref{fig_w0wa_lsst_des}). 

In the first case, we assess the improvements provided by the high‑redshift \sne{} expected in the \whichlsstsnrsample.
For this test, we impose a maximum redshift equal to that of the highest‑redshift \sne{} in the \desfiveyearsne{} catalog ($z_{\rm max}^{\rm DES} \sim 1$). The resulting constraints are shown in red.
In this scenario, the constraining power on $\wde$ and $\wa$ degrades by factors of $\times 1.9$ and $\times 1.6$, respectively, implying that the inclusion of high‑redshift objects yields an improvement of roughly $30\%-50\%$.

In the second case, we construct a \sne{} subsample from the full \whichlsstsnrsample{} that matches the redshift distribution of the \desfiveyearsne{} catalog.
Consequently, this subsample contains approximately the same number of objects as the \desfiveyearsne{} dataset.
The resulting constraints are shown in yellow in Fig.~\ref{fig_w0wa_lsstzmaxcutanddeslike_des_appendix}, and we note that they are similar to those shown in teal.

Together, these tests show that the factor of $\times2-\times2.5$ improvement in dark energy constraints for the full LSST sample is driven primarily by the increased number of supernovae and partially by the inclusion of high-redshift objects, rather than by calibration improvements.

\section{Dependence of cosmological constraints on the underlying cosmological model}
\label{app_constraints_and_cosmo_dep}

\begin{figure}
\centering
\includegraphics[width=0.48\textwidth, keepaspectratio]{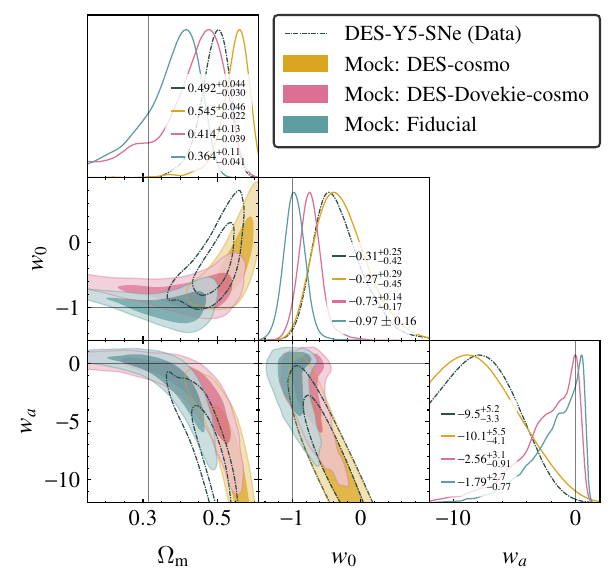}
\caption{
Dependence of the expected cosmological constraints from \desfiveyearsne-like \sne{} on the underlying cosmology: Open and filled teal contours show the constraints for the original DES data vector and the mock data vector for the fiducial $\Lambda$CDM cosmology (same as Fig.~\ref{fig_w0wa_lsst_des}). 
The yellow and pink contours correspond to mock data vectors generated assuming DES-cosmo \citep{des24_sne_cosmology} and DES-Dovekie-cosmo \citep{popovic25}, respectively. As expected, the constraints for the DES-cosmo mock data (yellow) roughly match the open teal contours. They are, however, significantly weaker than the baseline results, as DES-cosmo probes a different region of the $\wde - \wa$ posterior volume. In contrast, the DES-Dovekie-cosmo constraints (pink) are only slightly degraded (by $\sim10\% - 20\%$) relative to the baseline case.
}
\label{fig_w0wa_des_multiple_cosmo_appendix}
\end{figure}

In this section, we assess how the cosmological constraints depend on the assumed underlying cosmology.
As a reminder, throughout this work the fiducial cosmology is taken to be $\lcdm$, based on the \planck{} 2018 measurements described in \S\ref{sec_mock_data_vectors}. 
Here, we consider two alternative cosmologies motivated by recent results derived from the \desfiveyearsne{} sample. 
For the first case, we explore an extreme scenario using the inferred values $\omegam = 0.495$, $\wde = -0.36$, and $\wa = -8.8$ from the \sne-only solution of the \desfiveyearsne{} cosmology analysis \citep{des24_sne_cosmology}.
We refer to this model as DES-cosmo. 
In the second case, we use the updated parameter values from the reanalysis of the \desfiveyearsne{} cosmology incorporating the revised \sne{} calibration \citep{popovic25}, and including CMB and \desidata.
For this model, we set $\omegam = 0.33$, $\wde = -0.803$, and $\wa = -0.72$, and we refer to this cosmology as DES-Dovekie-cosmo. 
We consider this solution more realistically centred than \sne-only, where the posterior extends to high values of $\Omega_{\rm m}$ that are outside the bounds of other datasets (for example, cosmic shear).
We regenerate the data vectors for both these models. 

We present the results in Fig.~\ref{fig_w0wa_des_multiple_cosmo_appendix}, showing constraints from \sne-only.
In the figure, the original constraints from the \desfiveyearsne{} data-vector are shown as the dash-dotted open teal contours, while the constraints from the mock data vectors assuming the fiducial $\lcdm$ cosmology are shown as the filled teal contours.
These are identical to the results in Fig.~\ref{fig_w0wa_lsst_des} and are included here for reference.
The constraints for DES-cosmo are shown in yellow, and those for DES-Dovekie-cosmo are shown in pink.

As seen in the figure, the inferred best‑fit values shift to match the assumed underlying cosmology for the yellow and pink contours.
We also note that the uncertainties are larger compared to the baseline case (filled teal contours).
For DES‑cosmo (yellow), the errors worsen significantly ---by $\gtrsim \times 2$ for all parameters--- relative to the baseline results.
These constraints also roughly match those from the original DES analysis (open teal contours), as expected.
This arises from the fact that DES-cosmo probes a different region of the $\wde - \wa$ posterior volume compared to the fiducial case.
In contrast, for DES‑Dovekie‑cosmo (pink), the degradation is much smaller, at the level of only 10\%-20\% compared to the baseline. 
While not shown in the figure, we also repeated this test for the joint constraints. 
In this case, the significant degradation in the constraining power for DES-cosmo disappears for $\omegam$, although the trend remains the same for the dark energy equation-of-state parameters. 

\section{DESI data vectors}
\label{app_desi_data_vectors}
In Table~\ref{tab_desi_dr2_data_vectors}, we list the data vectors generated assuming the standard $\lcdm$ cosmology for DESI DR2. 
The covariance matrix is the same as the one given in \mbox{Table 1} of the original DR2 release paper \citep{abdulkarim25}. 
Table~\ref{tab_desi_dr3_data_vectors_lowz} and Table~\ref{tab_desi_dr3_data_vectors_highz} contain the DESI DR3 data vectors along with the correlations between different observables. 
The correlations are computed from the covariance matrices released in \citet{desidr3}.

\begin{deluxetable}{| c | c | c | c |}
\centering
\tabletypesize{\small}
\def\arraystretch{1.1}
\tablecaption{Data vectors for \desidata{} computed using the standard $\lcdm$ cosmology used in this work.
}
\label{tab_desi_dr2_data_vectors}
\tablehead{
Redshift $z$ & $D_{V}/r_{d}$ & $D_{M}/r_{d}$ & $D_{H}/r_{d}$
}
\startdata
\hline\hline
0.295 & 8.1 & \multicolumn{2}{c|}{-} \\\hline
0.51 & \multirow{5}{*}{-} & 13.51 & 22.8 \\\cline{1-1}\cline{3-4}
0.706 & & 17.7 & 20.2 \\\cline{1-1}\cline{3-4}
0.934 & & 22 & 17.6 \\\cline{1-1}\cline{3-4}
1.321 & & 28.1 & 14.1 \\\cline{1-1}\cline{3-4}
1.484 & & 30.3 & 12.9 \\\cline{1-1}\cline{3-4}
2.33 & & 6.6 & 39.2 \\\hline\hline
\enddata
\end{deluxetable}
\begin{deluxetable}{| c | c | c | c | c | c | c |}
\tabletypesize{\small}
\def\arraystretch{1.1}
\tablecaption{The data vectors and the correlations between observables for the low redshift \desidrthreedata{} data. 
The data vectors are calculated assuming $\lcdm$ cosmology, similar to Table~\ref{tab_desi_dr2_data_vectors}, and the correlations are computed using the covariance matrices released in \citep{desidr3}.
}\label{tab_desi_dr3_data_vectors_lowz}
\tablehead{
Redshift $z$ & $f{\sigma_{8}}$ & $D_{A}/r_{d}$ & $H_{z}r_{d}$ & $\rho(f{\sigma_{8}}, D_{A}/r_{d})$ & $\rho(f{\sigma_{8}}, H_{z}r_{d})$ & $\rho(D_{A}/r_{d}, H_{z}r_{d})$
}
\startdata
\hline\hline
0.05 & 0.44 & 1.42 & 10147.28 & 0.070 & 0.103 & 0.407 \\\hline
0.15 & 0.46 & 3.80 & 10686.55 & 0.069 & 0.102 & 0.407 \\\hline
0.25 & 0.47 & 5.68 & 11294.78 & 0.068 & 0.101 & 0.408 \\\hline
0.35 & 0.48 & 7.17 & 11969.21 & 0.070 & 0.104 & 0.409 \\\hline
0.45 & 0.48 & 8.36 & 12706.62 & 0.074 & 0.105 & 0.409 \\\hline
0.55 & 0.47 & 9.29 & 13503.56 & 0.073 & 0.104 & 0.409 \\\hline
0.65 & 0.47 & 10.04 & 14356.59 & 0.072 & 0.102 & 0.409 \\\hline
0.75 & 0.46 & 10.62 & 15262.39 & 0.071 & 0.101 & 0.410 \\\hline
0.85 & 0.45 & 11.08 & 16217.86 & 0.071 & 0.100 & 0.410 \\\hline
0.95 & 0.44 & 11.43 & 17220.11 & 0.074 & 0.103 & 0.410 \\\hline
1.05 & 0.43 & 11.69 & 18266.54 & 0.079 & 0.109 & 0.411 \\\hline
1.15 & 0.42 & 11.89 & 19354.79 & 0.073 & 0.109 & 0.420 \\\hline
1.25 & 0.40 & 12.03 & 20482.72 & 0.073 & 0.109 & 0.420 \\\hline
1.35 & 0.39 & 12.13 & 21648.43 & 0.073 & 0.108 & 0.420 \\\hline
1.45 & 0.38 & 12.18 & 22850.20 & 0.073 & 0.107 & 0.420 \\\hline
1.55 & 0.37 & 12.21 & 24086.50 & 0.072 & 0.105 & 0.421 \\\hline
1.65 & 0.36 & 12.20 & 25355.95 & 0.062 & 0.083 & 0.414 \\\hline
1.75 & 0.35 & 12.18 & 26657.29 & 0.061 & 0.082 & 0.414 \\\hline
1.85 & 0.34 & 12.14 & 27989.40 & 0.061 & 0.081 & 0.414 \\\hline
1.95 & 0.33 & 12.08 & 29351.27 & 0.060 & 0.080 & 0.414 \\\hline
2.05 & 0.32 & 12.01 & 30741.95 & 0.059 & 0.078 & 0.413 \\\hline\hline
\enddata
\end{deluxetable}
\begin{deluxetable}{| c | c | c | c |}
\tabletypesize{\small}
\def\arraystretch{1.1}
\tablecaption{Same as Table~\ref{tab_desi_dr3_data_vectors_lowz} but for the high redshift measurements expected in \desidrthreedata{} release. 
}
\label{tab_desi_dr3_data_vectors_highz}
\tablehead{
Redshift $z$ & $H_{z}r_{d}$ & $D_{A}/r_{d}$ & $\rho(H_{z}r_{d}, D_{A}/r_{d})$
}
\startdata
\hline\hline
2.12 & 31700.38 & 11.96 & 0.447 \\\hline
2.28 & 33991.91 & 11.82 & 0.456 \\\hline
2.43 & 36349.78 & 11.67 & 0.461 \\\hline
2.59 & 38771.20 & 11.51 & 0.466 \\\hline
2.75 & 41254.09 & 11.34 & 0.473 \\\hline
2.91 & 43796.63 & 11.16 & 0.474 \\\hline
3.07 & 46396.73 & 10.99 & 0.477 \\\hline
3.23 & 49053.12 & 10.81 & 0.477 \\\hline
3.39 & 51763.98 & 10.64 & 0.479 \\\hline
3.55 & 54528.16 & 10.46 & 0.484 \\\hline
\hline
\enddata
\end{deluxetable}

\ifdefined\PRformat
\bibliographystyle{apsrev4-1}
\fi
\bibliography{cmb_sne_bao}{}

\begin{thebibliography}{}
\expandafter\ifx\csname natexlab\endcsname\relax\def\natexlab#1{#1}\fi
\providecommand{\url}[1]{\href{#1}{#1}}
\providecommand{\dodoi}[1]{doi:~\href{http://doi.org/#1}{\nolinkurl{#1}}}
\providecommand{\doeprint}[1]{\href{http://ascl.net/#1}{\nolinkurl{http://ascl.net/#1}}}
\providecommand{\doarXiv}[1]{\href{https://arxiv.org/abs/#1}{\nolinkurl{https://arxiv.org/abs/#1}}}

\bibitem[{T.~M.~C. {Abbott} {et~al.}(2022){Abbott}, {Aguena}, {Alarcon},
  {Allam}, {Alves}, {Amon}, {Andrade-Oliveira}, {Annis}, {Avila}, {Bacon},
  {Baxter}, {Bechtol}, {Becker}, {Bernstein}, {Bhargava}, {Birrer}, {Blazek},
  {Brandao-Souza}, {Bridle}, {Brooks}, {Buckley-Geer}, {Burke}, {Camacho},
  {Campos}, {Carnero Rosell}, {Carrasco Kind}, {Carretero}, {Castander},
  {Cawthon}, {Chang}, {Chen}, {Chen}, {Choi}, {Conselice}, {Cordero},
  {Costanzi}, {Crocce}, {da Costa}, {da Silva Pereira}, {Davis}, {Davis}, {De
  Vicente}, {DeRose}, {Desai}, {Di Valentino}, {Diehl}, {Dietrich}, {Dodelson},
  {Doel}, {Doux}, {Drlica-Wagner}, {Eckert}, {Eifler}, {Elsner}, {Elvin-Poole},
  {Everett}, {Evrard}, {Fang}, {Farahi}, {Fernandez}, {Ferrero}, {Fert{\'e}},
  {Fosalba}, {Friedrich}, {Frieman}, {Garc{\'\i}a-Bellido}, {Gatti},
  {Gaztanaga}, {Gerdes}, {Giannantonio}, {Giannini}, {Gruen}, {Gruendl},
  {Gschwend}, {Gutierrez}, {Harrison}, {Hartley}, {Herner}, {Hinton},
  {Hollowood}, {Honscheid}, {Hoyle}, {Huff}, {Huterer}, {Jain}, {James},
  {Jarvis}, {Jeffrey}, {Jeltema}, {Kovacs}, {Krause}, {Kron}, {Kuehn},
  {Kuropatkin}, {Lahav}, {Leget}, {Lemos}, {Liddle}, {Lidman}, {Lima}, {Lin},
  {MacCrann}, {Maia}, {Marshall}, {Martini}, {McCullough}, {Melchior},
  {Mena-Fern{\'a}ndez}, {Menanteau}, {Miquel}, {Mohr}, {Morgan}, {Muir},
  {Myles}, {Nadathur}, {Navarro-Alsina}, {Nichol}, {Ogando}, {Omori},
  {Palmese}, {Pandey}, {Park}, {Paz-Chinch{\'o}n}, {Petravick}, {Pieres},
  {Plazas Malag{\'o}n}, {Porredon}, {Prat}, {Raveri}, {Rodriguez-Monroy},
  {Rollins}, {Romer}, {Roodman}, {Rosenfeld}, {Ross}, {Rykoff}, {Samuroff},
  {S{\'a}nchez}, {Sanchez}, {Sanchez}, {Sanchez Cid}, {Scarpine}, {Schubnell},
  {Scolnic}, {Secco}, {Serrano}, {Sevilla-Noarbe}, {Sheldon}, {Shin}, {Smith},
  {Soares-Santos}, {Suchyta}, {Swanson}, {Tabbutt}, {Tarle}, {Thomas}, {To},
  {Troja}, {Troxel}, {Tucker}, {Tutusaus}, {Varga}, {Walker}, {Weaverdyck},
  {Wechsler}, {Weller}, {Yanny}, {Yin}, {Zhang}, {Zuntz}, \& {DES
  Collaboration}}]{abbott22a}
{Abbott}, T.~M.~C., {Aguena}, M., {Alarcon}, A., {et~al.} 2022,
  \bibinfo{title}{{Dark Energy Survey Year 3 results: Cosmological constraints
  from galaxy clustering and weak lensing},} \prd, 105, 023520,
  \dodoi{10.1103/PhysRevD.105.023520}

\bibitem[{A.~G. {Adame} {et~al.}(2025){Adame}, {Aguilar}, {Ahlen}, {Alam},
  {Alexander}, {Allende Prieto}, {Alvarez}, {Alves}, {Anand}, {Andrade},
  {Armengaud}, {Avila}, {Aviles}, {Awan}, {Bahr-Kalus}, {Bailey}, {Baltay},
  {Bault}, {Behera}, {BenZvi}, {Beutler}, {Bianchi}, {Blake}, {Blum}, {Bonici},
  {Brieden}, {Brodzeller}, {Brooks}, {Buckley-Geer}, {Burtin}, {Calderon},
  {Canning}, {Carnero Rosell}, {Cereskaite}, {Cervantes-Cota}, {Chabanier},
  {Chaussidon}, {Chaves-Montero}, {Chebat}, {Chen}, {Chen}, {Claybaugh},
  {Cole}, {Cuceu}, {Davis}, {Dawson}, {de la Macorra}, {de Mattia}, {Deiosso},
  {Dey}, {Dey}, {Ding}, {Doel}, {Edelstein}, {Eftekharzadeh}, {Eisenstein},
  {Elbers}, {Elliott}, {Fagrelius}, {Fanning}, {Ferraro}, {Ereza}, {Findlay},
  {Flaugher}, {Font-Ribera}, {Forero-S{\'a}nchez}, {Forero-Romero}, {Frenk},
  {Garcia-Quintero}, {Garrison}, {Gazta{\~n}aga}, {Gil-Mar{\'\i}n}, {Gontcho},
  {Gonzalez-Morales}, {Gonzalez-Perez}, {Gordon}, {Green}, {Gruen}, {Gsponer},
  {Gutierrez}, {Guy}, {Hadzhiyska}, {Hahn}, {Hanif}, {Herrera-Alcantar},
  {Honscheid}, {Howlett}, {Huterer}, {Ir{\v{s}}i{\v{c}}}, {Ishak}, {Joyce},
  {Juneau}, {Kara{\c{c}}ayl{\i}}, {Kehoe}, {Kent}, {Kirkby}, {Kong}, {Koposov},
  {Kremin}, {Krolewski}, {Lahav}, {Lai}, {Lan}, {Landriau}, {Lang}, {Lasker},
  {Le Goff}, {Le Guillou}, {Leauthaud}, {Levi}, {Li}, {Lodha}, {Magneville},
  {Manera}, {Margala}, {Martini}, {Matthewson}, {Maus}, {McDonald},
  {Medina-Varela}, {Meisner}, {Mena-Fern{\'a}ndez}, {Miquel}, {Moon}, {Moore},
  {Moustakas}, {Mudur}, {Mueller}, {Mu{\~n}oz-Guti{\'e}rrez}, {Myers},
  {Nadathur}, {Napolitano}, {Neveux}, {Newman}, {Nguyen}, {Nie}, {Niz},
  {Noriega}, {Padmanabhan}, {Paillas}, {Palanque-Delabrouille}, {Pan},
  {Penmetsa}, {Percival}, {Pieri}, {Pinon}, {Poppett}, {Porredon}, {Prada},
  {P{\'e}rez-Fern{\'a}ndez}, {P{\'e}rez-R{\`a}fols}, {Rabinowitz}, {Raichoor},
  {Ram{\'\i}rez-P{\'e}rez}, {Ramirez-Solano}, {Rashkovetskyi}, {Ravoux},
  {Rezaie}, {Rich}, {Rocher}, {Rockosi}, {Roe}, {Rosado-Marin}, {Ross},
  {Rossi}, {Ruggeri}, {Ruhlmann-Kleider}, {Samushia}, {Sanchez}, {Saulder},
  {Schlafly}, {Schlegel}, {Schubnell}, {Seo}, {Shafieloo}, {Sharples},
  {Silber}, {Slosar}, {Smith}, {Sprayberry}, {Tan}, {Tarl{\'e}}, {Taylor},
  {Trusov}, {Vaisakh}, {Valcin}, {Valdes}, {Valogiannis}, {Vargas-Maga{\~n}a},
  {Verde}, {Walther}, {Wang}, {Wang}, {Weaver}, {Weaverdyck}, {Wechsler},
  {Weinberg}, {White}, {Wilson}, \& {Yi}}]{adame25}
{Adame}, A.~G., {Aguilar}, J., {Ahlen}, S., {et~al.} 2025,
  \bibinfo{title}{{DESI 2024 VII: cosmological constraints from the full-shape
  modeling of clustering measurements},} \jcap, 2025, 028,
  \dodoi{10.1088/1475-7516/2025/07/028}

\bibitem[{S. {Alam} {et~al.}(2017){Alam}, {Ata}, {Bailey}, {Beutler},
  {Bizyaev}, {Blazek}, {Bolton}, {Brownstein}, {Burden}, {Chuang}, {Comparat},
  {Cuesta}, {Dawson}, {Eisenstein}, {Escoffier}, {Gil-Mar{\'\i}n}, {Grieb},
  {Hand}, {Ho}, {Kinemuchi}, {Kirkby}, {Kitaura}, {Malanushenko},
  {Malanushenko}, {Maraston}, {McBride}, {Nichol}, {Olmstead}, {Oravetz},
  {Padmanabhan}, {Palanque-Delabrouille}, {Pan}, {Pellejero-Ibanez},
  {Percival}, {Petitjean}, {Prada}, {Price-Whelan}, {Reid},
  {Rodr{\'\i}guez-Torres}, {Roe}, {Ross}, {Ross}, {Rossi},
  {Rubi{\~n}o-Mart{\'\i}n}, {Saito}, {Salazar-Albornoz}, {Samushia},
  {S{\'a}nchez}, {Satpathy}, {Schlegel}, {Schneider}, {Sc{\'o}ccola}, {Seo},
  {Sheldon}, {Simmons}, {Slosar}, {Strauss}, {Swanson}, {Thomas}, {Tinker},
  {Tojeiro}, {Maga{\~n}a}, {Vazquez}, {Verde}, {Wake}, {Wang}, {Weinberg},
  {White}, {Wood-Vasey}, {Y{\`e}che}, {Zehavi}, {Zhai}, \& {Zhao}}]{alam17}
{Alam}, S., {Ata}, M., {Bailey}, S., {et~al.} 2017, \bibinfo{title}{{The
  clustering of galaxies in the completed SDSS-III Baryon Oscillation
  Spectroscopic Survey: cosmological analysis of the DR12 galaxy sample},}
  \mnras, 470, 2617, \dodoi{10.1093/mnras/stx721}

\bibitem[{S. {Alam} {et~al.}(2021){Alam}, {Aubert}, {Avila}, {Balland},
  {Bautista}, {Bershady}, {Bizyaev}, {Blanton}, {Bolton}, {Bovy}, {Brinkmann},
  {Brownstein}, {Burtin}, {Chabanier}, {Chapman}, {Choi}, {Chuang}, {Comparat},
  {Cousinou}, {Cuceu}, {Dawson}, {de la Torre}, {de Mattia}, {Agathe}, {des
  Bourboux}, {Escoffier}, {Etourneau}, {Farr}, {Font-Ribera}, {Frinchaboy},
  {Fromenteau}, {Gil-Mar{\'\i}n}, {Le Goff}, {Gonzalez-Morales},
  {Gonzalez-Perez}, {Grabowski}, {Guy}, {Hawken}, {Hou}, {Kong}, {Parker},
  {Klaene}, {Kneib}, {Lin}, {Long}, {Lyke}, {de la Macorra}, {Martini},
  {Masters}, {Mohammad}, {Moon}, {Mueller}, {Mu{\~n}oz-Guti{\'e}rrez}, {Myers},
  {Nadathur}, {Neveux}, {Newman}, {Noterdaeme}, {Oravetz}, {Oravetz},
  {Palanque-Delabrouille}, {Pan}, {Paviot}, {Percival}, {P{\'e}rez-R{\`a}fols},
  {Petitjean}, {Pieri}, {Prakash}, {Raichoor}, {Ravoux}, {Rezaie}, {Rich},
  {Ross}, {Rossi}, {Ruggeri}, {Ruhlmann-Kleider}, {S{\'a}nchez}, {S{\'a}nchez},
  {S{\'a}nchez-Gallego}, {Sayres}, {Schneider}, {Seo}, {Shafieloo}, {Slosar},
  {Smith}, {Stermer}, {Tamone}, {Tinker}, {Tojeiro}, {Vargas-Maga{\~n}a},
  {Variu}, {Wang}, {Weaver}, {Weijmans}, {Y{\`e}che}, {Zarrouk}, {Zhao},
  {Zhao}, \& {Zheng}}]{alam21}
{Alam}, S., {Aubert}, M., {Avila}, S., {et~al.} 2021,
  \bibinfo{title}{{Completed SDSS-IV extended Baryon Oscillation Spectroscopic
  Survey: Cosmological implications from two decades of spectroscopic surveys
  at the Apache Point Observatory},} \prd, 103, 083533,
  \dodoi{10.1103/PhysRevD.103.083533}

\bibitem[{A.~J. {Anderson} {et~al.}(2022){Anderson}, {Barry}, {Bender},
  {Benson}, {Bleem}, {Carlstrom}, {Cecil}, {Chang}, {Crawford}, {Dibert},
  {Dobbs}, {Fichman}, {Halverson}, {Holzapfel}, {Hryciuk}, {Karkare}, {Li},
  {Lisovenko}, {Marrone}, {McMahon}, {Montgomery}, {Natoli}, {Pan},
  {Raghunathan}, {Reichardt}, {Rouble}, {Shirokoff}, {Smecher}, {Stark},
  {Vieira}, \& {Young}}]{anderson22}
{Anderson}, A.~J., {Barry}, P., {Bender}, A.~N., {et~al.} 2022,
  \bibinfo{title}{{SPT-3G+: mapping the high-frequency cosmic microwave
  background using kinetic inductance detectors},} in Society of Photo-Optical
  Instrumentation Engineers (SPIE) Conference Series, Vol. 12190, Millimeter,
  Submillimeter, and Far-Infrared Detectors and Instrumentation for Astronomy
  XI, ed. J.~{Zmuidzinas} \& J.-R. {Gao}, 1219003, \dodoi{10.1117/12.2629755}

\bibitem[{K. {Aylor} {et~al.}(2019){Aylor}, {Joy}, {Knox}, {Millea},
  {Raghunathan}, \& {Kimmy Wu}}]{aylor19}
{Aylor}, K., {Joy}, M., {Knox}, L., {et~al.} 2019, \bibinfo{title}{{Sounds
  Discordant: Classical Distance Ladder and {\ensuremath{\Lambda}}CDM-based
  Determinations of the Cosmological Sound Horizon},} \apj, 874, 4,
  \dodoi{10.3847/1538-4357/ab0898}

\bibitem[{L. {Balkenhol} {et~al.}(2021){Balkenhol}, {Dutcher}, {Ade}, {Ahmed},
  {Anderes}, {Anderson}, {Archipley}, {Avva}, {Aylor}, {Barry}, {Basu Thakur},
  {Benabed}, {Bender}, {Benson}, {Bianchini}, {Bleem}, {Bouchet}, {Bryant},
  {Byrum}, {Carlstrom}, {Carter}, {Cecil}, {Chang}, {Chaubal}, {Chen}, {Cho},
  {Chou}, {Cliche}, {Crawford}, {Cukierman}, {Daley}, {de Haan}, {Denison},
  {Dibert}, {Ding}, {Dobbs}, {Everett}, {Feng}, {Ferguson}, {Foster}, {Fu},
  {Galli}, {Gambrel}, {Gardner}, {Goeckner-Wald}, {Gualtieri}, {Guns}, {Gupta},
  {Guyser}, {Halverson}, {Harke-Hosemann}, {Harrington}, {Henning}, {Hilton},
  {Hivon}, {Holder}, {Holzapfel}, {Hood}, {Howe}, {Huang}, {Irwin}, {Jeong},
  {Jonas}, {Jones}, {Khaire}, {Knox}, {Kofman}, {Korman}, {Kubik}, {Kuhlmann},
  {Kuo}, {Lee}, {Leitch}, {Lowitz}, {Lu}, {Meyer}, {Michalik}, {Millea},
  {Montgomery}, {Nadolski}, {Natoli}, {Nguyen}, {Noble}, {Novosad}, {Omori},
  {Padin}, {Pan}, {Paschos}, {Pearson}, {Posada}, {Prabhu}, {Quan}, {Rahlin},
  {Reichardt}, {Riebel}, {Riedel}, {Rouble}, {Ruhl}, {Sayre}, {Schiappucci},
  {Shirokoff}, {Smecher}, {Sobrin}, {Stark}, {Stephen}, {Story}, {Suzuki},
  {Thompson}, {Thorne}, {Tucker}, {Umilta}, {Vale}, {Vanderlinde}, {Vieira},
  {Wang}, {Whitehorn}, {Wu}, {Yefremenko}, {Yoon}, {Young}, \& {SPT-3G
  Collaboration}}]{balkenhol21}
{Balkenhol}, L., {Dutcher}, D., {Ade}, P.~A.~R., {et~al.} 2021,
  \bibinfo{title}{{Constraints on {\ensuremath{\Lambda}} CDM extensions from
  the SPT-3G 2018 E E and T E power spectra},} \prd, 104, 083509,
  \dodoi{10.1103/PhysRevD.104.083509}

\bibitem[{A.~N. {Bender} {et~al.}(2018){Bender}, {Ade}, {Ahmed}, {Anderson},
  {Avva}, {Aylor}, {Barry}, {Basu Thakur}, {Benson}, {Bleem},
  {et~al.}}]{bender18}
{Bender}, A.~N., {Ade}, P.~A.~R., {Ahmed}, Z., {et~al.} 2018,
  \bibinfo{title}{{Year two instrument status of the SPT-3G cosmic microwave
  background receiver},} in SPIE Conference Series, Vol. 10708, Millimeter,
  Submillimeter, and Far-Infrared Detectors and Instrumentation for Astronomy
  IX, 1070803, \dodoi{10.1117/12.2312426}

\bibitem[{B.~A. {Benson} {et~al.}(2014){Benson}, {Ade}, {Ahmed}, {Allen},
  {Arnold}, {Austermann}, {Bender}, {Bleem}, {Carlstrom}, {Chang},
  {et~al.}}]{benson14}
{Benson}, B.~A., {Ade}, P.~A.~R., {Ahmed}, Z., {et~al.} 2014,
  \bibinfo{title}{{SPT-3G: a next-generation cosmic microwave background
  polarization experiment on the South Pole telescope},} in SPIE Conference
  Series, Vol. 9153, SPIE Conference Series, 1, \dodoi{10.1117/12.2057305}

\bibitem[{F. Bernardeau(1997)Bernardeau}]{bernardeau96}
Bernardeau, F. 1997, \bibinfo{title}{{Weak lensing detection in CMB maps},}
  Astron. Astrophys., 324, 15.
\newblock \doarXiv{astro-ph/9611012}

\bibitem[{R. {Besuner} {et~al.}(2025){Besuner}, {Dey}, {Drlica-Wagner},
  {Ebina}, {Fernandez Moroni}, {Ferraro}, {Forero-Romero}, {Honscheid},
  {Jelinsky}, {Lang}, {Levi}, {Martini}, {Myers}, {Palanque-Delabrouille},
  {Panda}, {Poppett}, {Sailer}, {Schlegel}, {Shafieloo}, {Silber}, {White},
  {Abbott}, {Allen}, {Avila}, {Avil{\'e}s}, {Bailey}, {Bault}, {Bouri},
  {Boutsia}, {Burtin}, {Chierchie}, {Coulton}, {Dawson}, {Dey}, {Dor{\'e}},
  {Dunlop}, {Eisenstein}, {Emanuele}, {Escoffier}, {Estrada}, {Fagrelius},
  {Fanning}, {Fanning}, {Font-Ribera}, {Frieman}, {Galal}, {Gluscevic},
  {Gontcho}, {Green}, {Gutierrez}, {Guy}, {Hashemi}, {Heathcote}, {Holland},
  {Hou}, {Huterer}, {Irigoyen Gimenez}, {Ivanov}, {Joyce}, {Jullo}, {Juneau},
  {Juramy}, {Karcher}, {Kent}, {Kirkby}, {Kneib}, {Krause}, {Krolewski},
  {Lahav}, {Lapi}, {Leauthaud}, {Lewandowski}, {Li}, {Lin}, {Loverde},
  {MacBride}, {Magneville}, {Marshall}, {McDonald}, {Miller}, {Moustakas},
  {M{\"u}nchmeyer}, {Najita}, {Newman}, {Percival}, {Philcox}, {Pires},
  {Raichoor}, {Roach}, {Rockosi}, {Rombach}, {Ross}, {Sanchez}, {Schmidt},
  {Schubnell}, {Sebok}, {Seljak}, {Silverstein}, {Slepian}, {Stone}, {Stupak},
  {Tarl{\'e}}, {Li}, {Tyas}, {Vargas-Maga{\~n}a}, {Walker}, {Wenner},
  {Y{\`e}che}, {Zhang}, \& {Zhou}}]{besuner25}
{Besuner}, R., {Dey}, A., {Drlica-Wagner}, A., {et~al.} 2025,
  \bibinfo{title}{{The Spectroscopic Stage-5 Experiment},} arXiv e-prints,
  arXiv:2503.07923, \dodoi{10.48550/arXiv.2503.07923}

\bibitem[{L.~E. {Bleem} {et~al.}(2012){Bleem}, {van Engelen}, {Holder}, {Aird},
  {Armstrong}, {Ashby}, {Becker}, {Benson}, {Biesiadzinski}, {Brodwin},
  {Busha}, {Carlstrom}, {Chang}, {Cho}, {Crawford}, {Crites}, {de Haan},
  {Desai}, {Dobbs}, {Dor{\'e}}, {Dudley}, {Geach}, {George}, {Gladders},
  {Gonzalez}, {Halverson}, {Harrington}, {High}, {Holden}, {Holzapfel},
  {Hoover}, {Hrubes}, {Joy}, {Keisler}, {Knox}, {Lee}, {Leitch}, {Lueker},
  {Luong-Van}, {Marrone}, {Martinez-Manso}, {McMahon}, {Mehl}, {Meyer}, {Mohr},
  {Montroy}, {Natoli}, {Padin}, {Plagge}, {Pryke}, {Reichardt}, {Rest}, {Ruhl},
  {Saliwanchik}, {Sayre}, {Schaffer}, {Shaw}, {Shirokoff}, {Spieler},
  {Stalder}, {Stanford}, {Staniszewski}, {Stark}, {Stern}, {Story},
  {Vallinotto}, {Vanderlinde}, {Vieira}, {Wechsler}, {Williamson}, \&
  {Zahn}}]{bleem12}
{Bleem}, L.~E., {van Engelen}, A., {Holder}, G.~P., {et~al.} 2012,
  \bibinfo{title}{{A Measurement of the Correlation of Galaxy Surveys with CMB
  Lensing Convergence Maps from the South Pole Telescope},} \apjl, 753, L9,
  \dodoi{10.1088/2041-8205/753/1/L9}

\bibitem[{S. {Bocquet} {et~al.}(2024){Bocquet}, {Grandis}, {Bleem}, {Klein},
  {Mohr}, {Schrabback}, {Abbott}, {Ade}, {Aguena}, {Alarcon}, {Allam}, {Allen},
  {Alves}, {Amon}, {Anderson}, {Annis}, {Ansarinejad}, {Austermann}, {Avila},
  {Bacon}, {Bayliss}, {Beall}, {Bechtol}, {Becker}, {Bender}, {Benson},
  {Bernstein}, {Bhargava}, {Bianchini}, {Brodwin}, {Brooks}, {Bryant},
  {Campos}, {Canning}, {Carlstrom}, {Carnero Rosell}, {Carrasco Kind},
  {Carretero}, {Castander}, {Cawthon}, {Chang}, {Chang}, {Chaubal}, {Chen},
  {Chiang}, {Choi}, {Chou}, {Citron}, {Corbett Moran}, {Cordero}, {Costanzi},
  {Crawford}, {Crites}, {da Costa}, {Pereira}, {Davis}, {Davis}, {DeRose},
  {Desai}, {de Haan}, {Diehl}, {Dobbs}, {Dodelson}, {Doux}, {Drlica-Wagner},
  {Eckert}, {Elvin-Poole}, {Everett}, {Everett}, {Ferrero}, {Fert{\'e}},
  {Flores}, {Frieman}, {Gallicchio}, {Garc{\'\i}a-Bellido}, {Gatti}, {George},
  {Giannini}, {Gladders}, {Gruen}, {Gruendl}, {Gupta}, {Gutierrez},
  {Halverson}, {Harrison}, {Hartley}, {Herner}, {Hinton}, {Holder},
  {Hollowood}, {Holzapfel}, {Honscheid}, {Hrubes}, {Huang}, {Hubmayr}, {Huff},
  {Huterer}, {Irwin}, {James}, {Jarvis}, {Khullar}, {Kim}, {Knox}, {Kraft},
  {Krause}, {Kuehn}, {Kuropatkin}, {K{\'e}ruzor{\'e}}, {Lahav}, {Lee}, {Leget},
  {Li}, {Lin}, {Lowitz}, {MacCrann}, {Mahler}, {Mantz}, {Marshall},
  {McCullough}, {McDonald}, {McMahon}, {Mena-Fern{\'a}ndez}, {Menanteau},
  {Meyer}, {Miquel}, {Montgomery}, {Myles}, {Natoli}, {Navarro-Alsina},
  {Nibarger}, {Noble}, {Novosad}, {Ogando}, {Omori}, {Padin}, {Pandey},
  {Paschos}, {Patil}, {Pieres}, {Plazas Malag{\'o}n}, {Porredon}, {Prat},
  {Pryke}, {Raveri}, {Reichardt}, {Roberson}, {Rollins}, {Romero}, {Roodman},
  {Ruhl}, {Rykoff}, {Saliwanchik}, {Salvati}, {S{\'a}nchez}, {Sanchez},
  {Sanchez Cid}, {Saro}, {Schaffer}, {Secco}, {Sevilla-Noarbe}, {Sharon},
  {Sheldon}, {Shin}, {Sievers}, {Smecher}, {Smith}, {Somboonpanyakul},
  {Sommer}, {Stalder}, {Stark}, {Stephen}, {Strazzullo}, {Suchyta}, {Tarle},
  {To}, {Troxel}, {Tucker}, {Tutusaus}, {Varga}, {Veach}, {Vieira},
  {Vikhlinin}, {von der Linden}, {Wang}, {Weaverdyck}, {Weller}, {Whitehorn},
  {Wu}, {Yanny}, {Yefremenko}, {Yin}, {Young}, {Zebrowski}, {Zhang}, {Zohren},
  {Zuntz}, {(SPT}, \& {DES Collaborations)}}]{bocquet24}
{Bocquet}, S., {Grandis}, S., {Bleem}, L.~E., {et~al.} 2024,
  \bibinfo{title}{{SPT clusters with DES and HST weak lensing. II. Cosmological
  constraints from the abundance of massive halos},} \prd, 110, 083510,
  \dodoi{10.1103/PhysRevD.110.083510}

\bibitem[{D. {Brout} {et~al.}(2021){Brout}, {Hinton}, \& {Scolnic}}]{brout21}
{Brout}, D., {Hinton}, S.~R., \& {Scolnic}, D. 2021, \bibinfo{title}{{Binning
  is Sinning (Supernova Version): The Impact of Self-calibration in
  Cosmological Analyses with Type Ia Supernovae},} \apjl, 912, L26,
  \dodoi{10.3847/2041-8213/abf4db}

\bibitem[{Y. {Cai} {et~al.}(2025){Cai}, {Ren}, {Qiu}, {Li}, \& {Zhang}}]{cai25}
{Cai}, Y., {Ren}, X., {Qiu}, T., {Li}, M., \& {Zhang}, X. 2025,
  \bibinfo{title}{{The Quintom theory of dark energy after DESI DR2},} arXiv
  e-prints, arXiv:2505.24732, \dodoi{10.48550/arXiv.2505.24732}

\bibitem[{R.~R. {Caldwell} \& E.~V. {Linder}(2025){Caldwell} \&
  {Linder}}]{caldwell25}
{Caldwell}, R.~R., \& {Linder}, E.~V. 2025, \bibinfo{title}{{Null Impact of the
  Null Energy Condition in Current Cosmology},} arXiv e-prints,
  arXiv:2511.07526, \dodoi{10.48550/arXiv.2511.07526}

\bibitem[{E. {Camphuis} {et~al.}(2025){Camphuis}, {Quan}, {Balkenhol},
  {Khalife}, {Ge}, {Guidi}, {Huang}, {Lynch}, {Omori}, {Trendafilova},
  {Anderson}, {Ansarinejad}, {Archipley}, {Barry}, {Benabed}, {Bender},
  {Benson}, {Bianchini}, {Bleem}, {Bouchet}, {Bryant}, {Campitiello},
  {Carlstrom}, {Chang}, {Chaubal}, {Chichura}, {Chokshi}, {Chou}, {Coerver},
  {Crawford}, {Daley}, {de Haan}, {Dibert}, {Dobbs}, {Doohan}, {Doussot},
  {Dutcher}, {Everett}, {Feng}, {Ferguson}, {Fichman}, {Foster}, {Galli},
  {Gambrel}, {Gardner}, {Goeckner-Wald}, {Gualtieri}, {Guns}, {Halverson},
  {Hivon}, {Holder}, {Holzapfel}, {Hood}, {Hryciuk}, {K{\'e}ruzor{\'e}},
  {Knox}, {Korman}, {Kornoelje}, {Kuo}, {Levy}, {Lowitz}, {Lu}, {Maniyar},
  {Martsen}, {Menanteau}, {Millea}, {Montgomery}, {Nakato}, {Natoli}, {Noble},
  {Ouellette}, {Pan}, {Paschos}, {Phadke}, {Pollak}, {Prabhu}, {Raghunathan},
  {Rahimi}, {Rahlin}, {Reichardt}, {Rouble}, {Ruhl}, {Schiappucci}, {Simpson},
  {Sobrin}, {Stark}, {Stephen}, {Tandoi}, {Thorne}, {Umilta}, {Vieira},
  {Vitrier}, {Wan}, {Whitehorn}, {Wu}, {Young}, \& {Zebrowski}}]{camphuis25}
{Camphuis}, E., {Quan}, W., {Balkenhol}, L., {et~al.} 2025,
  \bibinfo{title}{{SPT-3G D1: CMB temperature and polarization power spectra
  and cosmology from 2019 and 2020 observations of the SPT-3G Main field},}
  arXiv e-prints, arXiv:2506.20707, \dodoi{10.48550/arXiv.2506.20707}

\bibitem[{J.-F. {Cardoso} {et~al.}(2008){Cardoso}, {Le Jeune}, {Delabrouille},
  {Betoule}, \& {Patanchon}}]{cardoso08}
{Cardoso}, J.-F., {Le Jeune}, M., {Delabrouille}, J., {Betoule}, M., \&
  {Patanchon}, G. 2008, \bibinfo{title}{{Component Separation With Flexible
  Models{\textemdash}Application to Multichannel Astrophysical Observations},}
  IEEE Journal of Selected Topics in Signal Processing, 2, 735,
  \dodoi{10.1109/JSTSP.2008.2005346}

\bibitem[{M. {Chevallier} \& D. {Polarski}(2001){Chevallier} \&
  {Polarski}}]{chevallier01}
{Chevallier}, M., \& {Polarski}, D. 2001, \bibinfo{title}{{Accelerating
  Universes with Scaling Dark Matter},} International Journal of Modern Physics
  D, 10, 213, \dodoi{10.1142/S0218271801000822}

\bibitem[{ {CMB-S4 Collaboration}(2019){CMB-S4 Collaboration}}]{cmbs4collab19}
{CMB-S4 Collaboration}. 2019, \bibinfo{title}{{CMB-S4 Science Case, Reference
  Design, and Project Plan},} arXiv e-prints.
\newblock \doarXiv{1907.04473}

\bibitem[{ {CMB-S4 Collaboration} {et~al.}(2016){CMB-S4 Collaboration},
  {Abazajian}, {Adshead}, {Ahmed}, {Allen}, {Alonso}, {Arnold}, {Baccigalupi},
  {Bartlett}, {Battaglia}, {et~al.}}]{cmbs4-sb1}
{CMB-S4 Collaboration}, {Abazajian}, K.~N., {Adshead}, P., {et~al.} 2016,
  \bibinfo{title}{{CMB-S4 Science Book, First Edition},} ArXiv e-prints.
\newblock \doarXiv{1610.02743}

\bibitem[{A. {Cozzumbo} {et~al.}(2025){Cozzumbo}, {Atzori Corona}, {Murgia},
  {Archidiacono}, \& {Cadeddu}}]{cozzumbo25}
{Cozzumbo}, A., {Atzori Corona}, M., {Murgia}, R., {Archidiacono}, M., \&
  {Cadeddu}, M. 2025, \bibinfo{title}{{A short blanket for cosmology: the CMB
  lensing anomaly behind the preference for a negative neutrino mass},} arXiv
  e-prints, arXiv:2511.01967, \dodoi{10.48550/arXiv.2511.01967}

\bibitem[{N. {Craig} {et~al.}(2024){Craig}, {Green}, {Meyers}, \&
  {Rajendran}}]{craig24}
{Craig}, N., {Green}, D., {Meyers}, J., \& {Rajendran}, S. 2024,
  \bibinfo{title}{{No {\ensuremath{\nu}}s is Good News},} Journal of High
  Energy Physics, 2024, 97, \dodoi{10.1007/JHEP09(2024)097}

\bibitem[{ {Dark Energy Survey and Kilo-Degree Survey Collaboration}(2023){Dark
  Energy Survey and Kilo-Degree Survey Collaboration}}]{abbott23c}
{Dark Energy Survey and Kilo-Degree Survey Collaboration}. 2023,
  \bibinfo{title}{{DES Y3 + KiDS-1000: Consistent cosmology combining cosmic
  shear surveys},} The Open Journal of Astrophysics, 6, 36,
  \dodoi{10.21105/astro.2305.17173}

\bibitem[{R. {Datta} {et~al.}(2018){Datta}, {Aiola}, {Choi}, {Devlin},
  {Dunkley}, {D{\"u}nner}, {Gallardo}, {Gralla}, {Halpern}, {Hasselfield},
  {et~al.}}]{datta18}
{Datta}, R., {Aiola}, S., {Choi}, S.~K., {et~al.} 2018, \bibinfo{title}{{The
  Atacama Cosmology Telescope: Two-season ACTPol Extragalactic Point Sources
  and their Polarization properties},} \mnras, 2799,
  \dodoi{10.1093/mnras/sty2934}

\bibitem[{ {DES Collaboration} {et~al.}(2024{\natexlab{a}}){DES Collaboration},
  {Abbott}, {Acevedo}, {Aguena}, {Alarcon}, {Allam}, {Alves}, {Amon},
  {Andrade-Oliveira}, {Annis}, {Armstrong}, {Asorey}, {Avila}, {Bacon},
  {Bassett}, {Bechtol}, {Bernardinelli}, {Bernstein}, {Bertin}, {Blazek},
  {Bocquet}, {Brooks}, {Brout}, {Buckley-Geer}, {Burke}, {Camacho},
  {Camilleri}, {Campos}, {Carnero Rosell}, {Carollo}, {Carr}, {Carretero},
  {Castander}, {Cawthon}, {Chang}, {Chen}, {Choi}, {Conselice}, {Costanzi}, {da
  Costa}, {Crocce}, {Davis}, {DePoy}, {Desai}, {Diehl}, {Dixon}, {Dodelson},
  {Doel}, {Doux}, {Drlica-Wagner}, {Elvin-Poole}, {Everett}, {Ferrero},
  {Fert{\'e}}, {Flaugher}, {Foley}, {Fosalba}, {Friedel}, {Frieman},
  {Frohmaier}, {Galbany}, {Garc{\'\i}a-Bellido}, {Gatti}, {Gaztanaga},
  {Giannini}, {Glazebrook}, {Graur}, {Gruen}, {Gruendl}, {Gutierrez},
  {Hartley}, {Herner}, {Hinton}, {Hollowood}, {Honscheid}, {Huterer}, {Jain},
  {James}, {Jeffrey}, {Kasai}, {Kelsey}, {Kent}, {Kessler}, {Kim}, {Kirshner},
  {Kovacs}, {Kuehn}, {Lahav}, {Lee}, {Lee}, {Lewis}, {Li}, {Lidman}, {Lin},
  {Malik}, {Marshall}, {Martini}, {Mena-Fern{\'a}ndez}, {Menanteau}, {Miquel},
  {Mohr}, {Mould}, {Muir}, {M{\"o}ller}, {Neilsen}, {Nichol}, {Nugent},
  {Ogando}, {Palmese}, {Pan}, {Paterno}, {Percival}, {Pereira}, {Pieres},
  {Malag{\'o}n}, {Popovic}, {Porredon}, {Prat}, {Qu}, {Raveri},
  {Rodr{\'\i}guez-Monroy}, {Romer}, {Roodman}, {Rose}, {Sako}, {Sanchez},
  {Sanchez Cid}, {Schubnell}, {Scolnic}, {Sevilla-Noarbe}, {Shah}, {Smith},
  {Smith}, {Soares-Santos}, {Suchyta}, {Sullivan}, {Suntzeff}, {Swanson},
  {S{\'a}nchez}, {Tarle}, {Taylor}, {Thomas}, {To}, {Toy}, {Troxel}, {Tucker},
  {Tucker}, {Uddin}, {Vincenzi}, {Walker}, {Weaverdyck}, {Wechsler}, {Weller},
  {Wester}, {Wiseman}, {Yamamoto}, {Yuan}, {Zhang}, \&
  {Zhang}}]{des24_sne_cosmology}
{DES Collaboration}, {Abbott}, T.~M.~C., {Acevedo}, M., {et~al.}
  2024{\natexlab{a}}, \bibinfo{title}{{The Dark Energy Survey: Cosmology
  Results with {\ensuremath{\sim}}1500 New High-redshift Type Ia Supernovae
  Using the Full 5 yr Data Set},} \apjl, 973, L14,
  \dodoi{10.3847/2041-8213/ad6f9f}

\bibitem[{ {DES Collaboration} {et~al.}(2024{\natexlab{b}}){DES Collaboration},
  {Abbott}, {Acevedo}, {Aguena}, {Alarcon}, {Allam}, {Alves}, {Amon},
  {Andrade-Oliveira}, {Annis}, {Armstrong}, {Asorey}, {Avila}, {Bacon},
  {Bassett}, {Bechtol}, {Bernardinelli}, {Bernstein}, {Bertin}, {Blazek},
  {Bocquet}, {Brooks}, {Brout}, {Buckley-Geer}, {Burke}, {Camacho},
  {Camilleri}, {Campos}, {Carnero Rosell}, {Carollo}, {Carr}, {Carretero},
  {Castander}, {Cawthon}, {Chang}, {Chen}, {Choi}, {Conselice}, {Costanzi}, {da
  Costa}, {Crocce}, {Davis}, {DePoy}, {Desai}, {Diehl}, {Dixon}, {Dodelson},
  {Doel}, {Doux}, {Drlica-Wagner}, {Elvin-Poole}, {Everett}, {Ferrero},
  {Fert{\'e}}, {Flaugher}, {Foley}, {Fosalba}, {Friedel}, {Frieman},
  {Frohmaier}, {Galbany}, {Garc{\'\i}a-Bellido}, {Gatti}, {Gaztanaga},
  {Giannini}, {Glazebrook}, {Graur}, {Gruen}, {Gruendl}, {Gutierrez},
  {Hartley}, {Herner}, {Hinton}, {Hollowood}, {Honscheid}, {Huterer}, {Jain},
  {James}, {Jeffrey}, {Kasai}, {Kelsey}, {Kent}, {Kessler}, {Kim}, {Kirshner},
  {Kovacs}, {Kuehn}, {Lahav}, {Lee}, {Lee}, {Lewis}, {Li}, {Lidman}, {Lin},
  {Malik}, {Marshall}, {Martini}, {Mena-Fern{\'a}ndez}, {Menanteau}, {Miquel},
  {Mohr}, {Mould}, {Muir}, {M{\"o}ller}, {Neilsen}, {Nichol}, {Nugent},
  {Ogando}, {Palmese}, {Pan}, {Paterno}, {Percival}, {Pereira}, {Pieres},
  {Malag{\'o}n}, {Popovic}, {Porredon}, {Prat}, {Qu}, {Raveri},
  {Rodr{\'\i}guez-Monroy}, {Romer}, {Roodman}, {Rose}, {Sako}, {Sanchez},
  {Sanchez Cid}, {Schubnell}, {Scolnic}, {Sevilla-Noarbe}, {Shah}, {Smith},
  {Smith}, {Soares-Santos}, {Suchyta}, {Sullivan}, {Suntzeff}, {Swanson},
  {S{\'a}nchez}, {Tarle}, {Taylor}, {Thomas}, {To}, {Toy}, {Troxel}, {Tucker},
  {Tucker}, {Uddin}, {Vincenzi}, {Walker}, {Weaverdyck}, {Wechsler}, {Weller},
  {Wester}, {Wiseman}, {Yamamoto}, {Yuan}, {Zhang}, \& {Zhang}}]{abbott24}
{DES Collaboration}, {Abbott}, T.~M.~C., {Acevedo}, M., {et~al.}
  2024{\natexlab{b}}, \bibinfo{title}{{The Dark Energy Survey: Cosmology
  Results with {\ensuremath{\sim}}1500 New High-redshift Type Ia Supernovae
  Using the Full 5 yr Data Set},} \apjl, 973, L14,
  \dodoi{10.3847/2041-8213/ad6f9f}

\bibitem[{ {DESI Collaboration}(2023){DESI Collaboration}}]{desidr3}
{DESI Collaboration}. 2023, Validation of the Scientific Program for the Dark
  Energy Spectroscopic Instrument, Zenodo, \dodoi{10.5281/zenodo.10063934}

\bibitem[{ {DESI Collaboration} {et~al.}(2025){DESI Collaboration},
  {Abdul-Karim}, {Aguilar}, {Ahlen}, {Alam}, {et~al.}}]{abdulkarim25}
{DESI Collaboration}, {Abdul-Karim}, M., {Aguilar}, J., {et~al.} 2025,
  \bibinfo{title}{{DESI DR2 Results II: Measurements of Baryon Acoustic
  Oscillations and Cosmological Constraints},} arXiv e-prints,
  arXiv:2503.14738, \dodoi{10.48550/arXiv.2503.14738}

\bibitem[{ {DESI Collaboration} {et~al.}(2016){DESI Collaboration},
  {Aghamousa}, {Aguilar}, {Ahlen}, {Alam}, {Allen}, {Allende Prieto}, {Annis},
  {Bailey}, {Balland}, {Ballester}, {Baltay}, {Beaufore}, {Bebek}, {Beers},
  {Bell}, {Bernal}, {Besuner}, {Beutler}, {Blake}, {Bleuler}, {Blomqvist},
  {Blum}, {Bolton}, {Briceno}, {Brooks}, {Brownstein}, {Buckley-Geer},
  {Burden}, {Burtin}, {Busca}, {Cahn}, {Cai}, {Cardiel-Sas}, {Carlberg},
  {Carton}, {Casas}, {Castander}, {Cervantes-Cota}, {Claybaugh}, {Close},
  {Coker}, {Cole}, {Comparat}, {Cooper}, {Cousinou}, {Crocce}, {Cuby},
  {Cunningham}, {Davis}, {Dawson}, {de la Macorra}, {De Vicente}, {Delubac},
  {Derwent}, {Dey}, {Dhungana}, {Ding}, {Doel}, {Duan}, {Ealet}, {Edelstein},
  {Eftekharzadeh}, {Eisenstein}, {Elliott}, {Escoffier}, {Evatt}, {Fagrelius},
  {Fan}, {Fanning}, {Farahi}, {Farihi}, {Favole}, {Feng}, {Fernandez},
  {Findlay}, {Finkbeiner}, {Fitzpatrick}, {Flaugher}, {Flender}, {Font-Ribera},
  {Forero-Romero}, {Fosalba}, {Frenk}, {Fumagalli}, {Gaensicke}, {Gallo},
  {Garcia-Bellido}, {Gaztanaga}, {Pietro Gentile Fusillo}, {Gerard},
  {Gershkovich}, {Giannantonio}, {Gillet}, {Gonzalez-de-Rivera},
  {Gonzalez-Perez}, {Gott}, {Graur}, {Gutierrez}, {Guy}, {Habib}, {Heetderks},
  {Heetderks}, {Heitmann}, {Hellwing}, {Herrera}, {Ho}, {Holland}, {Honscheid},
  {Huff}, {Hutchinson}, {Huterer}, {Hwang}, {Illa Laguna}, {Ishikawa},
  {Jacobs}, {Jeffrey}, {Jelinsky}, {Jennings}, {Jiang}, {Jimenez}, {Johnson},
  {Joyce}, {Jullo}, {Juneau}, {Kama}, {Karcher}, {Karkar}, {Kehoe}, {Kennamer},
  {Kent}, {Kilbinger}, {Kim}, {Kirkby}, {Kisner}, {Kitanidis}, {Kneib},
  {Koposov}, {Kovacs}, {Koyama}, {Kremin}, {Kron}, {Kronig}, {Kueter-Young},
  {Lacey}, {Lafever}, {Lahav}, {Lambert}, {Lampton}, {Landriau}, {Lang},
  {Lauer}, {Le Goff}, {Le Guillou}, {Le Van Suu}, {Lee}, {Lee}, {Leitner},
  {Lesser}, {Levi}, {L'Huillier}, {Li}, {Liang}, {Lin}, {Linder}, {Loebman},
  {Luki{\'c}}, {Ma}, {MacCrann}, {Magneville}, {Makarem}, {Manera}, {Manser},
  {Marshall}, {Martini}, {Massey}, {Matheson}, {McCauley}, {McDonald},
  {McGreer}, {Meisner}, {Metcalfe}, {Miller}, {Miquel}, {Moustakas}, {Myers},
  {Naik}, {Newman}, {Nichol}, {Nicola}, {Nicolati da Costa}, {Nie}, {Niz},
  {Norberg}, {Nord}, {Norman}, {Nugent}, {O'Brien}, {Oh}, \& {Olsen}}]{desi16}
{DESI Collaboration}, {Aghamousa}, A., {Aguilar}, J., {et~al.} 2016,
  \bibinfo{title}{{The DESI Experiment Part I: Science,Targeting, and Survey
  Design},} arXiv e-prints, arXiv:1611.00036, \dodoi{10.48550/arXiv.1611.00036}

\bibitem[{W. {Elbers}(2025){Elbers}}]{elbers25c}
{Elbers}, W. 2025, \bibinfo{title}{{Rapid late-time reionization: constraints
  and cosmological implications},} arXiv e-prints, arXiv:2508.21069,
  \dodoi{10.48550/arXiv.2508.21069}

\bibitem[{W. {Elbers} {et~al.}(2025){Elbers}, {Frenk}, {Jenkins}, {Li}, \&
  {Pascoli}}]{elbers25a}
{Elbers}, W., {Frenk}, C.~S., {Jenkins}, A., {Li}, B., \& {Pascoli}, S. 2025,
  \bibinfo{title}{{Negative neutrino masses as a mirage of dark energy},} \prd,
  111, 063534, \dodoi{10.1103/PhysRevD.111.063534}

\bibitem[{W. {Fang} {et~al.}(2008){Fang}, {Hu}, \& {Lewis}}]{fang08}
{Fang}, W., {Hu}, W., \& {Lewis}, A. 2008, \bibinfo{title}{{Crossing the
  phantom divide with parametrized post-Friedmann dark energy},} \prd, 78,
  087303, \dodoi{10.1103/PhysRevD.78.087303}

\bibitem[{B. {Feng} {et~al.}(2005){Feng}, {Wang}, \& {Zhang}}]{feng05}
{Feng}, B., {Wang}, X., \& {Zhang}, X. 2005, \bibinfo{title}{{Dark energy
  constraints from the cosmic age and supernova},} Physics Letters B, 607, 35,
  \dodoi{10.1016/j.physletb.2004.12.071}

\bibitem[{S. {Ferraro} \& J.~C. {Hill}(2018){Ferraro} \& {Hill}}]{ferraro18}
{Ferraro}, S., \& {Hill}, J.~C. 2018, \bibinfo{title}{{Bias to CMB lensing
  reconstruction from temperature anisotropies due to large-scale galaxy
  motions},} \prd, 97, 023512, \dodoi{10.1103/PhysRevD.97.023512}

\bibitem[{E.~G.~M. {Ferreira} {et~al.}(2025){Ferreira}, {McDonough},
  {Balkenhol}, {Kallosh}, {Knox}, \& {Linde}}]{ferreira25}
{Ferreira}, E. G.~M., {McDonough}, E., {Balkenhol}, L., {et~al.} 2025,
  \bibinfo{title}{{The BAO-CMB Tension and Implications for Inflation},} arXiv
  e-prints, arXiv:2507.12459, \dodoi{10.48550/arXiv.2507.12459}

\bibitem[{C. Frohmaier {et~al.}(2025)Frohmaier {et~al.}}]{frohmaier25}
Frohmaier, C., {et~al.} 2025, \bibinfo{title}{{TiDES: The 4MOST Time Domain
  Extragalactic Survey},} Astrophys. J., \dodoi{10.3847/1538-4357/adff4e}

\bibitem[{S. {Galli} {et~al.}(2014){Galli}, {Benabed}, {Bouchet}, {Cardoso},
  {Elsner}, {Hivon}, {Mangilli}, {Prunet}, \& {Wandelt}}]{galli14}
{Galli}, S., {Benabed}, K., {Bouchet}, F., {et~al.} 2014, \bibinfo{title}{{CMB
  polarization can constrain cosmology better than CMB temperature},} \prd, 90,
  063504, \dodoi{10.1103/PhysRevD.90.063504}

\bibitem[{F. {Ge} {et~al.}(2025){Ge}, {Millea}, {Camphuis}, {Daley}, {Huang},
  {Omori}, {Quan}, {Anderes}, {Anderson}, {Ansarinejad}, {Archipley},
  {Balkenhol}, {Benabed}, {Bender}, {Benson}, {Bianchini}, {Bleem}, {Bouchet},
  {Bryant}, {Carlstrom}, {Chang}, {Chaubal}, {Chen}, {Chichura}, {Chokshi},
  {Chou}, {Coerver}, {Crawford}, {de Haan}, {Dibert}, {Dobbs}, {Doohan},
  {Doussot}, {Dutcher}, {Everett}, {Feng}, {Ferguson}, {Fichman}, {Foster},
  {Galli}, {Gambrel}, {Gardner}, {Goeckner-Wald}, {Gualtieri}, {Guidi}, {Guns},
  {Halverson}, {Hivon}, {Holder}, {Holzapfel}, {Hood}, {Howe}, {Hryciuk},
  {K{\'e}ruzor{\'e}}, {Khalife}, {Knox}, {Korman}, {Kornoelje}, {Kuo}, {Lee},
  {Levy}, {Lowitz}, {Lu}, {Maniyar}, {Martsen}, {Menanteau}, {Montgomery},
  {Nakato}, {Natoli}, {Noble}, {Pan}, {Paschos}, {Phadke}, {Pollak}, {Prabhu},
  {Rahimi}, {Rahlin}, {Reichardt}, {Riebel}, {Rouble}, {Ruhl}, {Schiappucci},
  {Sobrin}, {Stark}, {Stephen}, {Tandoi}, {Thorne}, {Trendafilova}, {Umilta},
  {Vieira}, {Vitrier}, {Wan}, {Whitehorn}, {Wu}, {Young}, {Zebrowski}, \&
  {SPT-3G Collaboration}}]{ge25}
{Ge}, F., {Millea}, M., {Camphuis}, E., {et~al.} 2025,
  \bibinfo{title}{{Cosmology from CMB lensing and delensed
  <inline-formula><mml:math><mml:mi>E</mml:mi><mml:mi>E</mml:mi></mml:math></inline-formula>
  power spectra using 2019{\textendash}2020 SPT-3G polarization data},} \prd,
  111, 083534, \dodoi{10.1103/PhysRevD.111.083534}

\bibitem[{T. {Giannantonio} {et~al.}(2016){Giannantonio}, {Fosalba}, {Cawthon},
  {Omori}, {Crocce}, {Elsner}, {Leistedt}, {Dodelson}, {Benoit-L{\'e}vy},
  {Gazta{\~n}aga}, {Holder}, {Peiris}, {Percival}, {Kirk}, {Bauer}, {Benson},
  {Bernstein}, {Carretero}, {Crawford}, {Crittenden}, {Huterer}, {Jain},
  {Krause}, {Reichardt}, {Ross}, {Simard}, {Soergel}, {Stark}, {Story},
  {Vieira}, {Weller}, {Abbott}, {Abdalla}, {Allam}, {Armstrong}, {Banerji},
  {Bernstein}, {Bertin}, {Brooks}, {Buckley-Geer}, {Burke}, {Capozzi},
  {Carlstrom}, {Carnero Rosell}, {Carrasco Kind}, {Castander}, {Chang},
  {Cunha}, {da Costa}, {D'Andrea}, {DePoy}, {Desai}, {Diehl}, {Dietrich},
  {Doel}, {Eifler}, {Evrard}, {Neto}, {Fernandez}, {Finley}, {Flaugher},
  {Frieman}, {Gerdes}, {Gruen}, {Gruendl}, {Gutierrez}, {Holzapfel},
  {Honscheid}, {James}, {Kuehn}, {Kuropatkin}, {Lahav}, {Li}, {Lima}, {March},
  {Marshall}, {Martini}, {Melchior}, {Miquel}, {Mohr}, {Nichol}, {Nord},
  {Ogando}, {Plazas}, {Romer}, {Roodman}, {Rykoff}, {Sako}, {Saliwanchik},
  {Sanchez}, {Schubnell}, {Sevilla-Noarbe}, {Smith}, {Soares-Santos},
  {Sobreira}, {Suchyta}, {Swanson}, {Tarle}, {Thaler}, {Thomas}, {Vikram},
  {Walker}, {Wechsler}, \& {Zuntz}}]{giannantonio16}
{Giannantonio}, T., {Fosalba}, P., {Cawthon}, R., {et~al.} 2016,
  \bibinfo{title}{{CMB lensing tomography with the DES Science Verification
  galaxies},} \mnras, 456, 3213, \dodoi{10.1093/mnras/stv2678}

\bibitem[{W. {Giar{\`e}} {et~al.}(2025{\natexlab{a}}){Giar{\`e}}, {Mahassen},
  {Valentino}, \& {Pan}}]{giare25}
{Giar{\`e}}, W., {Mahassen}, T., {Valentino}, E.~D., \& {Pan}, S.
  2025{\natexlab{a}}, \bibinfo{title}{{An overview of what current data can
  (and cannot yet) say about evolving dark energy},} Physics of the Dark
  Universe, 48, 101906, \dodoi{10.1016/j.dark.2025.101906}

\bibitem[{W. {Giar{\`e}} {et~al.}(2025{\natexlab{b}}){Giar{\`e}}, {Mena},
  {Specogna}, \& {Di Valentino}}]{giare25b}
{Giar{\`e}}, W., {Mena}, O., {Specogna}, E., \& {Di Valentino}, E.
  2025{\natexlab{b}}, \bibinfo{title}{{Neutrino mass tension or suppressed
  growth rate of matter perturbations?},} \prd, 112, 103520,
  \dodoi{10.1103/njfc-pd1w}

\bibitem[{M. {Glowacki} \& K.-G. {Lee}(2026){Glowacki} \& {Lee}}]{glowacki26}
{Glowacki}, M., \& {Lee}, K.-G. 2026, \bibinfo{title}{{Cosmology with fast
  radio bursts},} in Encyclopedia of Astrophysics, Volume 5, Vol.~5, 448--470,
  \dodoi{10.1016/B978-0-443-21439-4.00073-0}

\bibitem[{D. {Green} \& J. {Meyers}(2025){Green} \& {Meyers}}]{green25}
{Green}, D., \& {Meyers}, J. 2025, \bibinfo{title}{{Cosmological preference for
  a negative neutrino mass},} \prd, 111, 083507,
  \dodoi{10.1103/PhysRevD.111.083507}

\bibitem[{N. {Gupta} {et~al.}(2019){Gupta}, {Reichardt}, {Ade}, {Anderson},
  {Archipley}, {Austermann}, {Avva}, {Beall}, {Bender}, {Benson}, {Bianchini},
  {Bleem}, {Carlstrom}, {Chang}, {Chiang}, {Citron}, {Moran}, {Crawford},
  {Crites}, {de Haan}, {Dobbs}, {Everett}, {Feng}, {Gallicchio}, {George},
  {Gilbert}, {Halverson}, {Harrington}, {Henning}, {Hilton}, {Holder},
  {Holzapfel}, {Hou}, {Hrubes}, {Huang}, {Hubmayr}, {Irwin}, {Knox}, {Lee},
  {Li}, {Lowitz}, {Luong-Van}, {Marrone}, {McMahon}, {Meyer}, {Mocanu}, {Mohr},
  {Montgomery}, {Nadolski}, {Natoli}, {Nibarger}, {Noble}, {Novosad}, {Padin},
  {Patil}, {Pryke}, {Ruhl}, {Saliwanchik}, {Sayre}, {Schaffer}, {Shirokoff},
  {Sievers}, {Smecher}, {Staniszewski}, {Stark}, {Story}, {Switzer}, {Tucker},
  {Vanderlinde}, {Veach}, {Vieira}, {Wang}, {Whitehorn}, {Williamson}, {Wu},
  {Yefremenko}, \& {Zhang}}]{gupta19}
{Gupta}, N., {Reichardt}, C.~L., {Ade}, P.~A.~R., {et~al.} 2019,
  \bibinfo{title}{{Fractional polarization of extragalactic sources in the 500
  deg$^{2}$ SPTpol survey},} \mnras, 490, 5712, \dodoi{10.1093/mnras/stz2905}

\bibitem[{J. {Guy} {et~al.}(2007){Guy}, {Astier}, {Baumont}, {Hardin}, {Pain},
  {Regnault}, {Basa}, {Carlberg}, {Conley}, {Fabbro}, {Fouchez}, {Hook},
  {Howell}, {Perrett}, {Pritchet}, {Rich}, {Sullivan}, {Antilogus}, {Aubourg},
  {Bazin}, {Bronder}, {Filiol}, {Palanque-Delabrouille}, {Ripoche}, \&
  {Ruhlmann-Kleider}}]{guy07}
{Guy}, J., {Astier}, P., {Baumont}, S., {et~al.} 2007, \bibinfo{title}{{SALT2:
  using distant supernovae to improve the use of type Ia supernovae as distance
  indicators},} Astron. Astrophys., 466, 11, \dodoi{10.1051/0004-6361:20066930}

\bibitem[{C. {Heymans} {et~al.}(2021){Heymans}, {Tr{\"o}ster}, {Asgari},
  {Blake}, {Hildebrandt}, {Joachimi}, {Kuijken}, {Lin}, {S{\'a}nchez}, {van den
  Busch}, {Wright}, {Amon}, {Bilicki}, {de Jong}, {Crocce}, {Dvornik}, {Erben},
  {Fortuna}, {Getman}, {Giblin}, {Glazebrook}, {Hoekstra}, {Joudaki},
  {Kannawadi}, {K{\"o}hlinger}, {Lidman}, {Miller}, {Napolitano}, {Parkinson},
  {Schneider}, {Shan}, {Valentijn}, {Verdoes Kleijn}, \& {Wolf}}]{heymans20}
{Heymans}, C., {Tr{\"o}ster}, T., {Asgari}, M., {et~al.} 2021,
  \bibinfo{title}{{KiDS-1000 Cosmology: Multi-probe weak gravitational lensing
  and spectroscopic galaxy clustering constraints},} \aap, 646, A140,
  \dodoi{10.1051/0004-6361/202039063}

\bibitem[{R. {Hlo{\v{z}}ek} {et~al.}(2020){Hlo{\v{z}}ek}, {Ponder}, {Malz},
  {Dai}, {Narayan}, {Ishida}, {Allam}, {Bahmanyar}, {Biswas}, {Galbany}, {Jha},
  {Jones}, {Kessler}, {Lochner}, {Mahabal}, {Mandel}, {Mart{\'\i}nez-Galarza},
  {McEwen}, {Muthukrishna}, {Peiris}, {Peters}, \& {Setzer}}]{hlozek20}
{Hlo{\v{z}}ek}, R., {Ponder}, K.~A., {Malz}, A.~I., {et~al.} 2020,
  \bibinfo{title}{{Results of the Photometric LSST Astronomical Time-series
  Classification Challenge (PLAsTiCC)},} arXiv e-prints, arXiv:2012.12392.
\newblock \doarXiv{2012.12392}

\bibitem[{D.~W. {Hogg}(1999){Hogg}}]{hogg99}
{Hogg}, D.~W. 1999, \bibinfo{title}{{Distance measures in cosmology},} arXiv
  e-prints, astro, \dodoi{10.48550/arXiv.astro-ph/9905116}

\bibitem[{R. {Hounsell} {et~al.}(2023){Hounsell}, {Scolnic}, {Brout}, {Rose},
  {Fox}, {Sako}, {Macias}, {Joshi}, {Desutua}, {Rubin}, {Casertano},
  {Perlmutter}, {Aldering}, {Mandel}, {Sosey}, {Suzuki}, \&
  {Ryan}}]{hounsell23}
{Hounsell}, R., {Scolnic}, D., {Brout}, D., {et~al.} 2023,
  \bibinfo{title}{{Roman CCS White Paper: Measuring Type Ia Supernovae
  Discovered in the Roman High Latitude Time Domain Survey},} arXiv e-prints,
  arXiv:2307.02670, \dodoi{10.48550/arXiv.2307.02670}

\bibitem[{W. {Hu} \& T. {Okamoto}(2002){Hu} \& {Okamoto}}]{hu02}
{Hu}, W., \& {Okamoto}, T. 2002, \bibinfo{title}{{Mass Reconstruction with
  Cosmic Microwave Background Polarization},} \apj, 574, 566,
  \dodoi{10.1086/341110}

\bibitem[{S. {Joudaki} {et~al.}(2020){Joudaki}, {Hildebrandt}, {Traykova},
  {Chisari}, {Heymans}, {Kannawadi}, {Kuijken}, {Wright}, {Asgari}, {Erben},
  {Hoekstra}, {Joachimi}, {Miller}, {Tr{\"o}ster}, \& {van den
  Busch}}]{joudaki20}
{Joudaki}, S., {Hildebrandt}, H., {Traykova}, D., {et~al.} 2020,
  \bibinfo{title}{{KiDS+VIKING-450 and DES-Y1 combined: Cosmology with cosmic
  shear},} \aap, 638, L1, \dodoi{10.1051/0004-6361/201936154}

\bibitem[{G. {Jungman} {et~al.}(1996){Jungman}, {Kamionkowski}, {Kosowsky}, \&
  {Spergel}}]{jungman96}
{Jungman}, G., {Kamionkowski}, M., {Kosowsky}, A., \& {Spergel}, D.~N. 1996,
  \bibinfo{title}{{Cosmological-parameter determination with microwave
  background maps},} \prd, 54, 1332, \dodoi{10.1103/PhysRevD.54.1332}

\bibitem[{T. {Karwal} \& M. {Kamionkowski}(2016){Karwal} \&
  {Kamionkowski}}]{karwal16}
{Karwal}, T., \& {Kamionkowski}, M. 2016, \bibinfo{title}{{Dark energy at early
  times, the Hubble parameter, and the string axiverse},} \prd, 94, 103523,
  \dodoi{10.1103/PhysRevD.94.103523}

\bibitem[{R. {Kessler} {et~al.}(2010){Kessler}, {Cinabro}, {Bassett}, {Dilday},
  {Frieman}, {et~al.}}]{kessler10}
{Kessler}, R., {Cinabro}, D., {Bassett}, B., {et~al.} 2010,
  \bibinfo{title}{{Photometric Estimates of Redshifts and Distance Moduli for
  Type Ia Supernovae},} Astrophys. J., 717, 40,
  \dodoi{10.1088/0004-637X/717/1/40}

\bibitem[{R. {Kessler} \& D. {Scolnic}(2017){Kessler} \& {Scolnic}}]{kessler17}
{Kessler}, R., \& {Scolnic}, D. 2017, \bibinfo{title}{{Correcting Type Ia
  Supernova Distances for Selection Biases and Contamination in Photometrically
  Identified Samples},} Astrophys. J., 836, 56,
  \dodoi{10.3847/1538-4357/836/1/56}

\bibitem[{R. {Kessler} {et~al.}(2023){Kessler}, {Vincenzi}, \&
  {Armstrong}}]{kessler23}
{Kessler}, R., {Vincenzi}, M., \& {Armstrong}, P. 2023,
  \bibinfo{title}{{Binning is Sinning: Redemption for Hubble Diagram Using
  Photometrically Classified Type Ia Supernovae},} \apjl, 952, L8,
  \dodoi{10.3847/2041-8213/ace34d}

\bibitem[{R. {Kessler} {et~al.}(2019){Kessler}, {Narayan}, {Avelino},
  {Bachelet}, {Biswas}, {Brown}, {Chernoff}, {Connolly}, {Dai}, {Daniel}, {Di
  Stefano}, {Drout}, {Galbany}, {Gonz{\'a}lez-Gait{\'a}n}, {Graham},
  {Hlo{\v{z}}ek}, {Ishida}, {Guillochon}, {Jha}, {Jones}, {Mandel},
  {Muthukrishna}, {O'Grady}, {Peters}, {Pierel}, {Ponder}, {Pr{\v{s}}a},
  {Rodney}, {Villar}, {LSST Dark Energy Science Collaboration}, \& {Transient
  and Variable Stars Science Collaboration}}]{kessler19}
{Kessler}, R., {Narayan}, G., {Avelino}, A., {et~al.} 2019,
  \bibinfo{title}{{Models and Simulations for the Photometric LSST Astronomical
  Time Series Classification Challenge (PLAsTiCC)},} Publ. Astron. Soc. Pac.,
  131, 094501, \dodoi{10.1088/1538-3873/ab26f1}

\bibitem[{A.~R. {Khalife} {et~al.}(2024){Khalife}, {Zanjani}, {Galli},
  {G{\"u}nther}, {Lesgourgues}, \& {Benabed}}]{khalife23}
{Khalife}, A.~R., {Zanjani}, M.~B., {Galli}, S., {et~al.} 2024,
  \bibinfo{title}{{Review of Hubble tension solutions with new SH0ES and SPT-3G
  data},} \jcap, 2024, 059, \dodoi{10.1088/1475-7516/2024/04/059}

\bibitem[{L. {Knox} \& M. {Millea}(2020){Knox} \& {Millea}}]{knox20}
{Knox}, L., \& {Millea}, M. 2020, \bibinfo{title}{{Hubble constant hunter's
  guide},} \prd, 101, 043533, \dodoi{10.1103/PhysRevD.101.043533}

\bibitem[{A. {Krolewski} {et~al.}(2021){Krolewski}, {Ferraro}, \&
  {White}}]{krolewski21}
{Krolewski}, A., {Ferraro}, S., \& {White}, M. 2021,
  \bibinfo{title}{{Cosmological constraints from unWISE and Planck CMB lensing
  tomography},} \jcap, 2021, 028, \dodoi{10.1088/1475-7516/2021/12/028}

\bibitem[{A. {Lewis}(2013){Lewis}}]{lewis13}
{Lewis}, A. 2013, \bibinfo{title}{{Efficient sampling of fast and slow
  cosmological parameters},} \prd, 87, 103529,
  \dodoi{10.1103/PhysRevD.87.103529}

\bibitem[{A. {Lewis} {et~al.}(2000){Lewis}, {Challinor}, \&
  {Lasenby}}]{lewis00}
{Lewis}, A., {Challinor}, A., \& {Lasenby}, A. 2000, \bibinfo{title}{{Efficient
  Computation of Cosmic Microwave Background Anisotropies in Closed
  Friedmann-Robertson-Walker Models},} \apj, 538, 473, \dodoi{10.1086/309179}

\bibitem[{M.-X. {Lin} {et~al.}(2019){Lin}, {Raveri}, \& {Hu}}]{lin19}
{Lin}, M.-X., {Raveri}, M., \& {Hu}, W. 2019, \bibinfo{title}{{Phenomenology of
  modified gravity at recombination},} \prd, 99, 043514,
  \dodoi{10.1103/PhysRevD.99.043514}

\bibitem[{E.~V. {Linder}(2003){Linder}}]{linder03}
{Linder}, E.~V. 2003, \bibinfo{title}{{Exploring the Expansion History of the
  Universe},} \prl, 90, 091301, \dodoi{10.1103/PhysRevLett.90.091301}

\bibitem[{K. {Lodha} {et~al.}(2025){Lodha}, {Calderon}, {Matthewson},
  {Shafieloo}, {Ishak}, {Pan}, {Garcia-Quintero}, {Huterer}, {Valogiannis},
  {Ure{\~n}a-L{\'o}pez}, {Kamble}, {Parkinson}, {Kim}, {Zhao},
  {Cervantes-Cota}, {Rohlf}, {Lozano-Rodr{\'\i}guez}, {Rom{\'a}n-Herrera},
  {Abdul-Karim}, {Aguilar}, {Ahlen}, {Alves}, {Andrade}, {Armengaud}, {Aviles},
  {Behera}, {BenZvi}, {Bianchi}, {Brodzeller}, {Brooks}, {Burtin}, {Canning},
  {Rosell}, {Casas}, {Castander}, {Charles}, {Chaussidon}, {Chaves-Montero},
  {Chebat}, {Claybaugh}, {Cole}, {Cuceu}, {Dawson}, {de la Macorra}, {de
  Mattia}, {Deiosso}, {Demina}, {Dey}, {Dey}, {Ding}, {Doel}, {Eisenstein},
  {Elbers}, {Ferraro}, {Font-Ribera}, {Forero-Romero}, {Garrison},
  {Gazta{\~n}aga}, {Gil-Mar{\'\i}n}, {Gontcho}, {Gonzalez-Morales},
  {Gutierrez}, {Guy}, {Hahn}, {Herbold}, {Herrera-Alcantar}, {Honscheid},
  {Howlett}, {Juneau}, {Kehoe}, {Kirkby}, {Kisner}, {Kremin}, {Lahav},
  {Lamman}, {Landriau}, {Le Guillou}, {Leauthaud}, {Levi}, {Li}, {Magneville},
  {Manera}, {Martini}, {Meisner}, {Mena-Fern{\'a}ndez}, {Miquel}, {Moustakas},
  {Santos}, {Mu{\~n}oz-Guti{\'e}rrez}, {Myers}, {Nadathur}, {Niz}, {Noriega},
  {Paillas}, {Palanque-Delabrouille}, {Percival}, {Pieri}, {Poppett}, {Prada},
  {P{\'e}rez-Fern{\'a}ndez}, {P{\'e}rez-R{\`a}fols}, {Ram{\'\i}rez-P{\'e}rez},
  {Rashkovetskyi}, {Ravoux}, {Ross}, {Rossi}, {Ruhlmann-Kleider}, {Samushia},
  {Sanchez}, {Schlegel}, {Schubnell}, {Seo}, {Sinigaglia}, {Sprayberry}, {Tan},
  {Tarl{\'e}}, {Taylor}, {Turner}, {Vargas-Maga{\~n}a}, {Walther}, {Weaver},
  {Wolfson}, {Y{\`e}che}, {Zarrouk}, {Zhou}, {Zou}, \& {DESI
  Collaboration}}]{lodha25}
{Lodha}, K., {Calderon}, R., {Matthewson}, W.~L., {et~al.} 2025,
  \bibinfo{title}{{Extended dark energy analysis using DESI DR2 BAO
  measurements},} \prd, 112, 083511, \dodoi{10.1103/w4c6-1r5j}

\bibitem[{T. {Louis} {et~al.}(2025){Louis}, {La Posta}, {Atkins}, {Jense},
  {Abril-Cabezas}, {Addison}, {Ade}, {Aiola}, {Alford}, {Alonso}, {Amiri},
  {An}, {Austermann}, {Barbavara}, {Battaglia}, {Battistelli}, {Beall}, {Bean},
  {Beheshti}, {Beringue}, {Bhandarkar}, {Biermann}, {Bolliet}, {Bond},
  {Calabrese}, {Capalbo}, {Carrero}, {Chen}, {Chesmore}, {Cho}, {Choi},
  {Clark}, {Cothard}, {Coughlin}, {Coulton}, {Crichton}, {Crowley}, {Darwish},
  {Devlin}, {Dicker}, {Duell}, {Duff}, {Duivenvoorden}, {Dunkley}, {Dunner},
  {Embil Villagra}, {Fankhanel}, {Farren}, {Ferraro}, {Foster}, {Freundt},
  {Fuzia}, {Gallardo}, {Garrido}, {Gerbino}, {Giardiello}, {Gill}, {Givans},
  {Gluscevic}, {Goldstein}, {Golec}, {Gong}, {Guan}, {Halpern}, {Harrison},
  {Hasselfield}, {Healy}, {Henderson}, {Hensley}, {Herv{\'\i}as-Caimapo},
  {Hill}, {Hilton}, {Hilton}, {Hincks}, {Hlo{\v{z}}ek}, {Ho}, {Hood},
  {Hornecker}, {Huber}, {Hubmayr}, {Huffenberger}, {Hughes}, {Ikape}, {Irwin},
  {Isopi}, {Joshi}, {Keller}, {Kim}, {Knowles}, {Koopman}, {Kosowsky},
  {Kramer}, {Kusiak}, {Lague}, {Lakey}, {Lee}, {Li}, {Li}, {Limon}, {Lokken},
  {Lungu}, {MacCrann}, {MacInnis}, {Madhavacheril}, {Maldonado}, {Maldonado},
  {Mallaby-Kay}, {Marques}, {van Marrewijk}, {McCarthy}, {McMahon}, {Mehta},
  {Menanteau}, {Moodley}, {Morris}, {Mroczkowski}, {Naess}, {Namikawa}, {Nati},
  {Nerval}, {Newburgh}, {Nicola}, {Niemack}, {Nolta}, {Orlowski-Scherer},
  {Pagano}, {Page}, {Pandey}, {Partridge}, {Perez Sarmiento}, {Prince},
  {Puddu}, {Qu}, {Ragavan}, {Ried Guachalla}, {Rogers}, {Rojas}, {Sakuma},
  {Schaan}, {Schmitt}, {Sehgal}, {Shaikh}, {Sherwin}, {Sierra}, {Sievers},
  {Sif{\'o}n}, {Simon}, {Sonka}, {Spergel}, {Staggs}, {Storer}, {Surrao},
  {Switzer}, {Tampier}, {Thornton}, {Trac}, {Tucker}, {Ullom}, {Vale}, {Van
  Engelen}, {Van Lanen}, {Vargas}, {Vavagiakis}, {Wagoner}, {Wang}, {Wenzl},
  {Wollack}, \& {Zheng}}]{louis25}
{Louis}, T., {La Posta}, A., {Atkins}, Z., {et~al.} 2025, \bibinfo{title}{{The
  Atacama Cosmology Telescope: DR6 Power Spectra, Likelihoods and $Λ$CDM
  Parameters},} arXiv e-prints, arXiv:2503.14452,
  \dodoi{10.48550/arXiv.2503.14452}

\bibitem[{M. {Loverde} \& Z.~J. {Weiner}(2024){Loverde} \&
  {Weiner}}]{loverde24}
{Loverde}, M., \& {Weiner}, Z.~J. 2024, \bibinfo{title}{{Massive neutrinos and
  cosmic composition},} \jcap, 2024, 048, \dodoi{10.1088/1475-7516/2024/12/048}

\bibitem[{ {LSST Dark Energy Science Collaboration}(2012){LSST Dark Energy
  Science Collaboration}}]{lsstdesc12}
{LSST Dark Energy Science Collaboration}. 2012, \bibinfo{title}{{Large Synoptic
  Survey Telescope: Dark Energy Science Collaboration},} arXiv e-prints,
  arXiv:1211.0310, \dodoi{10.48550/arXiv.1211.0310}

\bibitem[{ {LSST Science Collaboration} {et~al.}(2009){LSST Science
  Collaboration}, {Abell}, {Allison}, {Anderson}, {Andrew}, {Angel}, {Armus},
  {Arnett}, {Asztalos}, {Axelrod}, {Bailey}, {Ballantyne}, {Bankert},
  {Barkhouse}, {Barr}, {Barrientos}, {Barth}, {Bartlett}, {Becker}, {Becla},
  {Beers}, {Bernstein}, {Biswas}, {Blanton}, {Bloom}, {Bochanski}, {Boeshaar},
  {Borne}, {Bradac}, {Brandt}, {Bridge}, {Brown}, {Brunner}, {Bullock},
  {Burgasser}, {Burge}, {Burke}, {Cargile}, {Chandrasekharan}, {Chartas},
  {Chesley}, {Chu}, {Cinabro}, {Claire}, {Claver}, {Clowe}, {Connolly}, {Cook},
  {Cooke}, {Cooray}, {Covey}, {Culliton}, {de Jong}, {de Vries}, {Debattista},
  {Delgado}, {Dell'Antonio}, {Dhital}, {Di Stefano}, {Dickinson}, {Dilday},
  {Djorgovski}, {Dobler}, {Donalek}, {Dubois-Felsmann}, {Durech},
  {Eliasdottir}, {Eracleous}, {Eyer}, {Falco}, {Fan}, {Fassnacht}, {Ferguson},
  {Fernandez}, {Fields}, {Finkbeiner}, {Figueroa}, {Fox}, {Francke}, {Frank},
  {Frieman}, {Fromenteau}, {Furqan}, {Galaz}, {Gal-Yam}, {Garnavich},
  {Gawiser}, {Geary}, {Gee}, {Gibson}, {Gilmore}, {Grace}, {Green}, {Gressler},
  {Grillmair}, {Habib}, {Haggerty}, {Hamuy}, {Harris}, {Hawley}, {Heavens},
  {Hebb}, {Henry}, {Hileman}, {Hilton}, {Hoadley}, {Holberg}, {Holman},
  {Howell}, {Infante}, {Ivezic}, {Jacoby}, {Jain}, {R}, {Jedicke}, {Jee},
  {Garrett Jernigan}, {Jha}, {Johnston}, {Jones}, {Juric}, {Kaasalainen},
  {Styliani}, {Kafka}, {Kahn}, {Kaib}, {Kalirai}, {Kantor}, {Kasliwal},
  {Keeton}, {Kessler}, {Knezevic}, {Kowalski}, {Krabbendam}, {Krughoff},
  {Kulkarni}, {Kuhlman}, {Lacy}, {Lepine}, {Liang}, {Lien}, {Lira}, {Long},
  {Lorenz}, {Lotz}, {Lupton}, {Lutz}, {Macri}, {Mahabal}, {Mandelbaum},
  {Marshall}, {May}, {McGehee}, {Meadows}, {Meert}, {Milani}, {Miller},
  {Miller}, {Mills}, {Minniti}, {Monet}, {Mukadam}, {Nakar}, {Neill}, {Newman},
  {Nikolaev}, {Nordby}, {O'Connor}, {Oguri}, {Oliver}, {Olivier}, {Olsen},
  {Olsen}, {Olszewski}, {Oluseyi}, {Padilla}, {Parker}, {Pepper}, {Peterson},
  {Petry}, {Pinto}, {Pizagno}, {Popescu}, {Prsa}, {Radcka}, {Raddick},
  {Rasmussen}, {Rau}, {Rho}, {Rhoads}, {Richards}, {Ridgway}, {Robertson},
  {Roskar}, {Saha}, {Sarajedini}, {Scannapieco}, {Schalk}, {Schindler}, \&
  {Schmidt}}]{lsst09}
{LSST Science Collaboration}, {Abell}, P.~A., {Allison}, J., {et~al.} 2009,
  \bibinfo{title}{{LSST Science Book, Version 2.0},} arXiv e-prints,
  arXiv:0912.0201, \dodoi{10.48550/arXiv.0912.0201}

\bibitem[{G.~P. {Lynch} \& L. {Knox}(2025){Lynch} \& {Knox}}]{lynch25}
{Lynch}, G.~P., \& {Knox}, L. 2025, \bibinfo{title}{{What's the matter with
  {\ensuremath{\Sigma}}m{\ensuremath{\nu}}?},} \prd, 112, 083543,
  \dodoi{10.1103/613p-pph2}

\bibitem[{M.~S. {Madhavacheril} {et~al.}(2019){Madhavacheril}, {Battaglia},
  {Smith}, \& {Sievers}}]{madhavacheril19}
{Madhavacheril}, M.~S., {Battaglia}, N., {Smith}, K.~M., \& {Sievers}, J.~L.
  2019, \bibinfo{title}{{Cosmology with the kinematic Sunyaev-Zeldovich effect:
  Breaking the optical depth degeneracy with fast radio bursts},} \prd, 100,
  103532, \dodoi{10.1103/PhysRevD.100.103532}

\bibitem[{M.~S. {Madhavacheril} \& J.~C. {Hill}(2018){Madhavacheril} \&
  {Hill}}]{madhavacheril18}
{Madhavacheril}, M.~S., \& {Hill}, J.~C. 2018, \bibinfo{title}{{Mitigating
  Foreground Biases in CMB Lensing Reconstruction Using Cleaned Gradients},}
  \prd, 023534, \dodoi{10.1103/PhysRevD.98.023534}

\bibitem[{M.~S. {Madhavacheril} {et~al.}(2020){Madhavacheril}, {Smith},
  {Sherwin}, \& {Naess}}]{madhavacheril20b}
{Madhavacheril}, M.~S., {Smith}, K.~M., {Sherwin}, B.~D., \& {Naess}, S. 2020,
  \bibinfo{title}{{CMB lensing power spectrum estimation without instrument
  noise bias},} arXiv e-prints, arXiv:2011.02475,
  \dodoi{10.48550/arXiv.2011.02475}

\bibitem[{M.~S. {Madhavacheril} {et~al.}(2024){Madhavacheril}, {Qu}, {Sherwin},
  {MacCrann}, {Li}, {Abril-Cabezas}, {Ade}, {Aiola}, {Alford}, {Amiri},
  {Amodeo}, {An}, {Atkins}, {Austermann}, {Battaglia}, {Battistelli}, {Beall},
  {Bean}, {Beringue}, {Bhandarkar}, {Biermann}, {Bolliet}, {Bond}, {Cai},
  {Calabrese}, {Calafut}, {Capalbo}, {Carrero}, {Challinor}, {Chesmore}, {Cho},
  {Choi}, {Clark}, {C{\'o}rdova Rosado}, {Cothard}, {Coughlin}, {Coulton},
  {Crowley}, {Dalal}, {Darwish}, {Devlin}, {Dicker}, {Doze}, {Duell}, {Duff},
  {Duivenvoorden}, {Dunkley}, {D{\"u}nner}, {Fanfani}, {Fankhanel}, {Farren},
  {Ferraro}, {Freundt}, {Fuzia}, {Gallardo}, {Garrido}, {Givans}, {Gluscevic},
  {Golec}, {Guan}, {Hall}, {Halpern}, {Han}, {Harrison}, {Hasselfield},
  {Healy}, {Henderson}, {Hensley}, {Herv{\'\i}as-Caimapo}, {Hill}, {Hilton},
  {Hilton}, {Hincks}, {Hlo{\v{z}}ek}, {Ho}, {Huber}, {Hubmayr}, {Huffenberger},
  {Hughes}, {Irwin}, {Isopi}, {Jense}, {Keller}, {Kim}, {Knowles}, {Koopman},
  {Kosowsky}, {Kramer}, {Kusiak}, {La Posta}, {Lague}, {Lakey}, {Lee}, {Li},
  {Limon}, {Lokken}, {Louis}, {Lungu}, {MacInnis}, {Maldonado}, {Maldonado},
  {Mallaby-Kay}, {Marques}, {McMahon}, {Mehta}, {Menanteau}, {Moodley},
  {Morris}, {Mroczkowski}, {Naess}, {Namikawa}, {Nati}, {Newburgh}, {Nicola},
  {Niemack}, {Nolta}, {Orlowski-Scherer}, {Page}, {Pandey}, {Partridge},
  {Prince}, {Puddu}, {Radiconi}, {Robertson}, {Rojas}, {Sakuma}, {Salatino},
  {Schaan}, {Schmitt}, {Sehgal}, {Shaikh}, {Sierra}, {Sievers}, {Sif{\'o}n},
  {Simon}, {Sonka}, {Spergel}, {Staggs}, {Storer}, {Switzer}, {Tampier},
  {Thornton}, {Trac}, {Treu}, {Tucker}, {Ullom}, {Vale}, {Van Engelen}, {Van
  Lanen}, {van Marrewijk}, {Vargas}, {Vavagiakis}, {Wagoner}, {Wang}, {Wenzl},
  {Wollack}, {Xu}, {Zago}, \& {Zheng}}]{madhavacheril24}
{Madhavacheril}, M.~S., {Qu}, F.~J., {Sherwin}, B.~D., {et~al.} 2024,
  \bibinfo{title}{{The Atacama Cosmology Telescope: DR6 Gravitational Lensing
  Map and Cosmological Parameters},} \apj, 962, 113,
  \dodoi{10.3847/1538-4357/acff5f}

\bibitem[{V. {Mainieri} {et~al.}(2024){Mainieri}, {Anderson}, {Brinchmann},
  {Cimatti}, {Ellis}, {Hill}, {Kneib}, {McLeod}, {Opitom}, {Roth},
  {Sanchez-Saez}, {Smiljanic}, {Tolstoy}, {Bacon}, {Randich}, {Adamo},
  {Annibali}, {Arevalo}, {Audard}, {Barsanti}, {Battaglia}, {Bayo Aran},
  {Belfiore}, {Bellazzini}, {Bellini}, {Beltran}, {Berni}, {Bianchi}, {Biazzo},
  {Bisero}, {Bisogni}, {Bland-Hawthorn}, {Blondin}, {Bodensteiner}, {Boffin},
  {Bonito}, {Bono}, {Bouche}, {Bowman}, {Braga}, {Bragaglia}, {Branchesi},
  {Brucalassi}, {Bryant}, {Bryson}, {Busa}, {Camera}, {Carbone}, {Casali},
  {Casali}, {Casasola}, {Castro}, {Catelan}, {Cavallo}, {Chiappini}, {Cioni},
  {Colless}, {Colzi}, {Contarini}, {Couch}, {D'Ammando}, {d'Assignies D.},
  {D'Orazi}, {da Silva}, {Dainotti}, {Damiani}, {Danielski}, {De Cia}, {de
  Jong}, {Dhawan}, {Dierickx}, {Driver}, {Dupletsa}, {Escoffier}, {Escorza},
  {Fabrizio}, {Fiorentino}, {Fontana}, {Fontani}, {Forero Sanchez}, {Franois},
  {Galindo-Guil}, {Gallazzi}, {Galli}, {Garcia}, {Garcia-Rojas}, {Garilli},
  {Grand}, {Guarcello}, {Hazra}, {Helmi}, {Herrero}, {Iglesias}, {Ilic},
  {Irsic}, {Ivanov}, {Izzo}, {Jablonka}, {Joachimi}, {Kakkad}, {Kamann},
  {Koposov}, {Kordopatis}, {Kovacevic}, {Kraljic}, {Kuncarayakti}, {Kwon}, {La
  Forgia}, {Lahav}, {Laigle}, {Lazzarin}, {Leaman}, {Leclercq}, {Lee}, {Lee},
  {Lehnert}, {Lira}, {Loffredo}, {Lucatello}, {Magrini}, {Maguire}, {Mahler},
  {Zahra Majidi}, {Malavasi}, {Mannucci}, {Marconi}, {Martin}, {Marulli},
  {Massari}, {Matsuno}, {Mattheee}, {McGee}, {Merc}, {Merle}, {Miglio},
  {Migliorini}, {Minchev}, {Minniti}, {Miret-Roig}, {Monreal Ibero}, {Montano},
  {Montet}, {Moresco}, {Moretti}, {Moscardini}, {Moya}, {Mueller},
  {Nanayakkara}, {Nicholl}, {Nordlander}, {Onori}, {Padovani}, {Pala}, {Panda},
  {Pandey-Pommier}, {Pasquini}, {Pawlak}, {Pessi}, {Pisani}, {Popovic},
  {Prisinzano}, {Raddi}, {Rainer}, {Rebassa-Mansergas}, {Richard}, {Rigault},
  {Rocher}, {Romano}, {Rosati}, {Sacco}, {Sanchez-Janssen}, {Sander},
  {Sanders}, {Sargent}, {Sarpa}, {Schimd}, {Schipani}, {Sefusatti}, {Smith},
  {Spina}, {Steinmetz}, {Tacchella}, {Tautvaisiene}, {Theissen}, {Thomas},
  {Ting}, {Travouillon}, {Tresse}, {Trivedi}, {Tsantaki}, {Tsedrik}, {Urrutia},
  {Valenti}, {Van der Swaelmen}, {Van Eck}, {Verdiani}, {Verdier}, {Vergani},
  {Verhamme}, \& {Vernet}}]{mainieri24}
{Mainieri}, V., {Anderson}, R.~I., {Brinchmann}, J., {et~al.} 2024,
  \bibinfo{title}{{The Wide-field Spectroscopic Telescope (WST) Science White
  Paper},} arXiv e-prints, arXiv:2403.05398, \dodoi{10.48550/arXiv.2403.05398}

\bibitem[{A. Mitra {et~al.}(2023)Mitra, Kessler, More, \& Hlozek}]{mitra22}
Mitra, A., Kessler, R., More, S., \& Hlozek, R. 2023, \bibinfo{title}{{Using
  Host Galaxy Photometric Redshifts to Improve Cosmological Constraints with
  Type Ia Supernovae in the LSST Era},} Astrophys. J., 944, 212,
  \dodoi{10.3847/1538-4357/acb057}

\bibitem[{H. {Miyatake} {et~al.}(2023){Miyatake}, {Sugiyama}, {Takada},
  {Nishimichi}, {Li}, {Shirasaki}, {More}, {Kobayashi}, {Nishizawa}, {Rau},
  {Zhang}, {Takahashi}, {Dalal}, {Mandelbaum}, {Strauss}, {Hamana}, {Oguri},
  {Osato}, {Luo}, {Kannawadi}, {Hsieh}, {Armstrong}, {Bosch}, {Komiyama},
  {Lupton}, {Lust}, {MacArthur}, {Miyazaki}, {Murayama}, {Okura}, {Price},
  {Sunayama}, {Tait}, {Tanaka}, \& {Wang}}]{miyatake23}
{Miyatake}, H., {Sugiyama}, S., {Takada}, M., {et~al.} 2023,
  \bibinfo{title}{{Hyper Suprime-Cam Year 3 results: Cosmology from galaxy
  clustering and weak lensing with HSC and SDSS using the emulator based halo
  model},} \prd, 108, 123517, \dodoi{10.1103/PhysRevD.108.123517}

\bibitem[{T. {Namikawa} {et~al.}(2013){Namikawa}, {Hanson}, \&
  {Takahashi}}]{namikawa13}
{Namikawa}, T., {Hanson}, D., \& {Takahashi}, R. 2013,
  \bibinfo{title}{{Bias-hardened CMB lensing},} \mnras, 431, 609,
  \dodoi{10.1093/mnras/stt195}

\bibitem[{S. {Nojiri} \& S.~D. {Odintsov}(2005){Nojiri} \&
  {Odintsov}}]{nojiri05}
{Nojiri}, S., \& {Odintsov}, S.~D. 2005, \bibinfo{title}{{Inhomogeneous
  equation of state of the universe: Phantom era, future singularity, and
  crossing the phantom barrier},} \prd, 72, 023003,
  \dodoi{10.1103/PhysRevD.72.023003}

\bibitem[{T. {Okamoto} \& W. {Hu}(2003){Okamoto} \& {Hu}}]{okamoto03}
{Okamoto}, T., \& {Hu}, W. 2003, \bibinfo{title}{{Cosmic microwave background
  lensing reconstruction on the full sky},} \prd, 67, 083002,
  \dodoi{10.1103/PhysRevD.67.083002}

\bibitem[{Y. {Omori} {et~al.}(2017){Omori}, {Chown}, {Simard}, {Story},
  {Aylor}, {Baxter}, {Benson}, {Bleem}, {Carlstrom}, {Chang}, {Cho},
  {Crawford}, {Crites}, {de Haan}, {Dobbs}, {Everett}, {George}, {Halverson},
  {Harrington}, {Holder}, {Hou}, {Holzapfel}, {Hrubes}, {Knox}, {Lee},
  {Leitch}, {Luong-Van}, {Manzotti}, {Marrone}, {McMahon}, {Meyer}, {Mocanu},
  {Mohr}, {Natoli}, {Padin}, {Pryke}, {Reichardt}, {Ruhl}, {Sayre}, {Schaffer},
  {Shirokoff}, {Staniszewski}, {Stark}, {Vanderlinde}, {Vieira}, {Williamson},
  \& {Zahn}}]{omori17}
{Omori}, Y., {Chown}, R., {Simard}, G., {et~al.} 2017, \bibinfo{title}{{A 2500
  deg$^{2}$ CMB Lensing Map from Combined South Pole Telescope and Planck
  Data},} \apj, 849, 124, \dodoi{10.3847/1538-4357/aa8d1d}

\bibitem[{Y. {Omori} {et~al.}(2023){Omori}, {Baxter}, {Chang}, {Friedrich},
  {Alarcon}, {Alves}, {Amon}, {Andrade-Oliveira}, {Bechtol}, {Becker},
  {Bernstein}, {Blazek}, {Bleem}, {Camacho}, {Campos}, {Carnero Rosell},
  {Carrasco Kind}, {Cawthon}, {Chen}, {Choi}, {Cordero}, {Crawford}, {Crocce},
  {Davis}, {DeRose}, {Dodelson}, {Doux}, {Drlica-Wagner}, {Eckert}, {Eifler},
  {Elsner}, {Elvin-Poole}, {Everett}, {Fang}, {Fert{\'e}}, {Fosalba}, {Gatti},
  {Giannini}, {Gruen}, {Gruendl}, {Harrison}, {Herner}, {Huang}, {Huff},
  {Huterer}, {Jarvis}, {Krause}, {Kuropatkin}, {Leget}, {Lemos}, {Liddle},
  {MacCrann}, {McCullough}, {Muir}, {Myles}, {Navarro-Alsina}, {Pandey},
  {Park}, {Porredon}, {Prat}, {Raveri}, {Rollins}, {Roodman}, {Rosenfeld},
  {Ross}, {Rykoff}, {S{\'a}nchez}, {Sanchez}, {Secco}, {Sevilla-Noarbe},
  {Sheldon}, {Shin}, {Troxel}, {Tutusaus}, {Varga}, {Weaverdyck}, {Wechsler},
  {Wu}, {Yanny}, {Yin}, {Zhang}, {Zuntz}, {Abbott}, {Aguena}, {Allam}, {Annis},
  {Bacon}, {Benson}, {Bertin}, {Bocquet}, {Brooks}, {Burke}, {Carlstrom},
  {Carretero}, {Chang}, {Chown}, {Costanzi}, {da Costa}, {Crites}, {Pereira},
  {de Haan}, {De Vicente}, {Desai}, {Diehl}, {Dobbs}, {Doel}, {Everett},
  {Ferrero}, {Flaugher}, {Friedel}, {Frieman}, {Garc{\'\i}a-Bellido},
  {Gaztanaga}, {George}, {Giannantonio}, {Halverson}, {Hinton}, {Holder},
  {Hollowood}, {Holzapfel}, {Honscheid}, {Hrubes}, {James}, {Knox}, {Kuehn},
  {Lahav}, {Lee}, {Lima}, {Luong-Van}, {March}, {McMahon}, {Melchior},
  {Menanteau}, {Meyer}, {Miquel}, {Mocanu}, {Mohr}, {Morgan}, {Natoli},
  {Padin}, {Palmese}, {Paz-Chinch{\'o}n}, {Pieres}, {Plazas Malag{\'o}n},
  {Pryke}, {Reichardt}, {Romer}, {Ruhl}, {Sanchez}, {Schaffer}, {Schubnell},
  {Serrano}, {Shirokoff}, {Smith}, {Staniszewski}, {Stark}, {Suchyta}, {Tarle},
  {Thomas}, {To}, {Vieira}, {Weller}, {Williamson}, {DES}, \& {SPT
  Collaborations}}]{omori23}
{Omori}, Y., {Baxter}, E.~J., {Chang}, C., {et~al.} 2023,
  \bibinfo{title}{{Joint analysis of Dark Energy Survey Year 3 data and CMB
  lensing from SPT and Planck. I. Construction of CMB lensing maps and modeling
  choices},} \prd, 107, 023529, \dodoi{10.1103/PhysRevD.107.023529}

\bibitem[{D.~D.~Y. {Ong} {et~al.}(2026){Ong}, {Yallup}, \& {Handley}}]{ong26}
{Ong}, D. D.~Y., {Yallup}, D., \& {Handley}, W. 2026, \bibinfo{title}{{The
  Bayesian view of DESI DR2: Evidence and tension in a combined analysis with
  CMB and supernovae across cosmological models},} arXiv e-prints,
  arXiv:2603.05472.
\newblock \doarXiv{2603.05472}

\bibitem[{S.~J. {Osborne} {et~al.}(2014){Osborne}, {Hanson}, \&
  {Dor{\'e}}}]{osborne14}
{Osborne}, S.~J., {Hanson}, D., \& {Dor{\'e}}, O. 2014,
  \bibinfo{title}{{Extragalactic foreground contamination in temperature-based
  CMB lens reconstruction},} \jcap, 2014, 024,
  \dodoi{10.1088/1475-7516/2014/03/024}

\bibitem[{Z. {Pan} {et~al.}(2023){Pan}, {Bianchini}, {Wu}, {Ade}, {Ahmed},
  {Anderes}, {Anderson}, {Ansarinejad}, {Archipley}, {Aylor}, {Balkenhol},
  {Barry}, {Basu Thakur}, {Benabed}, {Bender}, {Benson}, {Bleem}, {Bouchet},
  {Bryant}, {Byrum}, {Camphuis}, {Carlstrom}, {Carter}, {Cecil}, {Chang},
  {Chaubal}, {Chen}, {Chichura}, {Cho}, {Chou}, {Cliche}, {Coerver},
  {Crawford}, {Cukierman}, {Daley}, {de Haan}, {Denison}, {Dibert}, {Ding},
  {Dobbs}, {Doussot}, {Dutcher}, {Everett}, {Feng}, {Ferguson}, {Fichman},
  {Foster}, {Fu}, {Galli}, {Gambrel}, {Gardner}, {Ge}, {Goeckner-Wald},
  {Gualtieri}, {Guidi}, {Guns}, {Gupta}, {Halverson}, {Harke-Hosemann},
  {Harrington}, {Henning}, {Hilton}, {Hivon}, {Holder}, {Holzapfel}, {Hood},
  {Howe}, {Huang}, {Irwin}, {Jeong}, {Jonas}, {Jones}, {K{\'e}ruzor{\'e}},
  {Khaire}, {Knox}, {Kofman}, {Korman}, {Kubik}, {Kuhlmann}, {Kuo}, {Lee},
  {Leitch}, {Levy}, {Lowitz}, {Lu}, {Maniyar}, {Menanteau}, {Meyer},
  {Michalik}, {Millea}, {Montgomery}, {Nadolski}, {Nakato}, {Natoli}, {Nguyen},
  {Noble}, {Novosad}, {Omori}, {Padin}, {Paschos}, {Pearson}, {Posada},
  {Prabhu}, {Quan}, {Raghunathan}, {Rahimi}, {Rahlin}, {Reichardt}, {Riebel},
  {Riedel}, {Ruhl}, {Sayre}, {Schiappucci}, {Shirokoff}, {Smecher}, {Sobrin},
  {Stark}, {Stephen}, {Story}, {Suzuki}, {Takakura}, {Tandoi}, {Thompson},
  {Thorne}, {Trendafilova}, {Tucker}, {Umilta}, {Vale}, {Vanderlinde},
  {Vieira}, {Wang}, {Whitehorn}, {Yefremenko}, {Yoon}, {Young}, \&
  {Zebrowski}}]{pan23}
{Pan}, Z., {Bianchini}, F., {Wu}, W.~L.~K., {et~al.} 2023,
  \bibinfo{title}{{Measurement of gravitational lensing of the cosmic microwave
  background using SPT-3G 2018 data},} \prd, 108, 122005,
  \dodoi{10.1103/PhysRevD.108.122005}

\bibitem[{J. {Peloton} {et~al.}(2017){Peloton}, {Schmittfull}, {Lewis},
  {Carron}, \& {Zahn}}]{peloton17}
{Peloton}, J., {Schmittfull}, M., {Lewis}, A., {Carron}, J., \& {Zahn}, O.
  2017, \bibinfo{title}{{Full covariance of CMB and lensing reconstruction
  power spectra},} \prd, 95, 043508, \dodoi{10.1103/PhysRevD.95.043508}

\bibitem[{ {Planck Collaboration} {et~al.}(2020{\natexlab{a}}){Planck
  Collaboration}, {Aghanim}, {Akrami}, {Ashdown}, {Aumont}, {Baccigalupi},
  {Ballardini}, {Banday}, {Barreiro}, {Bartolo}, {Basak}, {Battye}, {Benabed},
  {Bernard}, {Bersanelli}, {Bielewicz}, {Bock}, {Bond}, {Borrill}, {Bouchet},
  {Boulanger}, {Bucher}, {Burigana}, {Butler}, {Calabrese}, {Cardoso},
  {Carron}, {Challinor}, {Chiang}, {Chluba}, {Colombo}, {Combet}, {Contreras},
  {Crill}, {Cuttaia}, {de Bernardis}, {de Zotti}, {Delabrouille}, {Delouis},
  {Di Valentino}, {Diego}, {Dor{\'e}}, {Douspis}, {Ducout}, {Dupac}, {Dusini},
  {Efstathiou}, {Elsner}, {En{\ss}lin}, {Eriksen}, {Fantaye}, {Farhang},
  {Fergusson}, {Fernandez-Cobos}, {Finelli}, {Forastieri}, {Frailis},
  {Fraisse}, {Franceschi}, {Frolov}, {Galeotta}, {Galli}, {Ganga},
  {G{\'e}nova-Santos}, {Gerbino}, {Ghosh}, {Gonz{\'a}lez-Nuevo}, {G{\'o}rski},
  {Gratton}, {Gruppuso}, {Gudmundsson}, {Hamann}, {Handley}, {Hansen},
  {Herranz}, {Hildebrandt}, {Hivon}, {Huang}, {Jaffe}, {Jones}, {Karakci},
  {Keih{\"a}nen}, {Keskitalo}, {Kiiveri}, {Kim}, {Kisner}, {Knox},
  {Krachmalnicoff}, {Kunz}, {Kurki-Suonio}, {Lagache}, {Lamarre}, {Lasenby},
  {Lattanzi}, {Lawrence}, {Le Jeune}, {Lemos}, {Lesgourgues}, {Levrier},
  {Lewis}, {Liguori}, {Lilje}, {Lilley}, {Lindholm}, {L{\'o}pez-Caniego},
  {Lubin}, {Ma}, {Mac{\'\i}as-P{\'e}rez}, {Maggio}, {Maino}, {Mandolesi},
  {Mangilli}, {Marcos-Caballero}, {Maris}, {Martin}, {Martinelli},
  {Mart{\'\i}nez-Gonz{\'a}lez}, {Matarrese}, {Mauri}, {McEwen}, {Meinhold},
  {Melchiorri}, {Mennella}, {Migliaccio}, {Millea}, {Mitra},
  {Miville-Desch{\^e}nes}, {Molinari}, {Montier}, {Morgante}, {Moss}, {Natoli},
  {N{\o}rgaard-Nielsen}, {Pagano}, {Paoletti}, {Partridge}, {Patanchon},
  {Peiris}, {Perrotta}, {Pettorino}, {Piacentini}, {Polastri}, {Polenta},
  {Puget}, {Rachen}, {Reinecke}, {Remazeilles}, {Renzi}, {Rocha}, {Rosset},
  {Roudier}, {Rubi{\~n}o-Mart{\'\i}n}, {Ruiz-Granados}, {Salvati}, {Sandri},
  {Savelainen}, {Scott}, {Shellard}, {Sirignano}, {Sirri}, {Spencer},
  {Sunyaev}, {Suur-Uski}, {Tauber}, {Tavagnacco}, {Tenti}, {Toffolatti},
  {Tomasi}, {Trombetti}, {Valenziano}, {Valiviita}, {Van Tent}, {Vibert},
  {Vielva}, {Villa}, {Vittorio}, {Wandelt}, {Wehus}, {White}, {White},
  {Zacchei}, \& {Zonca}}]{planck20cosmo}
{Planck Collaboration}, {Aghanim}, N., {Akrami}, Y., {et~al.}
  2020{\natexlab{a}}, \bibinfo{title}{{Planck 2018 results. VI. Cosmological
  parameters},} \aap, 641, A6, \dodoi{10.1051/0004-6361/201833910}

\bibitem[{ {Planck Collaboration} {et~al.}(2020{\natexlab{b}}){Planck
  Collaboration}, {Aghanim}, {Akrami}, {Ashdown}, {Aumont}, {Baccigalupi},
  {Ballardini}, {Banday}, {Barreiro}, {Ba rtolo}, {Basak}, {Battye}, {Benabed},
  {Bernard}, {Bersanelli}, {Bielewicz}, {Bock}, {Bond}, {Borrill}, {Bouchet},
  {Bucher}, {Burigana}, {Butler}, {Calabrese}, {Cardoso}, {Carron},
  {Challinor}, {Chiang}, {Chluba}, {Colombo }, {Combet}, {Contreras}, {Crill},
  {Cuttaia}, {de Bernardis}, {de Zotti}, {Delabrouille}, {Delouis}, {Di
  Valentino}, {D iego}, {Dor{\'e}}, {Douspis}, {Ducout}, {Dupac}, {Dusini},
  {Efstathiou}, {Elsner}, {En{\ss}lin}, {Eriksen}, {Fantaye}, {Fergusson},
  {Fernandez-Cobos}, {Finelli}, {Forastieri}, {Frailis}, {Fraisse},
  {Franceschi}, {Frolov}, {Galeotta}, {Galli}, {Ganga}, {G{\'e}nova-Santos},
  {Gerbino}, {Ghosh}, {Gonz{\'a}lez-Nuevo}, {G{\'o}rski}, {Gratton},
  {Gruppuso}, {Gudmundsson}, {Hamann}, {Handley}, {Hansen}, {Herranz},
  {Hildebrandt}, {Hivon}, {Huang}, {Jaffe}, {Jones}, {Karakci}, {Keih{\"a}nen},
  {Keskitalo}, {Kiiveri}, {Kim}, {Kisner}, {Knox}, {Krachmalnicoff}, {Kunz},
  {Kurki-Suonio}, {Lagache}, {Lamarre}, {Lasenby}, {Lattanzi}, {Lawrence}, {Le
  Jeune}, {Lemos}, {Lesgourgues}, {Levrier}, {Lewis}, {Liguori}, {Lilje},
  {Lilley}, {Lindholm}, {L{\'o}pez-Caniego}, {Lubin}, {Ma},
  {Mac{\'\i}as-P{\'e}rez}, {Maggio}, {Maino}, {Mandolesi}, {Mangilli},
  {Marcos-Caballero}, {Maris}, {Martin}, {Martinelli},
  {Mart{\'\i}nez-Gonz{\'a}lez}, {Matarrese}, {Mauri}, {McEwen}, {Meinhold},
  {Melchiorri}, {Mennella}, {Migliaccio}, {Millea}, {Mitra},
  {Miville-Desch{\^e}nes}, {Molinari}, {Montier}, {Morgante}, {Moss}, {Natoli},
  {N{\o}rgaard-Nielsen}, {Pagano}, {Paoletti}, {Partridge}, {Patanchon},
  {Peiris}, {Perrotta}, {Pettorino}, {Piacentini}, {Polastri}, {Polenta},
  {Puget}, {Rachen}, {Reinecke}, {Remazeilles}, {Renzi}, {Rocha}, {Rosset},
  {Roudier}, {Rubi{\~n}o-Mart{\'\i}n}, {Ruiz-Granados}, {Salvati}, {Sandri},
  {Savelainen}, {Scott}, {Shellard}, {Sirignano}, {Sirri}, {Spencer},
  {Sunyaev}, {Suur-Uski}, {Tauber}, {Tavagnacco}, {Tenti}, {Toffolatti},
  {Tomasi}, {Trombetti}, {Valenziano}, {Valiviita}, {Van Tent}, {Vibert},
  {Vielva}, {Villa}, {Vittorio}, {Wandelt}, {Wehus}, {White}, {White},
  {Zacchei}, \& {Zonca}}]{planck20_2018cosmo}
{Planck Collaboration}, {Aghanim}, N., {Akrami}, Y., {et~al.}
  2020{\natexlab{b}}, \bibinfo{title}{{Planck 2018 results. VI. Cosmological
  parameters},} \aap, 641, A6, \dodoi{10.1051/0004-6361/201833910}

\bibitem[{B. {Popovic} {et~al.}(2021){Popovic}, {Brout}, {Kessler}, {Scolnic},
  \& {Lu}}]{popovic21}
{Popovic}, B., {Brout}, D., {Kessler}, R., {Scolnic}, D., \& {Lu}, L. 2021,
  \bibinfo{title}{{Improved Treatment of Host-galaxy Correlations in
  Cosmological Analyses with Type Ia Supernovae},} \apj, 913, 49,
  \dodoi{10.3847/1538-4357/abf14f}

\bibitem[{B. {Popovic} {et~al.}(2025){Popovic}, {Shah}, {Kenworthy}, {Kessler},
  {Davis}, {Goobar}, {Scolnic}, {Vincenzi}, {Wiseman}, {Chen}, {Charleton},
  {Acevedo}, {Armstrong}, {Boyd}, {Brout}, {Camilleri}, {Frieman}, {Galbany},
  {Grayling}, {Kelsey}, {Rose}, {S{\'a}nchez}, {Lee}, {M{\"o}ller}, {Smith},
  {Sullivan}, {Shiamtanis}, {Alarcon}, {Allam}, {Andrade-Oliveira}, {Avila},
  {Bacon}, {Blazek}, {Bocquet}, {Brooks}, {Burke}, {Carnero Rosell},
  {Carretero}, {Cawthon}, {da Costa}, {da Silva Pereira}, {Diehl}, {Dodelson},
  {Doel}, {Everett}, {Frohmaier}, {Garc{\'\i}a-Bellido}, {Gruen}, {Gutierrez},
  {Herner}, {Hinton}, {Hollowood}, {Honscheid}, {Huterer}, {James}, {Jeffrey},
  {Kuehn}, {Lahav}, {Lee}, {Lidman}, {Marshall}, {Mena-Fern{\'a}ndez},
  {Menanteau}, {Miquel}, {Muir}, {Myles}, {Ogando}, {Paterno}, {Plazas
  Malag{\'o}n}, {Porredon}, {Prat}, {Nichol}, {Romer}, {Roodman}, {Sanchez},
  {Sanchez Cid}, {Sevilla-Noarbe}, {Suchyta}, {Swanson}, {To}, {Tucker},
  {Walker}, \& {Weaverdyck}}]{popovic25}
{Popovic}, B., {Shah}, P., {Kenworthy}, W.~D., {et~al.} 2025,
  \bibinfo{title}{{The Dark Energy Survey Supernova Program: A Reanalysis Of
  Cosmology Results And Evidence For Evolving Dark Energy With An Updated Type
  Ia Supernova Calibration},} arXiv e-prints, arXiv:2511.07517,
  \dodoi{10.48550/arXiv.2511.07517}

\bibitem[{V. {Poulin} {et~al.}(2019){Poulin}, {Smith}, {Karwal}, \&
  {Kamionkowski}}]{poulin19}
{Poulin}, V., {Smith}, T.~L., {Karwal}, T., \& {Kamionkowski}, M. 2019,
  \bibinfo{title}{{Early Dark Energy can Resolve the Hubble Tension},} \prl,
  122, 221301, \dodoi{10.1103/PhysRevLett.122.221301}

\bibitem[{K. {Prabhu} {et~al.}(2024){Prabhu}, {Raghunathan}, {Millea}, {Lynch},
  {Ade}, {Anderes}, {Anderson}, {Ansarinejad}, {Archipley}, {Balkenhol},
  {Benabed}, {Bender}, {Benson}, {Bianchini}, {Bleem}, {Bouchet}, {Bryant},
  {Camphuis}, {Carlstrom}, {Cecil}, {Chang}, {Chaubal}, {Chichura}, {Chokshi},
  {Chou}, {Coerver}, {Crawford}, {Cukierman}, {Daley}, {de Haan}, {Dibert},
  {Dobbs}, {Doussot}, {Dutcher}, {Everett}, {Feng}, {Ferguson}, {Fichman},
  {Foster}, {Galli}, {Gambrel}, {Gardner}, {Ge}, {Goeckner-Wald}, {Gualtieri},
  {Guidi}, {Guns}, {Halverson}, {Hivon}, {Holder}, {Holzapfel}, {Hood},
  {Hryciuk}, {Huang}, {K{\'e}ruzor{\'e}}, {Knox}, {Korman}, {Kornoelje}, {Kuo},
  {Lee}, {Levy}, {Lowitz}, {Lu}, {Maniyar}, {Menanteau}, {Montgomery},
  {Nakato}, {Natoli}, {Noble}, {Novosad}, {Omori}, {Padin}, {Pan}, {Paschos},
  {Phadke}, {Pollak}, {Quan}, {Rahimi}, {Rahlin}, {Reichardt}, {Rouble},
  {Ruhl}, {Schiappucci}, {Smecher}, {Sobrin}, {Stark}, {Stephen}, {Suzuki},
  {Tandoi}, {Thompson}, {Thorne}, {Trendafilova}, {Tucker}, {Umilta},
  {Vitrier}, {Vieira}, {Wan}, {Wang}, {Whitehorn}, {Wu}, {Yefremenko}, {Young},
  \& {Zebrowski}}]{prabhu24}
{Prabhu}, K., {Raghunathan}, S., {Millea}, M., {et~al.} 2024,
  \bibinfo{title}{{Testing the {\ensuremath{\Lambda}}CDM Cosmological Model
  with Forthcoming Measurements of the Cosmic Microwave Background with
  SPT-3G},} \apj, 973, 4, \dodoi{10.3847/1538-4357/ad5ff1}

\bibitem[{F.~J. {Qu} {et~al.}(2025){Qu}, {Ge}, {Kimmy Wu}, {Abril-Cabezas},
  {Madhavacheril}, {Millea}, {Anderes}, {Anderson}, {Ansarinejad}, {Archipley},
  {Atkins}, {Balkenhol}, {Battaglia}, {Benabed}, {Bender}, {Benson},
  {Bianchini}, {Bleem}, {Bolliet}, {Bond}, {Bouchet}, {Bryant}, {Calabrese},
  {Camphuis}, {Carlstrom}, {Carron}, {Challinor}, {Chang}, {Chaubal}, {Chen},
  {Chichura}, {Choi}, {Chokshi}, {Chou}, {Coerver}, {Coulton}, {Crawford},
  {Daley}, {Darwish}, {de Haan}, {Devlin}, {Dibert}, {Dobbs}, {Doohan},
  {Doussot}, {Duivenvoorden}, {Dunkley}, {Dunner}, {Dutcher}, {Embil Villagra},
  {Everett}, {Farren}, {Feng}, {Ferraro}, {Ferguson}, {Fichman}, {Finson},
  {Foster}, {Gallardo}, {Galli}, {Gambrel}, {Gardner}, {Goeckner-Wald},
  {Gualtieri}, {Guidi}, {Guns}, {Halpern}, {Halverson}, {Hill}, {Hilton},
  {Hivon}, {Holder}, {Holzapfel}, {Hood}, {Howe}, {Hryciuk}, {Huang},
  {Hubmayr}, {K{\'e}ruzor{\'e}}, {Khalife}, {Kim}, {Knox}, {Korman},
  {Kornoelje}, {Kosowsky}, {Kuo}, {Jense}, {La Posta}, {Levy}, {Lowitz},
  {Louis}, {Lu}, {Lynch}, {MacCrann}, {Maniyar}, {Martsen}, {McMahon},
  {Menanteau}, {Montgomery}, {Nakato}, {Moodley}, {Namikawa}, {Natoli},
  {Niemack}, {Noble}, {Omori}, {Ouellette}, {Page}, {Pan}, {Paschos}, {Phadke},
  {Pollak}, {Prabhu}, {Quan}, {Raghunathan}, {Rahimi}, {Rahlin}, {Reichardt},
  {Riebel}, {Rouble}, {Ruhl}, {Schaan}, {Schiappucci}, {Sehgal}, {Sierra},
  {Simpson}, {Sherwin}, {Sif{\'o}n}, {Spergel}, {Staggs}, {Sobrin}, {Stark},
  {Stephen}, {Tandoi}, {Thorne}, {Trendafilova}, {Umilta}, {Van Engelen},
  {Vieira}, {Vitrier}, {Wan}, {Whitehorn}, {Wollack}, {Young}, \&
  {Zebrowski}}]{qu25}
{Qu}, F.~J., {Ge}, F., {Kimmy Wu}, W.~L., {et~al.} 2025,
  \bibinfo{title}{{Unified and consistent structure growth measurements from
  joint ACT, SPT and \textbackslashtextit\{Planck\} CMB lensing},} arXiv
  e-prints, arXiv:2504.20038, \dodoi{10.48550/arXiv.2504.20038}

\bibitem[{S. Raghunathan(2022)Raghunathan}]{raghunathan22}
Raghunathan, S. 2022, \bibinfo{title}{Assessing the Importance of Noise from
  Thermal Sunyaev–Zel′dovich Signals for CMB Cluster Surveys and Cluster
  Cosmology,} The Astrophysical Journal, 928, 16,
  \dodoi{10.3847/1538-4357/ac510f}

\bibitem[{S. {Raghunathan} \& Y. {Omori}(2023){Raghunathan} \&
  {Omori}}]{raghunathan23}
{Raghunathan}, S., \& {Omori}, Y. 2023, \bibinfo{title}{{A Cross-internal
  Linear Combination Approach to Probe the Secondary CMB Anisotropies:
  Kinematic Sunyaev-Zel'dovich Effect and CMB Lensing},} \apj, 954, 83,
  \dodoi{10.3847/1538-4357/ace0c6}

\bibitem[{C.~L. {Reichardt} {et~al.}(2021){Reichardt}, {Patil}, {Ade},
  {Anderson}, {Austermann}, {Avva}, {Baxter}, {Beall}, {Bender}, {Benson},
  {Bianchini}, {Bleem}, {Carlstrom}, {Chang}, {Chaubal}, {Chiang}, {Chou},
  {Citron}, {Moran}, {Crawford}, {Crites}, {de Haan}, {Dobbs}, {Everett},
  {Gallicchio}, {George}, {Gilbert}, {Gupta}, {Halverson}, {Harrington},
  {Henning}, {Hilton}, {Holder}, {Holzapfel}, {Hrubes}, {Huang}, {Hubmayr},
  {Irwin}, {Knox}, {Lee}, {Li}, {Lowitz}, {Luong-Van}, {McMahon}, {Mehl},
  {Meyer}, {Millea}, {Mocanu}, {Mohr}, {Montgomery}, {Nadolski}, {Natoli},
  {Nibarger}, {Noble}, {Novosad}, {Omori}, {Padin}, {Pryke}, {Ruhl},
  {Saliwanchik}, {Sayre}, {Schaffer}, {Shirokoff}, {Sievers}, {Smecher},
  {Spieler}, {Staniszewski}, {Stark}, {Tucker}, {Vanderlinde}, {Veach},
  {Vieira}, {Wang}, {Whitehorn}, {Williamson}, {Wu}, \&
  {Yefremenko}}]{reichardt21}
{Reichardt}, C.~L., {Patil}, S., {Ade}, P.~A.~R., {et~al.} 2021,
  \bibinfo{title}{{An Improved Measurement of the Secondary Cosmic Microwave
  Background Anisotropies from the SPT-SZ + SPTpol Surveys},} \apj, 908, 199,
  \dodoi{10.3847/1538-4357/abd407}

\bibitem[{A.~G. {Riess} {et~al.}(2022){Riess}, {Yuan}, {Macri}, {Scolnic},
  {Brout}, {Casertano}, {Jones}, {Murakami}, {Anand}, {Breuval}, {Brink},
  {Filippenko}, {Hoffmann}, {Jha}, {D'arcy Kenworthy}, {Mackenty}, {Stahl}, \&
  {Zheng}}]{riess22}
{Riess}, A.~G., {Yuan}, W., {Macri}, L.~M., {et~al.} 2022, \bibinfo{title}{{A
  Comprehensive Measurement of the Local Value of the Hubble Constant with 1 km
  s$^{-1}$ Mpc$^{-1}$ Uncertainty from the Hubble Space Telescope and the SH0ES
  Team},} \apjl, 934, L7, \dodoi{10.3847/2041-8213/ac5c5b}

\bibitem[{S. {Roy Choudhury}(2025){Roy Choudhury}}]{roychoudhury25}
{Roy Choudhury}, S. 2025, \bibinfo{title}{{Cosmology in Extended Parameter
  Space with DESI Data Release 2 Baryon Acoustic Oscillations: A
  2{\ensuremath{\sigma}}+ Detection of Nonzero Neutrino Masses with an Update
  on Dynamical Dark Energy and Lensing Anomaly},} \apjl, 986, L31,
  \dodoi{10.3847/2041-8213/ade1cc}

\bibitem[{S. {Roy Choudhury} \& T. {Okumura}(2024){Roy Choudhury} \&
  {Okumura}}]{roychoudhury24}
{Roy Choudhury}, S., \& {Okumura}, T. 2024, \bibinfo{title}{{Updated
  Cosmological Constraints in Extended Parameter Space with Planck PR4, DESI
  Baryon Acoustic Oscillations, and Supernovae: Dynamical Dark Energy, Neutrino
  Masses, Lensing Anomaly, and the Hubble Tension},} \apjl, 976, L11,
  \dodoi{10.3847/2041-8213/ad8c26}

\bibitem[{M.~A. {Sabogal} \& R.~C. {Nunes}(2025){Sabogal} \&
  {Nunes}}]{sabogal25}
{Sabogal}, M.~A., \& {Nunes}, R.~C. 2025, \bibinfo{title}{{Robust Evidence for
  Dynamical Dark Energy from DESI Galaxy-CMB Lensing Cross-Correlation and
  Geometric Probes},} arXiv e-prints, arXiv:2505.24465.
\newblock \doarXiv{2505.24465}

\bibitem[{N. {Sailer} {et~al.}(2025){Sailer}, {Farren}, {Ferraro}, \&
  {White}}]{sailer25}
{Sailer}, N., {Farren}, G.~S., {Ferraro}, S., \& {White}, M. 2025,
  \bibinfo{title}{{Dispu$τ$able: the high cost of a low optical depth},} arXiv
  e-prints, arXiv:2504.16932, \dodoi{10.48550/arXiv.2504.16932}

\bibitem[{N. {Sailer} {et~al.}(2023){Sailer}, {Ferraro}, \&
  {Schaan}}]{sailer23}
{Sailer}, N., {Ferraro}, S., \& {Schaan}, E. 2023,
  \bibinfo{title}{{Foreground-immune CMB lensing reconstruction with
  polarization},} \prd, 107, 023504, \dodoi{10.1103/PhysRevD.107.023504}

\bibitem[{B.~O. {S{\'a}nchez} {et~al.}(2022){S{\'a}nchez}, {Kessler},
  {Scolnic}, {Armstrong}, {Biswas}, {Bogart}, {Chiang}, {Cohen-Tanugi},
  {Fouchez}, {Gris}, {Heitmann}, {Hlo{\v{z}}ek}, {Jha}, {Kelly}, {Liu},
  {Narayan}, {Racine}, {Rykoff}, {Sullivan}, {Walter}, {Wood-Vasey}, \& {LSST
  Dark Energy Science Collaboration (DESC)}}]{sanchez22}
{S{\'a}nchez}, B.~O., {Kessler}, R., {Scolnic}, D., {et~al.} 2022,
  \bibinfo{title}{{SNIa Cosmology Analysis Results from Simulated LSST Images:
  From Difference Imaging to Constraints on Dark Energy},} \apj, 934, 96,
  \dodoi{10.3847/1538-4357/ac7a37}

\bibitem[{E. {Schaan} \& S. {Ferraro}(2019){Schaan} \& {Ferraro}}]{schaan19}
{Schaan}, E., \& {Ferraro}, S. 2019, \bibinfo{title}{{Foreground-Immune Cosmic
  Microwave Background Lensing with Shear-Only Reconstruction},} \prl, 122,
  181301, \dodoi{10.1103/PhysRevLett.122.181301}

\bibitem[{N. {Sch{\"o}neberg} {et~al.}(2019){Sch{\"o}neberg}, {Lesgourgues}, \&
  {Hooper}}]{schoneberg19}
{Sch{\"o}neberg}, N., {Lesgourgues}, J., \& {Hooper}, D.~C. 2019,
  \bibinfo{title}{{The BAO+BBN take on the Hubble tension},} \jcap, 2019, 029,
  \dodoi{10.1088/1475-7516/2019/10/029}

\bibitem[{S. {Shaikh} {et~al.}(2024){Shaikh}, {Harrison}, {van Engelen},
  {Marques}, {Abbott}, {Aguena}, {Alves}, {Amon}, {An}, {Bacon}, {Battaglia},
  {Becker}, {Bernstein}, {Bertin}, {Blazek}, {Bond}, {Brooks}, {Burke},
  {Calabrese}, {Rosell}, {Carretero}, {Cawthon}, {Chang}, {Chen}, {Choi},
  {Choi}, {da Costa}, {Pereira}, {Darwish}, {Davis}, {Desai}, {Devlin},
  {Diehl}, {Doel}, {Doux}, {Elvin-Poole}, {Farren}, {Ferraro}, {Ferrero},
  {Fert{\'e}}, {Flaugher}, {Frieman}, {Garc{\'\i}a-Bellido}, {Gatti},
  {Giannini}, {Giardiello}, {Gruen}, {Gruendl}, {Gutierrez}, {Hill}, {Hinton},
  {Hollowood}, {Honscheid}, {Huffenberger}, {Huterer}, {James}, {Jarvis},
  {Jeffrey}, {Jense}, {Knowles}, {Kim}, {Kramer}, {Lahav}, {Lee}, {Lima},
  {MacCrann}, {Madhavacheril}, {Marshall}, {McCullough}, {Mehta},
  {Mena-Fern{\'a}ndez}, {Miquel}, {Mohr}, {Moodley}, {Myles}, {Navarro-Alsina},
  {Newburgh}, {Niemack}, {Omori}, {Pandey}, {Partridge}, {Pieres},
  {Malag{\'o}n}, {Porredon}, {Prat}, {Qu}, {Robertson}, {Rollins}, {Roodman},
  {Samuroff}, {S{\'a}nchez}, {Sanchez}, {Sanchez Cid}, {Secco}, {Sehgal},
  {Sheldon}, {Sherwin}, {Shin}, {Sif{\'o}n}, {Smith}, {Suchyta}, {Swanson},
  {Tarle}, {Troxel}, {Tutusaus}, {Vargas}, {Weaverdyck}, {Wiseman}, {Yamamoto},
  {Zuntz}, {(The ACT}, \& {DES Collaborations)}}]{shaikh24}
{Shaikh}, S., {Harrison}, I., {van Engelen}, A., {et~al.} 2024,
  \bibinfo{title}{{Cosmology from cross-correlation of ACT-DR4 CMB lensing and
  DES-Y3 cosmic shear},} \mnras, 528, 2112, \dodoi{10.1093/mnras/stad3987}

\bibitem[{C.~L. {Steinhardt} {et~al.}(2025){Steinhardt}, {Phillips}, \&
  {Wojtak}}]{steinhardt25}
{Steinhardt}, C.~L., {Phillips}, P., \& {Wojtak}, R. 2025,
  \bibinfo{title}{{Dark Energy Constraints and Joint Cosmological Inference
  from Mutually Inconsistent Observations},} arXiv e-prints, arXiv:2504.03829,
  \dodoi{10.48550/arXiv.2504.03829}

\bibitem[{M. {Tegmark} {et~al.}(1997){Tegmark}, {Taylor}, \&
  {Heavens}}]{tegmark97}
{Tegmark}, M., {Taylor}, A.~N., \& {Heavens}, A.~F. 1997,
  \bibinfo{title}{{Karhunen-Lo{\`e}ve Eigenvalue Problems in Cosmology: How
  Should We Tackle Large Data Sets?},} \apj, 480, 22, \dodoi{10.1086/303939}

\bibitem[{ {The CMB-HD Collaboration} {et~al.}(2022){The CMB-HD Collaboration},
  {Aiola}, {Akrami}, {Basu}, {Boylan-Kolchin}, {Brinckmann}, {Bryan}, {Casey},
  {Chluba}, {Clesse}, {Cyr-Racine}, {Di Mascolo}, {Dicker}, {Essinger-Hileman},
  {Farren}, {Fedderke}, {Ferraro}, {Fuller}, {Galitzki}, {Gluscevic}, {Grin},
  {Han}, {Hasselfield}, {Hlozek}, {Holder}, {Hotinli}, {Jain}, {Johnson},
  {Johnson}, {Klaassen}, {MacInnis}, {Madhavacheril}, {Mandal}, {Mauskopf},
  {Meerburg}, {Meyers}, {Miranda}, {Mroczkowski}, {Mukherjee}, {Munchmeyer},
  {Munoz}, {Naess}, {Nagai}, {Namikawa}, {Newburgh}, {Nguyen}, {Niemack},
  {Oppenheimer}, {Pierpaoli}, {Raghunathan}, {Schaan}, {Sehgal}, {Sherwin},
  {Simon}, {Slosar}, {Smith}, {Spergel}, {Switzer}, {Trivedi}, {Tsai}, {van
  Engelen}, {Wandelt}, {Wollack}, \& {Wu}}]{sehgal22}
{The CMB-HD Collaboration}, {Aiola}, S., {Akrami}, Y., {et~al.} 2022,
  \bibinfo{title}{{Snowmass2021 CMB-HD White Paper},} arXiv e-prints,
  arXiv:2203.05728, \dodoi{10.48550/arXiv.2203.05728}

\bibitem[{ {The Simons Observatory Collaboration} {et~al.}(2018){The Simons
  Observatory Collaboration}, {Ade}, {Aguirre}, {Ahmed}, {Aiola}, {Ali},
  {Alonso}, {Alvarez}, {Arnold}, {Ashton}, {et~al.}}]{SO18}
{The Simons Observatory Collaboration}, {Ade}, P., {Aguirre}, J., {et~al.}
  2018, \bibinfo{title}{{The Simons Observatory: Science goals and forecasts},}
  ArXiv e-prints.
\newblock \doarXiv{1808.07445}

\bibitem[{ {The Simons Observatory Collaboration} {et~al.}(2025){The Simons
  Observatory Collaboration}, {Abitbol}, {Abril-Cabezas}, {Adachi}, {Ade},
  {Adler}, {Agrawal}, {Aguirre}, {Ahmed}, {Aiola}, {Alford}, {Ali}, {Alonso},
  {Alvarez}, {An}, {Arnold}, {Ashton}, {Atkins}, {Austermann}, {Azzoni},
  {Baccigalupi}, {Baleato Lizancos}, {Barron}, {Barry}, {Bartlett},
  {Battaglia}, {Battye}, {Baxter}, {Bazarko}, {Beall}, {Bean}, {Beck},
  {Beckman}, {Begin}, {Beheshti}, {Beringue}, {Bhandarkar}, {Bhimani},
  {Bianchini}, {Biermann}, {Biquard}, {Bixler}, {Boada}, {Boettger}, {Bolliet},
  {Bond}, {Borrill}, {Borrow}, {Braithwaite}, {Brien}, {Brown}, {Bruno},
  {Bryan}, {Bustos}, {Cai}, {Calabrese}, {Calafut}, {Carl}, {Carones},
  {Carron}, {Challinor}, {Chanial}, {Chen}, {Cheung}, {Chiang}, {Chinone},
  {Chluba}, {Cho}, {Choi}, {Chu}, {Clancy}, {Clark}, {Clarke}, {Clements},
  {Connors}, {Contaldi}, {Coppi}, {Corbett}, {Cothard}, {Coulton}, {Crowley},
  {Crowley}, {Cukierman}, {D'Ewart}, {Dachlythra}, {Datta}, {Day-Weiss}, {de
  Haan}, {Devlin}, {Di Mascolo}, {Dicker}, {Dober}, {Doux}, {Dow}, {Doyle},
  {Duell}, {Duff}, {Duivenvoorden}, {Dunkley}, {Dutcher}, {D{\"u}nner},
  {Edenton}, {El Bouhargani}, {Errard}, {Fabbian}, {Fanfani}, {Farren},
  {Fergusson}, {Ferraro}, {Flauger}, {Foster}, {Freese}, {Frisch}, {Frolov},
  {Fuller}, {Galitzki}, {Gallardo}, {Galvez Ghersi}, {Ganga}, {Gao}, {Garrido},
  {Gawiser}, {Gerbino}, {Giardiello}, {Gill}, {Gilles}, {Giri}, {Gleave},
  {Gluscevic}, {Goeckner-Wald}, {Golec}, {Gordon}, {Gralla}, {Gratton},
  {Green}, {Groh}, {Groppi}, {Guan}, {Gupta}, {Gu{\dh}mundsson}, {Hagstotz},
  {Hargrave}, {Haridas}, {Harrington}, {Harrison}, {Hasegawa}, {Hasselfield},
  {Haynes}, {Hazumi}, {He}, {Healy}, {Henderson}, {Hensley}, {Hertig},
  {Herv{\'\i}as-Caimapo}, {Higuchi}, {Hill}, {Hill}, {Hilton}, {Hilton},
  {Hincks}, {Hinshaw}, {Hlo{\v{z}}ek}, {Ho}, {Ho}, {Ho}, {Hoang}, {Hoh},
  {Hornecker}, {Hornsby}, {Hoshino}, {Hotinli}, {Huang}, {Huber}, {Hubmayr},
  {Huffenberger}, {Hughes}, {Idicherian Lonappan}, {Ikape}, {Irwin}, {Iuliano},
  {Jaffe}, {Jain}, {Jense}, {Jeong}, {Johnson}, {Johnson}, {Johnson}, {Jones},
  {Jost}, {Kaneko}, {Karpel}, {Kasai}, {Katayama}, {Keating}, {Keller},
  {Keskitalo}, {Kim}, {Kisner}, \& {Kiuchi}}]{ASO25}
{The Simons Observatory Collaboration}, {Abitbol}, M., {Abril-Cabezas}, I.,
  {et~al.} 2025, \bibinfo{title}{{The Simons Observatory: Science Goals and
  Forecasts for the Enhanced Large Aperture Telescope},} arXiv e-prints,
  arXiv:2503.00636, \dodoi{10.48550/arXiv.2503.00636}

\bibitem[{J. {Torrado} \& A. {Lewis}(2019){Torrado} \& {Lewis}}]{torrado19}
{Torrado}, J., \& {Lewis}, A. 2019, {Cobaya: Bayesian analysis in cosmology},,
  Astrophysics Source Code Library, record ascl:1910.019 \doeprint{1910.019}

\bibitem[{C. {Trendafilova}(2023){Trendafilova}}]{trendafilova23}
{Trendafilova}, C. 2023, \bibinfo{title}{{The impact of cross-covariances
  between the CMB and reconstructed lensing power},} \jcap, 2023, 071,
  \dodoi{10.1088/1475-7516/2023/10/071}

\bibitem[{A. {Vikman}(2005){Vikman}}]{vikman05}
{Vikman}, A. 2005, \bibinfo{title}{{Can dark energy evolve to the phantom?},}
  \prd, 71, 023515, \dodoi{10.1103/PhysRevD.71.023515}

\bibitem[{M. {Vincenzi} {et~al.}(2024){Vincenzi}, {Brout}, {Armstrong},
  {Popovic}, {Taylor}, {Acevedo}, {Camilleri}, {Chen}, {Davis}, {Lee},
  {Lidman}, {Hinton}, {Kelsey}, {Kessler}, {M{\"o}ller}, {Qu}, {Sako},
  {Sanchez}, {Scolnic}, {Smith}, {Sullivan}, {Wiseman}, {Asorey}, {Bassett},
  {Carollo}, {Carr}, {Foley}, {Frohmaier}, {Galbany}, {Glazebrook}, {Graur},
  {Kovacs}, {Kuehn}, {Malik}, {Nichol}, {Rose}, {Tucker}, {Toy}, {Tucker},
  {Yuan}, {Abbott}, {Aguena}, {Alves}, {Allam}, {Andrade-Oliveira}, {Annis},
  {Bacon}, {Bechtol}, {Bernstein}, {Brooks}, {Burke}, {Carnero Rosell},
  {Carretero}, {Castander}, {Conselice}, {da Costa}, {Pereira}, {Desai},
  {Diehl}, {Doel}, {Ferrero}, {Flaugher}, {Friedel}, {Frieman},
  {Garc{\'\i}a-Bellido}, {Gatti}, {Giannini}, {Gruen}, {Gruendl}, {Hollowood},
  {Honscheid}, {Huterer}, {James}, {Kuropatkin}, {Lahav}, {Lee}, {Lin},
  {Marshall}, {Mena-Fern{\'a}ndez}, {Menanteau}, {Miquel}, {Palmese}, {Pieres},
  {Plazas Malag{\'o}n}, {Porredon}, {Romer}, {Roodman}, {Sanchez}, {Sanchez
  Cid}, {Schubnell}, {Sevilla-Noarbe}, {Suchyta}, {Swanson}, {Tarle}, {To},
  {Walker}, {Weaverdyck}, \& {Yamamoto}}]{vincenzi24}
{Vincenzi}, M., {Brout}, D., {Armstrong}, P., {et~al.} 2024,
  \bibinfo{title}{{The Dark Energy Survey Supernova Program: Cosmological
  Analysis and Systematic Uncertainties},} \apj, 975, 86,
  \dodoi{10.3847/1538-4357/ad5e6c}

\bibitem[{A. {Vitrier} {et~al.}(2025){Vitrier}, {Fichman}, {Balkenhol},
  {Camphuis}, {Guidi}, {Khalife}, {Anderson}, {Ansarinejad}, {Archipley},
  {Benabed}, {Bender}, {Benson}, {Bianchini}, {Bleem}, {Bouchet}, {Bryant},
  {Campitiello}, {Carlstrom}, {Chang}, {Chaubal}, {Chichura}, {Chokshi},
  {Chou}, {Coerver}, {Crawford}, {Daley}, {de Haan}, {Dibert}, {Dobbs},
  {Doohan}, {Doussot}, {Dutcher}, {Everett}, {Feng}, {Ferguson}, {Ferree},
  {Foster}, {Galli}, {Gambrel}, {Gardner}, {Ge}, {Goeckner-Wald}, {Gualtieri},
  {Guns}, {Halverson}, {Hivon}, {Holder}, {Holzapfel}, {Hood}, {Hryciuk},
  {Huang}, {K{\'e}ruzor{\'e}}, {Knox}, {Korman}, {Kornoelje}, {Kuo}, {Levy},
  {Li}, {Lowitz}, {Lu}, {Lynch}, {Maniyar}, {Martsen}, {Menanteau}, {Millea},
  {Montgomery}, {Nakato}, {Natoli}, {Noble}, {Omori}, {Ouellette}, {Pan},
  {Paschos}, {Phadke}, {Pollak}, {Prabhu}, {Quan}, {Rahimi}, {Rahlin},
  {Reichardt}, {Rouble}, {Ruhl}, {Schiappucci}, {Silva Oliveira}, {Simpson},
  {Sobrin}, {Stark}, {Stephen}, {Tandoi}, {Thorne}, {Trendafilova}, {Umilta},
  {Vieira}, {Wan}, {Whitehorn}, {Wu}, {Young}, \& {Zebrowski}}]{vitrier25}
{Vitrier}, A., {Fichman}, K., {Balkenhol}, L., {et~al.} 2025,
  \bibinfo{title}{{Towards constraining cosmological parameters with SPT-3G
  observations of 25\% of the sky},} arXiv e-prints, arXiv:2510.24669,
  \dodoi{10.48550/arXiv.2510.24669}

\bibitem[{B. {Wang} {et~al.}(2020){Wang}, {Zhu}, {Li}, \& {Zhao}}]{wang20}
{Wang}, B., {Zhu}, Z., {Li}, A., \& {Zhao}, W. 2020,
  \bibinfo{title}{{Comprehensive Analysis of the Tidal Effect in Gravitational
  Waves and Implication for Cosmology},} \apjs, 250, 6,
  \dodoi{10.3847/1538-4365/aba2f3}

\bibitem[{J. {Zuntz} {et~al.}(2021){Zuntz}, {Lanusse}, {Malz}, {Wright},
  {Slosar}, {Abolfathi}, {Alonso}, {Bault}, {Bom}, {Brescia}, {Broussard},
  {Campagne}, {Cavuoti}, {Cypriano}, {Fraga}, {Gawiser}, {Gonzalez}, {Green},
  {Hatfield}, {Iyer}, {Kirkby}, {Nicola}, {Nourbakhsh}, {Park}, {Teixeira},
  {Heitmann}, {Kovacs}, {Mao}, \& {LSST Dark Energy Science
  Collaboration}}]{zuntz21}
{Zuntz}, J., {Lanusse}, F., {Malz}, A.~I., {et~al.} 2021, \bibinfo{title}{{The
  LSST-DESC 3x2pt Tomography Optimization Challenge},} The Open Journal of
  Astrophysics, 4, 13, \dodoi{10.21105/astro.2108.13418}

\end{thebibliography}
\bibliographystyle{aasjournalv7}
\end{document}